\documentclass[aps,prd,twocolumn]{revtex4}
\usepackage{epsfig}
\usepackage{axodraw}
\newcommand{\newc}{\newcommand}
\newc{\wt}{\widetilde}
\newc{\cL}{{\cal L}}
\newc{\ra}{\rightarrow}
\newc{\eps}{\epsilon}
\newc{\bino}{\widetilde{\cal B}}
\newc{\wino}{\widetilde{\cal W}}
\newc{\gluino}{\widetilde{\cal G}}
\newc{\half}{\frac{1}{2}}
\newc{\third}{\frac{1}{3}}
\newc{\fourth}{\frac{1}{4}}
\newc{\eighth}{\frac{1}{8}}
\newc{\gev}{\mbox{~GeV}}
\newc{\lra}{\leftrightarrow}
\newc{\Dslash}{\not\!\! D}
\newc{\sg}{{\cal G}}
\newc{\ovl}{\overline}
\newc{\ok}{$\surd$}
\newc{\etal}{{\it et al.}\ }
\newc{\Hbar}{{\bar H}}
\newc{\hhbar}{{\overline h}}
\newc{\Ubar}{{\bar U}}
\newc{\Dbar}{{\bar D}}
\newc{\Ebar}{{\bar E}}
\newc{\eg}{{\it e.g.}\ }
\newc{\ie}{{\it i.e.}\ }
\newc{\nonum}{\nonumber}
\newc{\kap}{\kappa}
\newc{\Dt}{\frac{d}{dt}}
\newc{\rpv}{{\mbox{${\not\!\!R_p}$}}}
\newc{\bpv}{$\not\!\!B_p$}
\newc{\mpl}{$M_{Pl}$\ }
\newc{\mx}{$M_X$\ }
\newc{\tev}{\mbox{~TeV}}
\newc{\sect}[1]{\ref{sec:#1}}
\newc{\nonr}{\nonum}
\newc{\vev}[1]{\langle{#1}\rangle}
\newc{\eq}[1]{(\ref{eq:#1})}
\newc{\eqs}[2]{(\ref{eq:#1},\ref{eq:#2})}
\newc{\lab}[1]{\label{eq:#1}}
\newc{\Lam}{{\bf \Lambda}}
\newc{\ltau}{\lambda_\tau}
\newc{\lt}{\lambda_t}
\newc{\lb}{\lambda_b}
\newc{\lae}{{\Lam}_E}
\newc{\lad}{{\Lam}_D}
\newc{\lau}{{\Lam}_U}
\newc{\lame}[1]{{\Lam}_{E^{#1}}}
\newc{\lamhe}[1]{{\h}_{E^{#1}}}
\newc{\lamhed}[1]{{\h}_{E^{#1}}^\dagger}
\newc{\lamhd}[1]{{\h}_{D^{#1}}}
\newc{\lamhdd}[1]{{\h}_{D^{#1}}^\dagger}
\newc{\lamhu}[1]{{\h}_{U^{#1}}}
\newc{\lamhud}[1]{{\h}_{U^{#1}}^\dagger}
\newc{\lamd}[1]{{\Lam}_{D^{#1}}}
\newc{\lamu}[1]{{\Lam}_{U^{#1}}}
\newc{\lamet}[1]{{\Lam}_{E^{#1}}^T}
\newc{\lamdt}[1]{{\Lam}_{D^{#1}}^T}
\newc{\lamut}[1]{{\Lam}_{U^{#1}}^T}
\newc{\lames}[1]{{\Lam}_{E^{#1}}^*}
\newc{\lamds}[1]{{\Lam}_{D^{#1}}^*}
\newc{\lamus}[1]{{\Lam}_{U^{#1}}^*}
\newc{\lamed}[1]{{\Lam}_{E^{#1}}^\dagg}
\newc{\lamdd}[1]{{\Lam}_{D^{#1}}^\dagg}
\newc{\lamud}[1]{{\Lam}_{U^{#1}}^\dagg}
\newc{\lam}{{\bf \lambda}}
\newc{\lamp}{{\bf \lambda}^{\prime}}
\newc{\lampp}{{\bf \lambda}^{\prime\prime}}
\newc{\Y}{{\bf Y}}
\newc{\h}{{\bf h}}
\newc{\meee}{{{\rm {\bf  m}}_e}}
\newc{\mdee}{{{\rm {\bf  m}}_d}}
\newc{\myew}{{{\rm {\bf m}}_u}}
\newc{\ye}{{\Y}_E}
\newc{\he}{{\h}_E}
\newc{\hed}{{\h}_E^\dagger}
\newc{\yd}{{\Y}_D}
\newc{\hd}{{\h}_D}
\newc{\hdd}{{\h}_D^\dagger}
\newc{\yu}{{\Y}_U}
\newc{\hu}{{\h}_U}
\newc{\hud}{{\h}_U^\dagger}
\newc{\yes}{{\Y}_E^*}
\newc{\yds}{{\Y}_D^*}
\newc{\yus}{{\Y}_U^*}
\newc{\yet}{{\Y}_E^T}
\newc{\ydt}{{\Y}_D^T}
\newc{\yut}{{\Y}_U^T}
\newc{\yed}{{\Y}_E^\dagg}
\newc{\ydd}{{\Y}_D^\dagg}
\newc{\yud}{{\Y}_U^\dagg}
\newc{\dagg}{\dagger}
\newc{\lp}{\left(}
\newc{\rp}{\right)}
\newc{\inv}{\frac{1}{16\pi^2}}
\newc{\invsq}{\frac{1}{(16\pi^2)^2}}
\newc{\ggam}[2]{\gamma_{#2}^{#1}}
\newc{\yukgam}[2]{\inv \gamma_{#1}^{(1){#2}}+\invsq\gamma_{{#1}}^{(2){#2}}}
\newc{\susyunif}{ohman,nirpaul,marcelacarlos,susyunif}
\newc{\lsim}{\stackrel{<}{\sim}}
\newc{\gsim}{\stackrel{>}{\sim}}
\newc{\Tr}{{~\rm Tr}}
\newc{\me}{{(\bf m_{\tilde{E}}}^2)}
\newc{\mh}[1]{m_{H_{#1}}^2}
\newc{\ml}{{(\bf m_{\tilde{L}}}^2)}
\newc{\md}{{(\bf m_{\tilde{D}}}^2)}
\newc{\mup}{{(\bf m_{\tilde{U}}}^2)}
\newc{\mq}{{(\bf m_{\tilde{Q}}}^2)}
\newc{\mlh}[1]{({\bf m}_{ \tilde{L}_{#1} H_1}^2)}
\newc{\mhl}[1]{({\bf m}_{ H_1 \tilde{L}_{#1}}^2)}
\newc{\del}{\partial}
\newc{\beq}{\begin{equation}}
\newc{\eeq}{\end{equation}}
\newc{\barr}{\begin{eqnarray}}
\newc{\earr}{\end{eqnarray}}
\newc{\dspl}{\displaystyle}
\newc{\phmin}{\phantom{-}}
\newc{\stau}{{\tilde\tau}}

\begin{document}

\title{The  R-Parity Violating Minimal  
Supergravity Model\footnote{{Preprint number : TUM-522/03, LAPTH-997/TH,
BONN-TH-2003-04}}}

\author{B.~C.~Allanach\footnote{Current permanent address: 
DAMTP, CMS, Wilberforce Road, Cambridge, CB3 0WA, UK}}

\affiliation{LAPTH, 9 chemin de Bellevue, Annecy 74941, France}

\author{A.~Dedes\footnote{{Permanent address after 1$^{\rm st}$ October 2003: 
Institute for Particle Physics Phenomenology, University of Durham, 
DH1 3LE, UK }}}

\affiliation{Technische Universit\"at M\"unchen, Physik Department,
  D-85748 Garching, Germany}

\author{H.~K.~Dreiner}

\affiliation{Physikalisches Institut der Universit\"at Bonn, Nussallee
  12, D-53115 Bonn, Germany}

\begin{abstract}
We present the minimal supersymmetric standard model with general
broken R-parity, focusing on minimal supergravity (mSUGRA). We discuss the
origins of lepton number violation in supersymmetry.  We have computed
the full set of coupled one-loop renormalization group equations for
the gauge couplings, the superpotential parameters and for all the
soft supersymmetry breaking parameters. We provide analytic
formul{\ae} for the scalar potential minimization conditions which may
be iterated to arbitrary precision.  We compute the low-energy
spectrum of the superparticles and the neutrinos as a function of the
small set of parameters at the unification scale in the general
basis. Specializing to mSUGRA, we use the neutrino masses to set new
bounds on the R-parity violating couplings. These bounds are up-to
five orders of magnitude stricter than the previously existing
ones. In addition, new bounds on the R-parity violating couplings are
also derived demanding a non-tachyonic sneutrino spectrum.  We
investigate the nature of the lightest supersymmetric particle and
find extensive regions in parameter space, where it is {\it not} the
neutralino. This leads to a novel set of supersymmetric signatures,
which we classify.
\end{abstract}

\maketitle

\section{Introduction}
\label{intro}
The most widely studied supersymmetric scenario is the minimal
supersymmetric standard model (MSSM) with {\it conserved} R-parity
\cite{Nilles:1983ge,Kane,Martin:1997ns}.  The unification of the three
Standard Model gauge couplings, $g_i$, at the scale $M_X={\cal
  O}(10^{16}\gev)$ \cite{Amaldi:1991cn}, is a strong indication that
supersymmetry (SUSY) is embedded in a unified model.  In the simplest
such model \cite{Nilles:1983ge}, SUSY breaking occurs in a hidden
sector (decoupled from the Standard Model gauge interactions), and is
communicated to our visible sector via gravity \cite{foot1}.  The
scale of SUSY breaking in the visible sector is thus the Planck scale,
$M_P=10^{19}\gev$.

The large number of parameters in the MSSM is restricted by making
well-motivated simplifying assumptions at the unification scale.  In
the special case of the minimal supergravity model (mSUGRA), there are
five parameters beyond those of the Standard Model:
\begin{equation}
M_0,\;M_{1/2},\;A_0,\,\tan\beta,\,sgn(\mu)\;.
\label{parameters}
\end{equation}
These are the universal scalar mass, $M_0$, gaugino mass, $M_{1/2}$,
and trilinear scalar coupling, $A_0$, respectively, as well as the ratio
of the Higgs vacuum expectation values (vev's), $\tan\beta$, and the
sign of the bi-linear Higgs mixing parameter, $\mu$. Given these 5
parameters at the unification scale, we can predict the full mass
spectrum as well as the couplings of the particles at the weak scale
via the supersymmetric renormalization group equations (RGEs). This
is the most widely used model for extensive phenomenological and
experimental tests of supersymmetry.  It is the purpose of this paper
to create an analogous model in the case of supersymmetry with {\it
broken} R-parity (\rpv): the R-parity violating minimal supergravity model
(\rpv-mSUGRA).

The mSUGRA model with universal boundary conditions was first extended
to include bi-linear \rpv\ by Hempfling~\cite{Hempfling:1995wj},
focusing on the neutrino sector. A further detailed analysis in this
framework was performed by Hirsch {\it et al.} \cite{Hirsch:2000ef}. 
de Carlos and White were the first to go beyond bi-linear \rpv\ and
consider the full set of \rpv-couplings \cite{deCarlos:1996du,CW}.
However, they restricted themselves to the third generation
Higgs-Yukawa couplings and used an approximate method to minimize the
scalar potential. We detail below how we go beyond this work.

We shall consider the chiral superfield particle content
\begin{equation}
Q_i^x,\;{\bar D}_i^x,\;{\bar U}_i^x,\;L_i^a,\;{\bar E}_i,\;
H_1^a,\;H_2^a\,.
\end{equation}
Here $i=1,2,3$ is a generation index, $x=1,2,3$ and $a=1,2$ are $SU(3)
$ and $SU(2)$ gauge indices, respectively.  In supersymmetry, the
lepton doublet superfields $L_i^a$ and the Higgs doublet superfield
coupling to the down-like quarks, $H_1$, have the same gauge and
Lorentz quantum numbers (This is an essential feature in our
discussion below.).  When appropriate we shall combine them into the
chiral superfields ${\cal L}^a_{\alpha=0,\ldots,3}=(H_1^a,\,L_{i=1,2,
3}^a)$.  The gauge quantum numbers of the chiral superfields and the
vector superfields are given in Table~\ref{table1}.

\begin{table*}[t]
\caption{
The particle content of the mSUGRA \rpv-model in terms of superfields 
and their decomposition into components with their $SU(3)_c\times 
SU(2)_L\times U(1)_Y$ quantum numbers. $x,X$ are SU(3) representation
and generator indices, $a,A$ are SU(2) representation and generator 
indices. $\alpha=0,..,3$ is the family index of the lepton superfield,
and $i=1,..3$ the usual family index of quarks, leptons and their 
superpartners. The fermionic components of the superfields are two 
component Weyl spinors.}
\begin{tabular}{|c|c|c|}
\hline
Chiral Superfields & $SU(3)_c\times SU(2)_L\times U(1)_Y$ & Components 
\\ \hline
$Q^{a,x}_i$ & $(\bf{3},{\bf{2}},\frac{1}{6})$ & 
$\biggl ($ \begin{tabular}{c}  $\tilde{u}_L$ \\ $\tilde{d}_L$
\end{tabular}
$\biggr )$
$\; ,\;$ $\biggl ($ \begin{tabular}{c}  $u_L$ \\ $d_L$
\end{tabular}$\biggr )$
\\[4mm] 
$\bar{D}^{x}_i$ &  $(\bf{\bar{3}},{\bf{1}},\frac{1}{3})$ &
$\tilde{d}_R^* \;,\; d_R$ \\[2mm] 
$\bar{U}^{x}_i$ &  $(\bf{\bar{3}},{\bf{1}},-\frac{2}{3})$ &
$\tilde{u}_R^* \;,\; u_R$ \\[2mm] 
$\cL_\alpha^{a}=\{H_1^a,L_i^a\}$ & $(1,{\bf{2}},-\frac{1}{2})$ & 
$\biggl ($ \begin{tabular}{c}  $\tilde{\nu}_\alpha$ \\ $\tilde{e}_{L\alpha}$
\end{tabular}
$\biggr )= \biggl \{ \biggl ($ \begin{tabular}{c} 
$h_1^0$ \\ $h_1^-$ \end{tabular} 
$\biggr )\;,\; \biggl ($
\begin{tabular}{c} $\tilde{\nu}_i$ \\ $\tilde{e}_{Li}$ \end{tabular}
$\biggr ) \biggr \}$
 $\;,\;$ $ \biggl ($ 
\begin{tabular}{c}  $\nu_\alpha$ \\ $e_{L\alpha}$
\end{tabular}$\biggr )=\biggl \{ \biggl ($ \begin{tabular}{c} 
$\tilde{h}_1^0$ \\ $\tilde{h}_1^-$ \end{tabular} 
$\biggr )\;,\; \biggl ($
\begin{tabular}{c} $\nu_i$ \\ $e_{Li}$ \end{tabular}
$\biggr ) \biggr \}$
\\[4mm] 
$\bar{E}_i$ &  $(\bf{{1}},{\bf{1}},1)$ &
$\tilde{e}_R^* \;,\; e_R$ \\[2mm]
$H_2^a$ & $(\bf{1},{\bf{2}},\frac{1}{2})$ & 
$\biggl ($ \begin{tabular}{c}  $h_2^+$ \\ $h_2^0$
\end{tabular}
$\biggr )$
$\; ,\;$ $\biggl ($ \begin{tabular}{c}  $\tilde{h}_2^+$ \\ 
$\tilde{h}_2^0$ \end{tabular} $\biggr )$
\\[2mm] \hline
Vector Superfields & $SU(3)_c\times SU(2)_L\times U(1)_Y$ & Components 
\\ \hline 
$V_1$ & $(\bf{1},{\bf{1}},0)$ & $\bino$ $\;,\,$ $B_\mu$ \\[3mm]
$V_2$ & $(\bf{1},{\bf{3}},0)$ & $\wino^{(A)}$ 
$\;,\,$ $W^{(A)}_\mu$ \\[3mm]
$V_3$ & $(\bf{8},{\bf{1}},0)$ & $\gluino^{(X)}$ $\;,\,$ 
$G^{(X)}_\mu$ \\ \hline
\end{tabular}
\label{table1}
\end{table*}

\subsection{R-Parity Violation}

R-parity is defined as the discrete multiplicative symmetry{~\cite{Fayet}}
\begin{equation}
R_p=({\bf -1})^{2S+3B+L}\;,
\end{equation}
where $S$ is the spin, $B$ the baryon number and $L$ the lepton number
of the particle. All Standard Model particles including the two scalar
Higgs doublets have $R_p=+1$, their superpartners have $R_p=-1$.  When
allowing for R-parity violation, the full renormalizable
superpotential is given by{~\cite{Weinberg81}}
\begin{widetext}
\begin{eqnarray} 
W&=& \eps_{ab} \left[ (\ye)_{ij} L_i^a
H_1^b {\bar E}_j + (\yd)_{ij} Q_i^{ax} H_1^b {\bar D}_{jx} +
(\yu)_{ij} Q_i^{ax} H_2^b {\bar U}_{jx} \right]\nonr \\ &&+
\eps_{ab}\left[ \frac{1}{2} \lam_{ijk} L_i^a L_j^b{\bar E}_k +
\lamp_{ijk} L_i^a Q_j^{xb} {\bar D}_{kx} \right]+
\frac{1}{2}\eps_{xyz} \lampp_{ijk} {\bar U}_i^x{\bar
D}_j^y{\bar D}^z_k 
- \eps_{ab} \left[\mu H_1^a H_2^b
+\kap^i L_i^a H_2^b \right].  \label{superpot1} 
\end{eqnarray} 
\end{widetext}
Here {\bf Y}$_{E,D,U}$ are $3\times3$ matrices of Yukawa couplings;
$\lam_{ijk},\,\lam'_{ijk},\,\lam''_{ijk}$ are Yukawa couplings and
$\kap_i$ are mass-dimension one parameters.  $\eps_{ab}$ and $\eps
_{xyz}$ are the totally anti-symmetric tensors, with $\eps_{12}=\eps
_{123}=+1$. The terms proportional to $\lam,\,\lam',\,\lam'',$ and
$\kap_i$ violate R-parity explicitly and it is their effect that we
investigate in detail in this paper.  The terms proportional to
$\lam''$ violate baryon-number, whereas the terms proportional to
$\lam,\,\lam',$ and $\kap_i$ violate lepton number. Baryon- and
lepton-number violation can not be simultaneously present in the
theory, otherwise the proton will decay rapidly
\cite{Dreiner:1997uz,Smirnov:1995ey}. We discuss in detail in
Sect.~\ref{origins} how this can be guaranteed.

When extending mSUGRA to allow for R-parity violation, the particle
content remains the same but we have additional interactions in the
superpotential, Eq.~(\ref{superpot1}), as well as the soft-breaking
scalar potential ({\it c.f.} Eq.~(\ref{softhaber})). Thus within the
\rpv-mSUGRA the RGEs must be modified.  The running of the gauge
couplings is only affected at the two-loop level and the effects have
been discussed in Ref.~\cite{add}. Ref.~\cite{add} also contains the
\rpv\ two-loop RGEs for the superpotential parameters. Here we
restrict ourselves to the one-loop RGEs. In order to fix the notation,
we present the RGEs for the superpotential couplings as well as the
gauge couplings in Appendix~\ref{app2}.  Due to the flavour indices
the RGEs for the soft supersymmetry breaking terms are highly coupled
to each other. In Appendix~\ref{app3}, we discuss a very elegant
method developed by Jack and Jones \cite{jones} to derive the full set
of RGEs for the soft-supersymmetry breaking terms and apply it to the
case of the \rpv-mSUGRA. As we discuss, Jack and Jones' method is more
easily implemented in a numerical computation.  We also independently
calculate the $\beta$-functions of the theory by using the formulae
from Ref.~\cite{mv}. The resulting RGEs for the soft-supersymmetry
breaking terms are given explicitly in Appendix~\ref{app4}.  We have
checked that our results for the $\beta$-functions in Appendices C and
D are in full agreement. Furthermore, where relevant, they agree with
previous (subsets of) results which have been computed by the standard
method \cite{deCarlos:1996du,Barger:1995qe}.

Given the RGEs, we can compute the full model at the weak scale,
including the mass spectrum and the couplings of all the particles as
a function of our unified scale ($M_X$) boundary conditions. In our
numerical results for the \rpv-mSUGRA, we extend the parameters given
in Eq.~(\ref{parameters}) by only {\it one} \rpv-coupling. We thus
have
\begin{equation}
\{\lam,\lam',\lam''\}_1,\;M_0,\;M_{1/2},\;A_0,\,\tan\beta,\,sgn(\mu)\;,
\label{rpv-parameters}
\end{equation}
as our free parameters at $M_X$. $\{\lam,\lam',\lam''\}_1$ indicates
that only one \rpv-coupling is non-zero at $M_X$. We note that through
the coupled RGEs many couplings can be non-zero at $M_Z$ and this
is taken into account in the numerical implementation of our RGEs.

Due to existing experimental bounds on the $(\lam,\lam',\lam'')$
\cite{Allanach:1999ic,Bhattacharyya:1996nj}, the couplings are 
typically small and we thus expect the deviations from mSUGRA due to
\rpv\ to be small. However, besides the RGEs discussed above there are
four important aspects where there are significant changes and which
we dicuss in detail in this paper: {\it (i)}~the origin of lepton
number violation, {\it (ii)}~minimizing the scalar potential, {\it
(iii)}~neutrino masses, and {\it (iv)}~the nature of the lightest
supersymmetric particle (LSP).

\begin{itemize}
  
\item[{\it (i)}] Since the discovery of neutrino oscillations, we know
  that lepton {\it flavour} is violated. If the observed neutrinos
  have Majorana masses, then lepton {\it number} is violated as well.
  In the \rpv-MSSM, lepton number is naturally violated in the
  superpotential by the Yukawa couplings $(\lam,\lam')$ as well as the
  mass terms $\kap_i$. In Sect.~\ref{origins}, we discuss the origin
  of these terms in high energy unified theories and argue that they
  are just as well motivated as in the R-parity conserving case. For
  this we reanalyze the seminal work on $Z_2$ and $Z_3$ discrete gauge
  symmetries by Ibanez and Ross \cite{Ibanez:1991pr}. We find a
  slightly different set of allowed operators, but the conclusions
  remain the same.
  
  We argue that within supergravity, with gravity mediated
  supersymmetry breaking, it is natural to have both $\kap_i=0$ and
  ${\widetilde D}_i=0$ at the unification scale, $M_X$. This has not
  been taken into account in previous \rpv-RGE studies. (Here
  ${\widetilde D}_i$ is the to $\kap_i$ corresponding soft
  supersymmetry breaking bi-linear term, {\it c.f.}
  Eq.~(\ref{softhaber}).) This reduces the number of parameters we
  must consider to the set given in Eq.~(\ref{rpv-parameters}). At the
  weak-scale however, in general $\kap_i,\,{\tilde D}_i\not=0$, but
  these are then derived quantities.

\item[{\it (ii)}] Since the lepton doublet superfields $L^a_i$ have 
the same gauge and Lorentz quantum numbers as the down-like Higgs
doublet $H_1$, we effectively have a five Higgs doublet model for
which we must minimize the scalar potential. Within our RGE framework,
this must be done in a consistent approach while maintaining the value
of $\tan\beta$ given at the weak scale and also obtaining the correct
radiative electroweak symmetry breaking \cite{REWSB}. In
Ref.~\cite{Hirsch:2000ef} (bi-linear R-parity violation), points were
tested to see if they minimize the potential for the case (in their
notation) ${\widetilde B}= {\widetilde D}_i=A_0-1$. We have directly
minimized the potential and do not make the latter additional
assumption. Instead, we determine ${\widetilde B},\,{\widetilde D}_i$
via electroweak radiative breaking. If we obtain a point with
radiative breaking of colour or electric charge we disregard it. We
also go beyond the numerical approximations made in
Ref.~\cite{deCarlos:1996du} to obtain the full result. The technical
details of the iterative procedure are given in Sect.~\ref{rad-break}.

\item[{\it (iii)}] Due to the coupled \rpv-RGEs, a non-zero $\lam$ or
  $\lam'$ together with $\mu(M_X)\not=0$ will generate non-zero $\kap_
  i$'s at the weak scale
  \cite{Dreiner:1995hu,deCarlos:1996du,Nardi:1996iy}. The $\kap _i$'s
  lead to mixing between the neutrinos and neutralinos resulting in
  one non-zero neutrino mass at tree-level
  \cite{Hall:1983id,Ellis:gi,Banks:1995by}. Thus one or more
  non-vanishing $(\lam,\lam')$ at $M_X$ will result in one massive
  neutrino at the weak-scale via the RGEs and the $\kap_i$. Requiring
  this neutrino to be less than the cosmological bound on the sum of
  the neutrino masses determined by the WMAP collaboration \cite{wmap}
  using their data combined with the 2dFGRS data \cite{2dFGRS}
\begin{equation}
\sum m_{\nu_i} < 0.71\, {\rm eV}\,,
\label{eq:wmap}
\end{equation}
thus gives a bound on the $(\lam,\lam')$ at $M_X$. These bounds are
determined in Sect.~\ref{bounds} and are very strict for the specific
mSUGRA point SPS1a, but are fairly sensitive to the precise choice
of parameters at $M_X$. The bounds are summarized in
Table~\ref{tab:bounds}. In Refs.~\cite{deCarlos:1996du,Nardi:1996iy},
it was argued that such bounds exist, however no explicit bounds were
determined and the full flavour effects were also not considered. Here
we present for the first time a complete analysis of the corresponding
bounds. In a future publication we will address the possibility of
solving the atmospheric and solar neutrino problems within our
framework.

\item[{\it (iv)}] In the MSSM and mSUGRA the LSP is stable due to
conserved R-parity. It can thus have a significant cosmological relic
density \cite{Lee:1977ua,Goldberg:1983nd,Ellis:1983ew}.  Observational
bounds require the LSP to be charge and colour neutral
\cite{Ellis:1983ew} with a strong preference for the lightest
neutralino, ${\tilde\chi}^0_1$. In \rpv, the LSP is not stable and
thus not constrained by the observational bounds on relic
particles~\cite{foot2}. Therefore any supersymmetric particle can be
the LSP:
\begin{equation}
{\tilde {\cal G}},\;{\tilde\chi}^0_1,\;{\tilde\chi}_1^\pm,\;
{\tilde q}_{i=1,\dots,6},\;{\tilde\ell}_{i=1,\ldots,6}^\pm,\;
{\tilde\nu}_{j=1,2,3}\;,
\label{lsp-s}
\end{equation}
where ${\tilde\chi}^0_1,\,{\tilde\chi}_1^\pm$ denote the lightest
neutralino and chargino, and ${\tilde q}_i,\,{\tilde\ell}_i^\pm,\,{\tilde
\nu}_j,$ denote the right- and left-handed squarks, and charged sleptons 
as well as the left-handed sneutrinos, respectively.

Depending on the nature of the LSP, the collider phenomenology will be
completely different \cite{Dreiner:pe}. It is not feasible to study the
full range of signatures resulting from the different possible LSPs in
Eq.~(\ref{lsp-s}) or the different possible mass spectra. It is thus
mandatory to have a well motivated mass spectrum, including the LSP,
as in the MSSM and mSUGRA. Below in Sect.~\ref{results}, we determine
the nature of the LSP as well as the rest of the mass spectrum as a
function of our input parameters.  In the no-scale supergravity
models, we find significant ranges where the $\stau$ is the
LSP. In Sect.~\ref{non-chi-lsp} we discuss the phenomenology of a
$\stau$ LSP. 

The case of a stau LSP has to our knowledge first been discussed in
Ref.~\cite{Akeroyd:1997iq}, in the framework of third generation
bi-linear R-parity violation. In Ref.~\cite{Akeroyd:2001pm} the case
of tri-linear \rpv\ was considered, with the focus on the comparison
between charged Higgs and stau-LSP phenomenology. We go beyond this to
present a systematic analysis of all possible stau decays depending on
the dominant \rpv-coupling and classify the resulting signatures. For
a recent analysis on charged slepton LSP decays in the presence of
trilinear or bilinear \rpv-couplings, see Ref.~\cite{Porod}. There
only two-body decays are considered and the parameters are restricted
to the simultaneous solution of the solar and atmospheric neutrino
problems. In Sect.~\ref{non-chi-lsp} we present the general analysis.

Very recently, in Ref.~\cite{hirsch-porod}, the nature of the LSP in
correlation with the neutrino properties is studied in bi-linear \rpv,
\ie the tri-linear couplings $\lam_{ijk},\,\lam^ \prime_{ijk},\,\lam^{
  \prime\prime}_{ijk}$ are all set to zero by hand. Since also the
dependence on the supersymmetry breaking parameters is not the focus
of investigation, this work is complementary to our's.

A stau LSP with R-parity {\it conservation} on the lab scale, {\it
i.e.}  the stau is stable in collider experiments, has been discussed
in Ref.~\cite{deGouvea:1998yp}. For completeness we mention that
within R-parity conservation several authors have considered the case
of a gluino LSP \cite{gluino}.

\end{itemize}

\subsection{Outline}
In Sect.~\ref{origins}, we present the motivation for supersymmetry
with broken R-parity and discuss the possible origins of baryon- and
lepton-number violation. We focus in particular on the origin of the
$L_iH_2$ mixing. In Sect.~\ref{rpvmsugra}, we present the full set of
parameters and interactions in the mSUGRA model with broken R-parity,
including the SUSY breaking parameters. In Sect.~\ref{rad-break} we
discuss the radiative electroweak symmetry breaking including the full
minimization of the Higgs potential. In Sect.~\ref{sec:pspm} we
determine the complete mass spectrum as a function of our parameters.
In Sect.~\ref{boundary} we discuss the boundary conditions we impose
at $M_X$ and their numerical effect. In Sect.~\ref{results} we present
our main results including the bounds we obtain on the \rpv-Yukawa
couplings from the WMAP constraint on the neutrino masses.  In
Sect.~\ref{non-chi-lsp} we discuss the phenomenology of the stau LSP,
classifying possible final state signatures at colliders and computing
the stau decay length. We offer our summary and conclusions in
Sect.~\ref{summary}.

We present two methods for computing these equations in the Appendices
\ref{app2},\ref{app3} and \ref{app4}.  We present the complete set of
RGEs at one-loop in Appendix \ref{app4}. In Appendix D we compute the
four-body decay of the stau.

\section{Origins of Lepton- and Baryon-Number Violation}
\label{origins}

In this section we investigate general aspects of the origin of
baryon- and lepton-number violation in supersymmetry and thus the
motivation for R-parity violation \cite{Dreiner:1997uz}. We then
discuss in more detail the origin of the $\kap_iL_iH_2$ terms in the
context of only lepton-number violation.  In particular, for the
following, we would like to know under what conditions {\it after}
supersymmetry breaking we can rotate away both the $\kap_i L_iH_2$
terms and the corresponding soft breaking terms ${\widetilde
D}_i{\tilde L}_iH_2$.

\subsection{Discrete Symmetries}
\label{DS}

In the MSSM in terms of the resulting superpotential, R-parity is
equivalent to requiring invariance under the discrete symmetry matter
parity
\cite{Dreiner:1997uz}. If instead we require invariance under 
baryon-parity
\begin{eqnarray}
\begin{array}{ccl}
(Q,{\bar U},{\bar D}) &\longrightarrow & -(Q,{\bar U},{\bar D})\,,\\
(L,{\bar E},H_1,H_2)&\longrightarrow & \phmin (L,{\bar E},H_1,H_2)\,,
\end{array}
\end{eqnarray}
we allow for the terms $LL{\bar E}$, $LQ{\bar D}$, and $LH_2$ in the
superpotential, while maintaining a stable proton. Similarly lepton
parity only allows for the ${\bar U}{\bar D}{\bar D}$ terms. Thus when
allowing for a subset of R-parity violating interactions which ensure
proton stability, we must employ a discrete symmetry which treats
quark and lepton superfields differently.  In grand unified theories
(GUTs) this is unnatural, as we discuss below.

In string theories, we need not have a simple GUT gauge group. Thus
models exist for both lepton- and baryon-number violation
\cite{Bento:mu}, and there is no preference for $R_p$-conservation or
\rpv.  However, discrete symmetries can be problematic when gravity is
included. Unless it is a remnant of a broken gauge symmetry the
discrete symmetry will be broken by quantum gravity effects
\cite{Krauss:1988zc}. The requirement that the original gauge symmetry
be anomaly free, can be translated into a set of conditions on the
charges of the discrete symmetry \cite{Ibanez:1991pr,Banks:1991xj}. 
Considering the complete set of ${\bf Z}_{2}$ and ${\bf Z}_{3}$
discrete symmetries, and the particle content given in
Table~\ref{table1}, only the ${\bf Z}_ {2}$ symmetry R-parity, $R_2$,
and the ${\bf Z}_{3}$ symmetry $B_3=R_3L_3$ \cite{foot3} are
discrete gauge anomaly-free \cite{Ibanez:1991pr}. $B_3$ is
baryon-parity and allows for the interactions $LL{\bar E},\,LQ{\bar
D}\,$ and $LH_2$ but prohibits ${\bar U}{\bar D}{\bar D}$.

\begin{table*}[t!]
\begin{center}
\begin{tabular}{|c|c|c|c|c|c|c|c|c|c|c|c|c|c|}
\hline
&$\;H_1H_2\;$& $LH_2$ & $\;QQQL\;$ & $\;{\bar U}{\bar U}{\bar D}{\bar E}\;$ 
& $\;QQQH_1\;$ & $\;Q{\bar U}{\bar E}H_1\;$ 
& $\;LL H_2H_2\;$  & $\;L H_1H_2H_2\;$ & 
$\;{H}^*_2H_1{\bar E}\;$ & $\;Q{\bar U}L^*\;$ & ${\bar U}{\bar D}^*{\bar E}$ 
& $\;QQ{\bar D}^*\;$ \\ \hline
{\underline {GMP}}:&&&&&&&&&&& &\\
$R_2$ & \ok& &\ok&\ok&&&\ok&&& &&\\
$A_2R_2$ & & \ok&&&\ok&\ok&\ok&&\ok&\ok&\ok &\ok\\
$R_3$ &\ok&&\ok&\ok&&&&&& &&\\
$R_3A_3$ &&\ok&&&\ok&\ok&\ok&&\ok&\ok&\ok &\ok\\
$R_3A_3L_3$ &&&&\ok&\ok&&&\ok&&& &\ok \\
$R_3L_3^2$ &\ok&&&&&&&&&& &\\
$A_3$ &&&&&&&&\ok&&& &\\
$A_3L_3^2$ &&\ok&\ok&&&\ok&\ok&&\ok&\ok&\ok &\\ &&&&&&&&&&&& \\
{\underline {GLP}}:&&&&&&&&&&& &\\
$L_2$ &\ok&&&&\ok&&\ok&&&& &\ok\\
$A_2L_2$ &&\ok&\ok&\ok&&\ok&\ok&&\ok&\ok&\ok &\\
$L_3$ &\ok&&&&\ok&&&&&& &\ok\\
$\;R_3A_3^2L_3\;$ &&&\ok&&&&&\ok&&& &\\
$R_3A_3^2L_3^2$ &&\ok&&\ok&&\ok&\ok&&\ok&\ok&\ok &\\ &&&&&&&&&&&& \\
{\underline {GBP}}&&&&&&&&&&& &\\
$R_2L_2$ &\ok&\ok&&&&\ok&\ok&\ok&\ok&\ok&\ok &\\
$R_2A_2L_2$ &&&\ok&\ok&\ok&&\ok&\ok&&& &\ok\\
$R_3L_3$ &\ok&\ok&&&&\ok&\ok&\ok&\ok&\ok&\ok &\\
$A_3L_3$ &&&&\ok&&&&&&& &\\
$R_3A_3L_3^2$ &&&\ok&&\ok&&&&&& &\ok\\
\hline
\end{tabular}
\caption{ In the left column we have the complete list of independent
  ${\bf Z}_2$ and ${\bf Z}_3$ discrete symmetries as in
  \cite{Ibanez:1991pr}.  {\it GMP, GLP, GBP} denote generalized matter
  parity, lepton parity and baryon parity, respectively. In the top
  row we have the complete list of dimension-five operators which
  violate baryon- or lepton-number ({\it c.f.} Eq.~(\ref{dim5})). We
  have also included the operator $H_1H_2$. The symbol \ok\ denotes
  that the corresponding operator is allowed by that discrete
  symmetry. There are a few discrepancies compared to
  Ref.~\cite{Ibanez:1991pr}.}
\label{tab:operators}
\end{center}
\end{table*}

This however does not completely solve the problem of proton decay.
In supersymmetry there are also dangerous dimension-five operators
which violate lepton- or baryon-number. The complete list is
\begin{eqnarray}
\begin{array}{ccllcl}
O_1&=& [QQQL]_{\rm F}\,,\qquad  & O_2&=& [{\bar U}{\bar U}{\bar D}
{\bar E}]_{\rm F}\,,\\
O_3&=& [QQQH_1]_{\rm F}\,, & O_4&=& [Q{\bar U}{\bar E}H_1]_{\rm F}\,,\\
O_5&= & [LLH_2H_2]_{\rm F}\,, & O_6&=& [L H_1H_2H_2]_{\rm F}
\,,\\
O_7&=& [{\bar U}{\bar D}^*{\bar E}]_{\rm D} \,, & O_8&= &
[H_2^*H_1{\bar E}]_{\rm D}\,,\\
O_9&=&[Q{\bar U}L^*]_{\rm D}\,, & O_{10}&=&
[QQ{\bar D}^*]_{\rm D}\,,
\end{array}
\label{dim5}
\end{eqnarray}
where we have dropped gauge and generation indices. The subscripts
$F,D$ refer to taking the $F$- or the $D$-term of the given product of
superfields.  We differ from Ref.~\cite{Ibanez:1991pr} in that we have
dropped the operator $[H_2 H_2e^*]_{\rm D}$, which vanishes
identically and included the operator $[QQd^*]_{\rm D}$. As in
Ref.~\cite{Ibanez:1991pr}, we have systematically studied which ${\bf
Z}_{2}$ or ${\bf Z}_{3}$ symmetry allows for which dangerous
dimension-five operators. Our results are summarized in
Table~\ref{tab:operators}. We find some slight discrepancies with
Ref.~\cite{Ibanez:1991pr}. Furthermore, we have added the bilinear
superpotential term $\kappa_i L_i H_2$ ($\kap$-term) not presented in
Ref.~\cite{Ibanez:1991pr}. As expected, the $\mu$-term and the
$\kap$-term go hand in hand in generalized baryon parity models (GBP)
but the opposite is true for the generalized matter (GMP) or lepton
(GLP) parity models: since the $\mu$-term should, phenomenologically,
be a non-zero parameter the GMP or GLP models containing the $\kap
$-term are experimentally excluded. The requirement of neutrino masses
excludes also the GMP and GLP models which do not allow for the
$\Delta L=2$-term: $L L H_2 H_2$. These models do not have any other
source, within perturbation theory to incorporate neutrino
masses. From the models left, \ie [GMP : $R_2$, GLP : $L_2$, $L_3$,
GBP : $R_2 L_2$, $R_3L_3$] only two can be induced from broken and
anomaly free gauge symmetries: these are the GMP: $R_2$ (the usual
R-parity case) and the GBP: $B_3= R_3 L_3$.

Thus what we see from Table~\ref{tab:operators} is that although the
MSSM R-parity is capable of eliminating the dimension-4 operators it
is {\it not} capable of eliminating those of dimension-5. Both
dimension-4 and dimension-5 baryon number violating operators {\it
are not} allowed if the $Z_3$-discrete symmetry $B_3=R_3 L_3$ is
imposed instead of the R-parity ($R_2$) symmetry. In this article we
study the phenomenology of the model based on the discrete $Z_3$
symmetry $B_3=R_3 L_3$ \cite{foot4}.

\subsection{Grand Unified Models}

In GUTs, quarks and leptons are in common multiplets and this simple
approach does not suffice. We consider the case of the gauge groups
$SU(5)$ and $SO(10)$ separately.

\subsubsection{SU(5)}
In $SU(5)$ models, the trilinear and bi-linear R-parity violating
terms are respectively given by
\begin{equation}
h_{ijk}\, \Psi_i\Psi_jX_k\,,\qquad k_i\,\Psi_i \Phi_5\,,
\label{su5}
\end{equation}
where $\Psi$ is the {\bf 5$^*$} representation containing the ${\bar
D}$ and $L$ superfields, $X$ is the {\bf 10} representation containing
the $Q,\,{\bar U},$ and ${\bar E}$ superfields and $\Phi_5$ is the
Higgs superfield in the {\bf 5} representation. $h_{ijk}$ are Yukawa
couplings and $k_i$ dimension-one couplings. $i,j,k$ are generation
indices. Unless $h_{ijk}\lsim10^{-13}$, this leads to unacceptably
rapid proton decay. Thus this term must be forbidden by an additional
symmetry. The generalization of matter parity where now $\Psi$ and $X$
change sign prohibits both terms in Eq.~(\ref{su5}) and guarantees
that \rpv-terms are not generated once $SU(5)$ is broken.

Alternative (discrete) symmetries can also be considered. In
Ref.~\cite{Hall:1983id}, a discrete five-fold R-symmetry is
constructed which prohibits the terms in Eq.~(\ref{su5}). However,
after breaking $SU(5)\ra SU(3)\times SU(2)\times U(1)$ and integrating
out the heavy fields the operators $k_i\,\Psi_i \Phi_5$ are generated,
resulting in bi-linear R-parity violation. The size of the coupling
depends on the vacuum expectation values of the large dimensional
Higgs field representations which break $SU(5)$. Similar symmetries
can also be constructed to obtain tri-linear R-parity violation. This
was done in the case of ``flipped'' $SU(5)\times U(1)$ in
Ref.~\cite{Brahm:1989iy} and is easily transferred to the case of
$SU(5)$. The question of whether it is possible to obtain \ \rpv\ in
GUTs with a large ${\Delta L{/}\Delta B}$ hierarchy was also addressed
in Ref.~\cite{Tam1} employing a modified version of the minimal
$SU(5)$, where a built in Peccei-Quinn symmetry is broken at an
intermediate scale.

\subsubsection{$SO(10)$}
In $SO(10)$ GUTs \cite{Fritzsch:nn}, {\it B--L} is a gauge symmetry
and thus R-parity is conserved.  Explicitly, the matter fields of a
family are combined in a (spinorial) {\bf 16} representation and the
operators 
\begin{equation}
{\bf R}_{ijk}={\bf 16}_i\cdot{\bf 16}_j\cdot{\bf 16}_k\,,
\end{equation} 
are not $SO(10)$ invariant. (Again, $i,j,k$ are generation indices.) 
As in the $SU(5)$ case, one would now expect to generate R-parity
violating terms after breaking $SO(10)$ and {\it B--L}.  However, as
shown in Ref.~\cite{goran}, surprisingly, this strongly depends on the
Higgs representations chosen to perform the breaking.

If we include a {\bf 16}$_H$-Higgs representation to break $SO(10)$,
as well as higher dimensional Higgs representations we have the
non-renormalizable operators
\begin{equation}
{\bf N}_{ijkH}={\bf 16}_i\cdot{\bf 16}_j\cdot{\bf 16}_k
\cdot{\bf 16}_H\cdot G({\bf H})\,,
\end{equation}
where $G({\bf H})$ is a function of the higher dimensional Higgs
representations. When the Higgs fields get vacuum expectation values,
$SO(10)$ is broken and in general R-parity violating operators will be
generated. The exact nature of the resulting R-parity violation
depends on the employed Higgs fields and can be consistent with proton
decay experiments \cite{Giudice:1997wb}.

Instead, we can explicitly exclude a ${\bf 16}_H$ representation and
break $SO(10)$ by a {\bf 126}-Higgs representation \cite{goran}.
Since ${\bf R}_{ijk}$ is an odd product of spinorial representations it
is itself a spinorial representation. Without ${\bf 16}_H$ there is
now no spinorial Higgs representation and thus no $SO(10)$ invariant
combination
\begin{equation}
{\bf R}_{ijk}\cdot G'({\bf H})\,,
\end{equation}
where $G'({\bf H})$ is a general tensor product of Higgs
representations.  Thus after spontaneous symmetry breaking the
operators $R_{ijk}$ can not be generated and there is no explicit
R-parity violation in the theory.  However, in principle R-parity can
still be broken spontaneously with $\langle{\tilde \nu} \rangle\not=0$
or $\langle{\tilde \nu}^c\rangle \not=0$, where ${\tilde \nu}^c$ is a
right handed neutrino (which in this paper is only included in this
discussion of $SO(10)$). With the absence of a ${\bf 16}_H$ it was
shown in Ref.~\cite{goran} that F-flatness at the GUT scale requires
$\langle{\tilde \nu}^c\rangle=0$. This is also stable under the
renormalization group equations. At the GUT scale we must also have
$\langle{\tilde \nu} \rangle=0$, otherwise $SU(2)_L$ would be broken
at $M_{\rm GUT}$. Similarly at the weak scale, we must demand
$\langle{\tilde \nu} \rangle=0$ in order to avoid an unobserved
Majoran. Thus in this model R-parity is conserved at all energies and
guaranteed by a gauge symmetry \cite{goran}.

We conclude, that {\it a priori} there is no preference in
supersymmetric GUTs for or against R-parity violation. Finally, we
note in passing, that there exist few attempts in the literature to
construct superstring models which accommodate the lepton number \rpv
couplings~\cite{Alon}.

\subsection{Origin of the $\kap_iL_iH_2$ Terms}

It is well known, that through a field redefinition of the $L_i$ and
$H_1$ fields, the $\kap_i$ terms in the superpotential
Eq.~(\ref{superpot1}) can be rotated away at any scale 
\cite{Hall:1983id}. The full rotation matrix in the complex case was
only given recently in Ref.~\cite{Thor}. {\it After} supersymmetry
breaking, however, they can only be rotated away jointly with the
corresponding soft breaking terms ${\widetilde D}_i{\tilde L}_i{\tilde
H}_2$, if $\kap_ i$ and ${\widetilde D}_i$ are aligned
\cite{Dreiner:1995hu,Banks:1995by}. Even if they are aligned at a
given scale, this alignment is not stable under the renormalization
group equations \cite{Dreiner:1995hu,deCarlos:1996du,Nardi:1996iy}. 
However, if $\kap_i$ and ${\widetilde D}_i$ are aligned after
supersymmetry breaking then we can choose a basis where $\kap_i=
{\widetilde D}_i=0$ at the supersymmetry breaking scale. At the
electroweak scale, we then have a {\it prediction} for both $\kap_i$
and ${\widetilde D}_i$ through the renormalization group equations
(RGEs), given the initial choice of basis. We are thus interested in
the conditions for alignment after supersymmetry breaking in various
unification scenarios, in order to predict $\kap_i(M_Z)$ and
${\widetilde D}_i(M_Z)$.

We first consider the general superpotential of Eq.~(\ref{superpot1}),
restricted for the case $\mu=\kap_i=0$). It is invariant under a
discrete R-symmetry \cite{Weinberg1}, where the chiral {\it
superfields} have the following R-quantum numbers \cite{Rsymm}.
\vspace{-0.18cm}
\begin{center}
\begin{tabular}{|c|c|c|c|c|c|c|} \hline
$\;\;L_i\;\;$ & $\;\;{\bar E}_i\;\;$ & $\;\;Q_i\;\;$ & $\;\;{\bar U}_i\;\;$ 
& $\;\;{\bar D}_i\;\;$ & $\;\;H_1\;\;$ & $\;\;H_2\;\;$ \\
\hline
0 & -2 & -1 & -1 & -1 & 0 & 0 \\ \hline
\end{tabular}
\end{center}
The vector superfields have zero charge.  Each term in the
superpotential must have R-charge -2, which is canceled by the charges
of the Grassman coordinates.  Thus all tri-linear terms except ${\bar
U}{\bar D}{\bar D}$ are allowed.  Note, that since this is an
R-symmetry the fermionic components of the chiral and vector
superfields have a different charge than the superfield.  In
particular, the R-parity even components of the chiral superfields
have the quantum numbers of conventional lepton-number.  With this
somewhat unusual symmetry we have ensured lepton number {\it
conservation} for the SM fields
\cite{flavour}.

However, the phenomenology of this superpotential is unacceptable.
Below we show that if $\mu,\,\kap_i,\,{\widetilde B},\,{\widetilde
D}_i=0$, the CP-odd Higgs boson mass, $m_A=0$ and the lightest
chargino mass $M_{{\tilde\chi}^\pm_1}\lsim{\cal O}(30\,GeV)$, both in
disagreement with observation. $m_A=0$ due to the Peccei-Quinn
symmetry of the superpotential. We thus demand $\kap_i,\,\mu\not=0$,
in order to get consistent $SU(2)\times U(1)$ breaking and a
sufficiently heavy chargino. This in turn introduces {\it
lepton-number violation} for the low-energy SM fields.

The parameters $\kap_i$ and $\mu$ are dimensionful and in principle
present before supersymmetry breaking. The only mass scale in the
theory is the Planck scale $(M_P)$, and we thus expect $\kap_i,\,\mu =
{\cal O}(M_{P})$. Experiment requires $\mu={\cal O}(M_Z)$ and
$\kap_i\ll M_Z$. (The latter strict requirement is due to neutrino
masses, as we discuss in detail below.) This is the well-known
$\mu$-problem \cite{Kim:1983dt}, modified by the presence of the
$\kap_i$.  In the following, we discuss the origin of the weak-scale
$\mu$ and $\kap_i$ terms and their corresponding soft terms.  We can
then determine under what conditions the $\kap_i$ and ${\widetilde
D}_i$ can be simultaneously rotated away at the unification scale.  We
begin by discussing supergravity theories where there are several
proposed solutions to the $\mu$-problem
\cite{Kim:1983dt,Giudice:1988yz,Chamseddine:1995gb,muproblem}.  We
review them here in the light of the additional $\kap_i$ terms.

\subsection{Supergravity}
We consider a set of real scalar fields $z_i$ for the hidden sector
and a set $y_a$ for the observable sector \cite{Nilles:1983ge}.
Collectively we denote them $Z_A$.  The supergravity Lagrangian
depends only on the dimensionless scalar function (K\"ahler potential)
\cite{notation2}
\begin{eqnarray} 
{\cal
  G}(z_i,z^{i*};y_a,y^{a*})&=& -d(z_i,z^{i*};y_a,y^{a*})/M_P^2
\nonum \\
&& -\log(f(z_i;y_a)/M^3_{P}) \,.
\end{eqnarray}
Here $d$ determines the K\"ahler metric and $f$ is the superpotential,
which is a holomorphic function. The scalar potential is given by
\begin{eqnarray}
  V&=&-M_{P}^4\exp(-{\cal G})\left[3+ {\cal G}_A({\cal G}^{-1})^A_B
{\cal G}^B \right] 
+\half D_\alpha D^\alpha\,,\nonum \\
&=&\exp\left(\frac{d(Z_A,Z^{A*})}{M_P^2}\right)\times \\
&& \hspace{-0.5cm} \times\left[
(d^{-1})^A_B F^{A\dagger} F_B-3f^\dagger(Z^{A*}) f(Z_A)/M^2_P\right]+ \half
  D_\alpha D^\alpha\;. \nonum 
\end{eqnarray}
Here ${\cal G}^A\equiv\partial {\cal G}/\partial Z_A$, and
\begin{eqnarray}
\!\!\!\!({\cal
  G}^{-1})^A_B&\equiv&\frac{\partial^2 {\cal G}^{-1}}{\partial Z_A\partial
Z^{B*}} \\
F_A&\equiv& \frac{\partial f(Z_A)}{\partial Z_A}+M^{-2}\frac{\partial
  d(Z_A,Z^{A*})}{\partial Z_A} f(Z_A), 
\end{eqnarray}
and $D^\alpha$ is the auxiliary field of the vector superfield. The
derivatives of $d^{-1}$ are defined analogously.

The most general form of the low-energy scalar potential after
supersymmetry breaking is
\cite{Soni:1983rm}
\begin{eqnarray}
V&=& \left(\frac{\partial g(y)}{\partial y_a}\right)^\dagger \left(
\frac{\partial g(y)}{\partial y_a}\right)+m_{3/2}^2 S_{ab} y_a y_b^\dagger 
\nonum \\
&& + m_{3/2} [h(y)+h^\dagger(y)] +  \half D^\alpha D^\alpha\;.
\end{eqnarray}
Here $g(y_a)$ is the superpotential for the {\it low-energy} fields
derived from $f(Z_A)$ and $m_{3/2}$ is the gravitino mass.  The first
and the last terms are the usual $F$- and $D$-term contributions to
the scalar potential. The second and third terms arise from
supersymmetry breaking. The general constant matrix $S_{ab}$ has in
principle arbitrary entries, {\it i.e.} the soft scalar masses can be
non-universal.

$h(y)$ is a superpotential, \ie\ a holomorphic function of the $y_a$.
In the renormalizable case, it is at most trilinear in the fields $y_a$
and contains the supersymmetry breaking $A$ and $B$-terms
\cite{nilles-a-term}.  $g(y)$ and $h(y)$ are superpotentials of the
same fields and due to gauge invariance thus contain the same terms.
However in general, the coefficients are independent and thus in
particular the $A$- and $B$-terms need not be proportional to the
corresponding terms in $g(y)$.  But if the superpotential satisfies
\begin{equation}
f(z_i;y_a)=f_1(z_i)+f_2(y_a)\;,
\label{eq:superpot-sum}
\end{equation}
then the soft-breaking term $h(y)$ is a linear com\-bination of the
superpotential $g(y)$ and $y_a \partial g(y_a)/\partial y_a$
\cite{Soni:1983rm} and thus each term is proportional to the
corresponding term in $g(y)$.  The condition \eq{superpot-sum} is
quite natural.  If the $z_i$ all transform non-trivially under only
the hidden-sector gauge group and the $y_a$ transform non-trivially
only under the observable sector gauge group, then combined with the
requirement of renormalizability we obtain the condition
\eq{superpot-sum}.

We now consider the {\it observable} sector superpotential given in
Eq.~(\ref{superpot1}). If our superpotential at the unification scale
satisfies Eq.~\eq{superpot-sum}, the ${\widetilde D}_i$ will be
aligned with the $\kap_i$ after supersymmetry breaking and they can be
simultaneously rotated away. Or looked at differently: before
supersymmetry breaking we can always rotate the fields ${\cal
L}^a_\alpha$ such that $\kap_i =0$. If we then break supersymmetry at
this scale, while obeying Eq.~\ref{eq:superpot-sum}, we automatically
obtain ${\widetilde D}_i=0$ as well, since the coefficients in
$h(y_a)$ are proportional to those in $g(y_a)$.  Thus in the case of a
renormalizable superpotential we expect universal $A$ and $B$ terms
and thus an alignment of $\kap_i$ and ${\widetilde D}_i$ at the
unification scale.

\subsection{Implementing a Solution to the $\mu$-Problem}

The most widely discussed solution to the $\mu$-problem is to prohibit
the $\mu H_1H_2$ in the superpotential via a symmetry, for example an
R-symmetry, and instead introduce a non-renormalizable term into the
K\"ahler potential, ${\cal G}$, which results in the $\mu$-term after
supersymmetry breaking.  By using the mass scale inherent in
supersymmetry breaking one then obtains $\mu={\cal O}(M_Z)$. This was
first proposed by Kim and Nilles \cite{Kim:1983dt} who introduced the
non-renormalizable term into the superpotential $f$. The R-symmetry
was global and the resulting axion was phenomenologically acceptable.
Giudice and Masiero \cite{Giudice:1988yz} introduced a non-holomorphic
term into the K\"ahler metric function $d$ instead, also invoking an
R-symmetry to prohibit terms in the superpotential. The details of the
axion were not considered. In certain cases the two mechanisms are
equivalent \cite{Weinberg1}.  In the following, we briefly consider
the implications of Ref.~\cite{Kim:1983dt} for the $\kap_i$ terms and
extend this to Ref.~\cite{Giudice:1988yz}.

In the context of R-parity violation we have both a $\mu$ and a $\kap
_i$ problem. As an example, we introduce the following
non-renormalizable terms into the superpotential,
\begin{equation}
f'=\frac{1}{M_{P}} (a z_1z_2H_1H_2 + b_i z_3z_4 L_iH_2),
\label{eq:mu-potential}
\end{equation}
assuming them to be invariant under the symmetries of the model. In
general, we could have higher powers of the $z_i$.  If the Peccei-Quinn
\cite{Peccei:1977hh} charges which prohibit the bilinear terms in the
superpotential are lepton-flavour blind but distinguish $H_1$ and
$L_i$ then we would expect the general form shown above. $a,\,b_i$ are
dimensionless constants. Due to the independent fields $z_i$ we can
not rotate away the $b_i$ terms. After supersymmetry breaking we get
\begin{eqnarray}
\mu&=&\frac{\vev{z_1}\vev{z_2}}{M_P}={\cal O}(M_Z),\\
\kap_i&=&\frac{\vev{z_3}\vev{z_4}}{M_P}={\cal O}(M_Z).
\end{eqnarray}

If the fields $z_i$ are hidden-sector fields and $f'$ mixes the hidden
and observable sectors then the soft supersymmetry bilinears are in
general not aligned with the $\kap_i$ since there are now the
additional terms
\begin{equation}
\frac{1}{M_P}\left(
\frac{\partial f_1}{\partial z_4}\vev{z_3}+
\frac{\partial f_1}{\partial z_3}\vev{z_4} 
\right) b_i {\tilde{\cal L}}_iH_2,
\end{equation}
which have independent coefficients from the purely hidden sector.
Here we have made use of the hidden-sector function $f_1$ of
Eq.~\eq{superpot-sum}. The resulting $\kap_i$ terms are still ${\cal
O}(M_Z)$. If ${\partial f_1}/{\partial z_{i=1,2,3,4}}=0$ then we have
alignment. 

Alternatively, the Peccei-Quinn charges can be such that $H_1$ has the
same charge as the $L_i$. This is exactly the case of the $B_3=R_3
L_3$ discrete symmetry we discussed in some detail in section~\ref{DS}
and follow in this article.  The charge of $H_1$ and the $L_i$ under
this symmetry is $-2/3$.  In this case $z_1z_2=z_3z_4$ in
Eq.~\eq{mu-potential} and the $\kap_i$ terms can be rotated away {\it
before} supersymmetry breaking. No ${\widetilde D}_i$ soft terms are
generated in supersymmetry breaking then and we have $\kap_i=
{\widetilde D}_i=0$ at the high scale.

We conclude that it is possible to have alignment of the bilinear
terms at the supersymmetry breaking scale but not necessary. The
eventual answer will depend on the underlying unified theory. We shall
assume that we can rotate away the $\kap_i$ terms before supersymmetry
breaking.

\section{The Minimal  R-parity violating Supersymmetric Standard Model}
\label{rpvmsugra}

The model we consider has the particle content given in
Table~\ref{table1} and the superpotential given in
Eq.~(\ref{superpot1}). Within this superpotential we shall make the
assumption that at the unification scale, $M_X \simeq 10^{16}\gev$,
the terms $\kap_iL_iH_2$ have been rotated to zero. For real
parameters the orthogonal rotation on the fields ${\cal L}_\alpha$
which accomplishes this is given by
\begin{equation}
\cL_\alpha = {\cal O}_{\alpha\beta} \cL_\beta^\prime \;,
\end{equation}
and explicitly in components
\begin{eqnarray}
\left (\begin{array}{c} H_1 \\ L_1 \\ L_2 \\ L_3 \end{array}\right )=
\left ( \begin{array}{cccc} c_3 & -s_3 & 0 & 0 \\
                            c_2s_3 & c_2c_3 & -s_2 & 0 \\
                            c_1s_2s_3 &c_1s_2c_3 & c_1c_2 & -s_1 \\
                            s_1s_2s_3 & s_1s_2c_3 & s_1c_2 & c_1 
\end{array} \right ) \left ( \begin{array}{c} \cL_0^\prime \\
  \cL_1^\prime
 \\ \cL_2^\prime \\
\cL_3^\prime \end{array} \right ) , 
\end{eqnarray}
where $c_i=\cos\theta_i$ and $s_i=\sin\theta_i$, and 
\begin{eqnarray}
c_1&=&\frac{\kap_2}{\sqrt{\kap_2^2+\kap_3^2}}\;,\quad
s_1=\frac{\kap_3}{\sqrt{\kap_2^2+\kap_3^2}}\;,\nonum\\
c_2&=&\frac{\kap_1}{\sqrt{\vec{\kap}^2}}\;,\quad
\;\;\;\;\;\;\;\;s_2=\frac{\sqrt{\kap_2^2+\kap_3^2}}
{\sqrt{\vec{\kap}^2}}\;,\\
c_3&=&\frac{\mu}{\sqrt{\mu^2+\vec{\kap}^2}}\;,\quad
s_3=\frac{{\sqrt{\vec{\kap}^2}}}
{\sqrt{\mu^2+\vec{\kap}^2}}\;. \nonum
\label{angles}
\end{eqnarray}
Here we have introduced the notation $\vec{\kap}^2=\sum_i
\kap_i^2$.  The more general case of complex parameters is given in
Ref.~\cite{Thor}; we shall restrict ourselves to real parameters here.
After the above field redefinition, the only remaining superfield
bi-linear term is
\begin{equation}
\mu''H_1'H_2
\end{equation}
with $\mu''=\sqrt{\mu^2+\vec{\kap}^2}$ and $H_1'\equiv \cL_0'$. This
will be our starting bilinear superpotential term at $M_X$ in our RGE
studies below.

The RGEs for the $\kap_i$ are given by (see Appendix A)
\begin{equation}
16\pi^2\Dt\kap^i=\kap^i\,\ggam{H_2}{H_2}+\kap^p\,\ggam{L_i}{L_p}
+\mu\,\ggam{L_i}{H_1}, \label{kappa1}
\end{equation}
where at one-loop the anomalous dimension mixing $L_i$ and $H_1$ is
given by
\begin{equation}
\gamma^{H_1}_{L_i} = {\gamma^{L_i}_{H_1}}^* =-3 \lam^{'*}_{ijk}(\yd)_{jk}
- \lam^*_{ijk}(\ye)_{jk}\;,\label{LH-mix}
\end{equation}
with a summation over $j,k$ implied. (The remaining anomalous
dimensions are given in Appendix~\ref{app2}.) Therefore, given
$\mu\not=0$ at $M_X$ and a non-zero $\lam$ or $\lam'$, we will in
general generate a non-zero $\kap_i(M_Z)$
\cite{Dreiner:1995hu,deCarlos:1996du,Nardi:1996iy,add}. Below we
discuss special exceptional cases where this is not the case.

In order to fix all the parameters we also need to know
the general soft supersymmetry breaking Lagrangian which we denote
\begin{widetext}
  \begin{eqnarray}
-{\cal L}_{\rm SOFT} &=& {\tilde{\cL}}_\alpha^\dagger 
({\bf m_{\tilde{\cL}}}^2)_{\alpha\beta}
  {\tilde{\cL}}_\beta + \mh{2} H_2^\dagger
  H_2 + \widetilde{Q}^\dagger \mq \widetilde{Q} + \widetilde{\bar{E}}
  \me \widetilde{\bar{E}}^\dagger + \widetilde{\bar{D}} \md
  \widetilde{\bar{D}}^\dagger
  + \widetilde{\bar{U}} \mup \widetilde{\bar{U}}^\dagger \nonum \\[1mm]
  &+ & \eps_{ab} \left[ (\hu)_{ij} Q_i^a H_2^b \bar{U}_j+ \frac{1}{2}
    h_{\alpha\beta k}\tilde{\cL}_\alpha^a \tilde{\cL}_\beta^b
    \bar{E}_k +h^\prime_{\alpha jk} \tilde{\cL}_\alpha^a Q_j^b
    \bar{D}_k - b_\alpha \tilde{\cL}_\alpha^a H_2^b~+~{\rm H.c}
  \right]+ \frac{1}{2}\epsilon_{xyz}h^{\prime\prime}_{ijk} \bar{U}_i^x
  \bar{D}_j^y \bar{D}_k^z  ~+~{\rm H.c} \nonum \\[1mm]
  &+& \biggl [ \frac{1}{2} M_1 \bino\bino + \frac{1}{2} M_2
  \wino^{({\Gamma})}\wino^{({\Gamma})} + \frac{1}{2} M_3
  \gluino^{(R)}\gluino^{(R)}~+~{\rm H.c} \biggr ] \;.
 \label{softhaber}
\end{eqnarray}
\end{widetext}
Here, ${\tilde F}\in [{\tilde Q},\,{\tilde {\bar U}},\, {\tilde {\bar
D}},\,{\tilde {\bar E}},\,{\tilde{\cal L}}]$ denote the scalar
component of the corresponding chiral superfield. ${\bf m_{\tilde{F}
}}^2$ are the soft-breaking scalar masses. Note that these are
$3\times3$ matrices for the squarks and for the lepton
singlets. However, $( {\bf m_{\tilde{\cL}}} ^2)_{\alpha \beta}$ is a
$4\times4$ matrix. $(\hu)_{ij},\,h_ {\alpha \beta
k},\,h^\prime_{\alpha jk},$ and $h^{\prime\prime }_{ijk} $ as well as
$b_\alpha=(\widetilde{B},
\widetilde{D}_i)$ are the soft breaking trilinear and bilinear terms,
respectively.

The RGEs for the ${\widetilde D}_i$ are given at one-loop by
\begin{eqnarray}
16 \pi^2 \frac{d\widetilde{D}_i}{dt} &=&
\biggl [\gamma^{L_i}_{L_l}\widetilde{D}^l
+\gamma_{H_2}^{H_2}\widetilde{D}^i
 \biggr]
+\widetilde{B} \gamma_{H_1}^{L_i}  -2\, \mu (\gamma_1)_{H_1}^{L_i}\nonum \\
&-&2 \biggl [ (\gamma_1)^{L_i}_{L_l} \kap^l +  (\gamma_1)^{H_2}_{H_2}\kap^i
\biggr ] \;,
\label{RG-D2}
\end{eqnarray}
with the anomalous dimensions ($\gamma$) and the functions
($\gamma_1$) defined in Appendices~\ref{app3} and \ref{app4},
respectively. These RGEs are clearly distinct from those for $\kap_i$,
above. It is thus clear that given $\kap_i(M_X)=\widetilde{D}_i
(M_X)=0$ we will lose alignment between the two at the electroweak
scale \cite{Dreiner:1995hu,deCarlos:1996du,Nardi:1996iy,add}. In order
to describe the weak-scale physics, we thus require the full set of
parameters given in Eqs.~(\ref{superpot1}) and (\ref{softhaber}).

\section{Electroweak Symmetry Breaking}
\label{rad-break}

The full scalar potential is given by
\begin{eqnarray}
{\bf V}_{\rm SCALAR}={\bf V}_{\rm SUSY}+{\bf V}_{\rm SOFT}\;,
\label{vscalar}
\end{eqnarray}
with the supersymmetric F-term and D-term scalar potential given
by \cite{explain}
\begin{eqnarray}
&&{\bf V}_{\rm SUSY}={\bf V}_{\rm F}+{\bf V}_{\rm D}= \nonum \\[1cm]
&&\sum_\Phi \left|\frac{\partial W}{\partial\Phi}\right|^2 +
\sum_{\ell=1}^3 \frac{g_\ell^2}{2}\sum_A\biggl (\sum_{m,n}\Phi^*_m
T_{\ell,A}^{mn}\Phi_{n}\biggr )^2 \;, \label{vsusy}
\end{eqnarray}
respectively and ${\bf V}_{\rm SOFT}=-{\cal L}_{\rm SOFT}$. In
Eq.~(\ref{vsusy}), the fields $\Phi_{m,n}$ denote the scalar fields in
the theory, $g_{\ell =1,2,3}$ are the gauge couplings with $g_ 1$ for
$U(1)_Y$, $g_2$ for $SU(2)_L$ and $g_3$ for the $SU(3)_C$ gauge
group. In order to simplify the expressions, we shall use the coupling
$g\equiv\sqrt{\frac{3}{5}}
\, g_1$.  $m,n..$ and $A,B...$ are representation and gauge generator
indices, respectively.  The explicit expressions for ${\bf V}_{\rm F}$
and ${\bf V}_{\rm D}$ can be found in Ref.~\cite{Haber}.

In the following, we shall focus on the complex neutral scalar fields:
$h_2^0,\,{\tilde \nu}_\alpha\equiv(h_1^0,\,{\tilde \nu}_{i=1,2,3})$. For
these the scalar potential is given by
\begin{widetext}
\begin{eqnarray}
{\bf V}_{\rm
  Neutral}&=& (m_{H_2}^2+|\mu_\alpha|^2 )|h_2^0|^2 +
\biggl [(m_{\tilde{\cL}}^2)_{\alpha\beta}+
\mu_\alpha^*\mu_\beta \biggr ] \tilde{\nu}_\alpha^*\tilde{\nu}_\beta
-(b_\alpha \tilde{\nu}_\alpha h_2^0+b_\alpha^* \tilde{\nu}_\alpha^*
h_2^{0*}) \nonum \\
&+& \frac{1}{8}(g^2+g_2^2) \biggl (|h_2^0|^2-|\tilde{\nu}_\alpha|^2
\biggr )^2 + {\bf \Delta V}\;,
\label{vneutral}
\end{eqnarray}
\end{widetext}
where ${\bf \Delta V}$ denotes higher order corrections~\cite{Chun}.
In order to minimize this potential, it is convenient to write the
complex neutral scalar fields in terms of CP-even, $x_2, \, r_\alpha$,
and CP-odd $y_2,\, t_\alpha$ real field fluctuations
\begin{eqnarray}
h_2^0&=&x_2+iy_2\,, \\
{\tilde \nu}_\alpha&=& r_\alpha+it_\alpha\,.
\end{eqnarray}
At the minimum the scalar fields thus take on the values $\vev{x_2}=
v_u,\,\vev{r_\alpha}=v_\alpha$, ($v_\alpha=(v_d,v_1,v_2,v_3)$) and
$\vev{y_2}=\vev{ t_\alpha}=0$. The minimization conditions for ${\bf
V}_{\rm Neutral}$ can be written as,
\begin{eqnarray}
\left.
\frac{\partial {\bf V}_{\rm Neutral}}{\partial {x_2}}
\right|_{{\rm min} } =0\,\,\,\,\,\;\;,\,\,\,\,\;\;   
\left.\frac{\partial {\bf V}_{\rm Neutral}}{\partial r_\alpha}
\right|_{{\rm min}} =0\,, 
\end{eqnarray}
where ``min'' refers to setting the scalar fields to their values at
the minimum. We then derive the following five minimization
conditions, where $\alpha,\,\beta=0, 1, 2,3$ and there is an implied
sum over repeated indices
\begin{widetext}
\begin{eqnarray}
\Re e  \biggl [ (m_{\tilde{\cL}}^2)_{\alpha\beta}+
\mu_\alpha^*\mu_\beta \biggr ] v_\alpha-\Re e(b_\beta) v_u
-\frac{1}{4} (g^2+g_2^2) \biggl (|v_u|^2-|v_\alpha|^2\biggr ) v_\beta
+\frac{1}{2} \frac{\partial {\bf \Delta V}}{\partial v_\beta} &=& 0\;,
\nonum \\ (m_{H_2}^2+|\mu_\alpha|^2 )v_u -\Re e(b_\beta) v_\beta
+\frac{1}{4} (g^2+g_2^2) \biggl (|v_u|^2-|v_\alpha|^2\biggr ) v_u
+\frac{1}{2} \frac{\partial {\bf \Delta V}}{\partial v_u} &=& 0
\;. \label{mins}
\end{eqnarray}
\end{widetext}
Here $\Re e$ denotes the real value and we have written $(\partial
{\bf\Delta V}/\partial r_\alpha)|_{\rm min}$ as $\partial {\bf\Delta
V}/\partial v_\alpha$ and $(\partial {\bf\Delta V}/\partial x_2)|_
{\rm min}$ as $\partial {\bf\Delta V}/\partial v_u$. Next, we solve
this system of equations. We start by defining
\cite{Nowakowski:1995dx}
\begin{eqnarray}
\tan\beta \ \equiv \ \frac{v_u}{v_d} \;,
\label{tanbeta}
\end{eqnarray}
and
\begin{eqnarray}
v^2 \equiv v_u^2+v_d^2+\sum_{i=1}^3 v_i^2 = \frac{2 M_W^2}{g_2^2} \;,
\label{vevtotal}
\end{eqnarray}
where in our   convention $v=174\,{\rm GeV}$. Then the vev's $v_d$
and $v_u$ can be written 
\begin{eqnarray}
v_d^2 &=& \cos^2\beta \biggl (v^2-\sum_{i=1}^3 v_i^2 \biggr ) \;,
\label{vevd} \\
v_u^2 &=& \sin^2\beta \biggl (v^2-\sum_{i=1}^3 v_i^2 \biggr ) \;,
\label{vevu} 
\end{eqnarray}
with $v_i$ being the three sneutrino vev's.  The advantage of using the
definition given in Eqs.~(\ref{tanbeta},\,\ref{vevtotal}) is that
$\tan\beta$ is the same in the R-parity conserved (RPC) and
\rpv-models. This facilitates the direct comparison, in particular
when $v_i/v\ll 1$.

Using these definitions and the notation $(v_i^2)\equiv\sum_i v_i^2$,
the five minimization conditions in Eq.~(\ref{mins}) can be written as
(again there is an implied sum over repeated indices.)
\begin{widetext}
\begin{eqnarray}
(m_{H_1}^2 +\mu^2) v_d + \!\biggl [ \mlh{i}+\kap_i^*\mu \biggr ] v_i
-\widetilde{B} v_u + \frac{1}{2} M_Z^2 \cos2\beta v_d +\frac{1}{2}
(g^2+g_2^2) \sin^2\beta \: v_d\: (v_i^2) + \frac{1}{2} 
\frac{\partial {\bf \Delta V}}{\partial v_d} &=& 0 ,
\label{minI} \\
\!\!\!\!\!\!\biggl [ \mhl{i}+\mu^* \kap_i \biggr ] v_d + 
\! \biggl [ \ml_{ji}+
\kap_j^*\kap_i \biggr ] v_j -\!\widetilde{D}_i v_u
+\!\frac{1}{2} M_Z^2 \cos2\beta \: v_i +\!\frac{1}{2} (g^2+g_2^2)
\sin^2\beta \: v_i  \: (v_j^2) +\!\frac{1}{2}
\frac{\partial {\bf \Delta V}}{\partial v_i} &=& 0 ,
\label{minII} \\
(m_{H_2}^2+\mu^2+ |\kap_i|^2 )v_u -\widetilde{B} v_d
-\widetilde{D}_i v_i -\frac{1}{2} M_Z^2 \cos2\beta v_u
-\frac{1}{2} (g^2+g_2^2) \sin^2\beta \: v_u \: (v_i^2)
+\frac{1}{2}\frac{\partial {\bf \Delta V}}{\partial v_u} &=& 0\,.
\label{minIII}
\end{eqnarray}
\end{widetext}
In order to solve the above equations, we first derive $\mu$ in terms
of $v_u,\,v_d,$ and $v_i$ from Eqs.~(\ref{minI}) and (\ref{minIII}).
It is obtained after solving the quadratic equation
\begin{eqnarray}
A\mu^2+B\mu+ \Gamma =0 \;,\label{minIV}
\end{eqnarray}
with 
\begin{eqnarray}
A & \equiv & \tan^2\beta -1 \;, \qquad B \;\equiv\; 
 -\kap_i^* \frac{v_i}{v_d} \;, \\
\Gamma  & \equiv & \biggl \{ \biggl [ \ovl{m}_{H_2}^2+|\kap_i|^2
-\frac{(g^2+g_2^2)}{2} (v_i)^2 -\widetilde{D}_i \frac{v_i}{v_u} \biggr ]
\tan^2\beta \nonum \\
&-& \biggl [ \ovl{m}_{H_1}^2 + \mlh{i} \frac{v_i}{v_d} \biggr
] \biggr \} +\frac{1}{2} M_Z^2 (\tan^2\beta -1)\;. 
\label{Gam}
\end{eqnarray}
The solution to Eq.~(\ref{minIV}) can be written in a more familiar
form,
\begin{widetext}
\begin{eqnarray}
|\mu|^2 \ = \ \frac{\biggl [  \ovl{m}_{H_1}^2 + \mlh{i} \frac{v_i}{v_d} 
+\kap_i^* \mu \frac{v_i}{v_d} \biggr ] - \biggl [ \ovl{m}_{H_2}^2
+|\kap_i|^2 - \frac{1}{2} (g^2+g_2^2) v_i^2 - \widetilde{D}_i
\frac{v_i}{v_u} \biggr ] \tan^2\beta}{\tan^2\beta -1} \ -\ \frac{1}{2}
M_Z^2 \;. \label{muMSSM}
\end{eqnarray}
\end{widetext}
We recover the familiar minimization condition \cite{Dedes:2002dy} in
the RPC limit $\kap_i,$ $v_i,\,\widetilde{D}_i,\,\mlh{i} \to 0$.

Eq.~(\ref{minIV}) or equivalently Eq.~(\ref{muMSSM}), has two solutions
for the parameter $\mu$ : $\mu>0$ and $\mu<0$. We thus retain the sign
of $\mu$ as a free parameter. Furthermore, the factor $\kap_i^*\frac
{v_i}{v_d}$ that multiplies the $\mu$ parameter in Eq.~(\ref{muMSSM})
is small since, as we show below, $v_i \ll v_d$ to obtain a small
neutrino mass, $m_\nu\lsim{\cal O}({\rm eV})$.

We can now express ${\widetilde B}$ in terms of $\mu,\,v_u,\,
v_d,\,v_i$ from Eqs.~(\ref{minI}) and (\ref{minIII})
\begin{eqnarray}
\widetilde{B} &=& \frac{\sin2\beta}{2} \biggl \{\biggl [ \ovl{m}_{H_1}^2
+\ovl{m}_{H_2}^2 + 2 |\mu|^2 +|\kap_i|^2 \biggr ]\nonum \\ &+& \biggl
[ \mlh{i}+\kap_i^*\mu \biggr ]\frac{v_i}{v_d} -\widetilde{D}_i
\frac{v_i}{v_u}
\biggr \} \;, \label{minV}
\end{eqnarray}
where in both Eqs.~(\ref{Gam}) and (\ref{minV}) we have introduced
the simplifying notation
\begin{eqnarray}
\ovl{m}_{H_2}^2 &\equiv & m_{H_2}^2 +\frac{1}{2\: v_u} 
\frac{\partial {\bf \Delta V}}{\partial v_u} \;, \\
\ovl{m}_{H_1}^2 &\equiv & m_{H_1}^2 +\frac{1}{2\: v_d} 
\frac{\partial {\bf \Delta V}}{\partial v_d} \;. 
\end{eqnarray}
Eq.~(\ref{minII}) can now be cast in the form,
\begin{eqnarray}
(M_{\tilde{\nu}}^2)_{ij} v_j &=& -\biggl [ \mhl{i}+\mu^*\kap_i \biggr ]
v_d \nonumber \\&&\qquad
+ \widetilde{D}_i v_u -\frac{1}{2}\frac{\partial 
{\bf \Delta V}}{\partial v_i} \;, \label{minVI}
\end{eqnarray}
where
\begin{eqnarray}
(M_{\tilde{\nu}}^2)_{ij} &=& \ml_{ji}+\kap_i\kap_j^* +\frac{1}{2}
M_Z^2 \cos2\beta \: \delta_{ij} \nonum \\ &+& \frac{(g^2+g_2^2)}{2} 
\sin^2\beta\, (v^2-v_u^2-v_d^2)\: \delta_{ij} \;.
\end{eqnarray}

Here we outline the iterative numerical procedure we follow to obtain
the minimum of the potential for a given value of $\tan\beta$.
\begin{enumerate}
\item  We start in the RPC limit with $v_i=0$ and thus obtain from 
 Eqs.~(\ref{vevd},\,\ref{vevu}) initial values for $v_u$ and $v_d$ (in
 terms of $\tan\beta$).

\item We solve Eqs.~(\ref{minIV}) (or (\ref{muMSSM})) and (\ref{minV})
 also first in the RPC limit, $v_i=0,\kap_i=\widetilde{D}_i=\mhl{i}=0
 $, and thus obtain initial values for $\mu$ and $\widetilde{B}$. 

\item We treat $v_u,\,v_d$ and $\mu,\,\widetilde {B}$ as known and
 solve the system of Eqs.~(\ref{minVI}) in terms of the $v_i$. This
 system is linear and a lengthy analytical expression of the solution
 exists.

\item We return to the first step and compute the corrected values of
 $v_u,\,v_d$ including the $v_i$'s using
 Eqs.~(\ref{vevd},\,\ref{vevu}).  The reader should note that $\tan
 \beta= v_u/v_d$ remains exactly the same as in the R-parity
 conserving MSSM case (see Eqs.(\ref{vevd},\,\ref{vevu}). This is the
 advantage of this formulation--developed for the first time in
 Ref.~\cite{Nowakowski:1995dx}--and is used throughout this paper. In
 our calculation, we include the full one loop corrections and the
 dominant two loop ones as they have been calculated in the RPC case
 in Ref.~\cite{Dedes:2002dy} but not R-parity violating loop
 corrections~\cite{Chun}.

\item We repeat the second step but use the non-zero values of $v_i$
as well as the newly computed values of $v_u,\,v_d$. At this point we
now also include the non-zero values of $\kap_i,\,{\widetilde D}_i$.
The latter could have been included from the beginning but it is
computationally more convenient to do this in the second iteration.

\item We now iterate the procedure until convergence of $\mu,
\widetilde{B},\,v_u,\,v_d,v_i$ is reached.

\end{enumerate}
We have explicitly checked that our iteration procedure is 
very robust and for all the initial parameters we display in our
numerical results, we have found the iteration procedure to
converge. 

Finally, it is well known that the MSSM provides a mechanism of
breaking radiatively the electroweak $SU(2)_L\times U(1)_Y$ symmetry
down to $U(1)_{\rm em}$~\cite{REWSB}. Electroweak symmetry breaking in
the MSSM occurs when $m_{H_2}^2<0$ in Eq.~(\ref{minIII}). This is
indeed realized in the MSSM since $m_{H_2}^2$ is driven to negative
values by the large top Yukawa coupling once we employ the RGEs. As we
see from Eq.~(\ref{mh2}) the \rpv-couplings do not affect directly the
``running'' of $m_{H_2}^2$. However, they do affect the running of
$m_{H_1}^2$ in Eq.~(\ref{mh1}) through the mixed wave function
$H_1-L_i$. These corrections turn out to be small, since $m^2_{L_i
H_1}$ is small, in the minimal supergravity scenario we assume in this
article. Concluding, the radiative electroweak symmetry breaking in
the \rpv-case works in exactly the same way as in the RPC case.


\section{Particle and Superparticle Masses}
\label{sec:pspm}

In the literature, it is common to make a specific basis choice for
the CP-even neutral scalar fields $h_2^0,{\tilde\nu}_\alpha$, in
particular the basis where only $v_u,\,v_d\not=0$ and $v_i=0$. We
shall present our results for particle and superparticle masses in the
generic basis, where all vev's can be non-zero, $v_u,\,v_d,\,v_i\not=0
$. We shall strictly follow the conventions of Grossman and
Haber~\cite{Haber} which in the R-parity conserved limit co{\"\i}ncide
with those of Haber and Kane~\cite{Kane}. We list in turn the mass
matrices and show how they depend on our basic parameters, as well as
the minimum of the potential determined in the previous section.  It
is then straightforward for the reader to choose his/her favorite
basis or to work with the basis independent spectrum given below.

\subsection{Gauge Boson Masses}
For completeness and in order  to fix our notation below, we write here the 
masses of the Z and $W^\pm$ gauge bosons,
\begin{eqnarray}
M_W^2 &=& \frac{1}{2}\: g_2^2\: (v_u^2+v_\alpha^2)\;, \label{mw}  \\
M_Z^2 &=& \frac{1}{2}\: (g^2 +g_2^2)\: (v_u^2+v_\alpha^2)\;,
\label{mz}
\end{eqnarray}
where again $v_\alpha^2\equiv v_d^2+\sum_{i=1}^3 v_i^2$. The photon
and the gluons are of course massless. The reader should note the
participation of the sneutrino vev's $v_i$ in the masses of the Z- and
$W^\pm$-gauge bosons.

\subsection{CP-Even Higgs-Sneutrino Masses}

From Eq.~(\ref{vneutral}), we see that after electroweak symmetry
breaking, the sneutrinos, ${\tilde\nu}_i$, mix with the Higgs bosons
$h_2^0,\,h_1^0\equiv{\tilde\nu}_0$. If CP is conserved, the mass
eigenstates separate into CP-even and CP-odd states. Following
Grossman and Haber~\cite{Haber}, let us denote with $\tilde{\nu}_+$
($\tilde{\nu}_-$) the CP-even (CP-odd) sneutrino mass eigenstates. If
R-parity is broken, the mass of $\tilde{\nu}_+$ is in general different
from the mass of $\tilde{\nu}_-$, {\it i.e.} there is a sneutrino,
anti-sneutrino mass splitting. The CP-even Higgs-sneutrino mass
eigenstates are denoted by $h^0,H^0,\tilde{\nu}_{+}^i$, where the mass
$M_{h^0}<M_{H^0}$. They are obtained in the generic basis after the
diagonalization of a $5\times 5$ mass matrix
\begin{eqnarray}
{\cal L}=-\frac{1}{2} (x_2,r_\gamma) {\cal M}^2_{\rm CP-even}\left (
\begin{array}{c} x_2 \\  r_\delta \end{array} \right )\,,
\end{eqnarray}  
where 
\begin{eqnarray}
{\cal M}^2_{\rm CP-even}& =& 
\\
&&\nonum \\
&& \hspace{-2.3cm} 
\left ( \begin{array}{cc} 
\frac{\dspl{b_\alpha v_\alpha}}{\dspl{v_u}}
+\frac{\dspl{(g^2+g_2^2)}}{\dspl{2}} v_u^2 &
- b_\delta -\frac{\dspl{(g^2+g_2^2)}}{\dspl{2}} v_u v_\delta \\[3mm]
- b_\gamma -\frac{\dspl{(g^2+g_2^2)}}{\dspl{2}} v_u v_\gamma  &
(m_{\tilde{\nu}}^2)_{\gamma\delta}+ \frac{\dspl{(g^2+g_2^2)}}{\dspl{2}}
v_\gamma
v_\delta  \end{array} \right )\;,\nonumber
\label{cpeven}
\end{eqnarray}
with
\begin{eqnarray}
(m_{\tilde{\nu}}^2)_{\alpha\beta}&\equiv
& [(m_{\tilde{\cL}}^2)_{\alpha\beta}+ \mu^*_\alpha\mu_\beta
] \nonumber  \\
&&\quad -\frac{\dspl{(g^2+g_2^2)}}{\dspl{4}}
(v_u^2-v_\gamma^2)\delta_{\alpha\beta} \;, \label{snu}
\end{eqnarray}
and where $v_\gamma^2\equiv \sum_\gamma v_\gamma^2$.  Recall that $b_
\alpha=({\tilde B}, {\widetilde D}_i)$.

\subsection{CP-Odd Higgs-Antisneutrino Masses}

The CP-odd Higgs-sneutrino mass eigenstates $A,\tilde{\nu}_{-}^i$ (and
the massless Goldstone boson in the unitary gauge) are obtained in the
generic basis after the diagonalization of a $5\times 5$ mass matrix
\begin{eqnarray}
{\cal L}=-\frac{1}{2} (y_2,y_\gamma) {\cal M}^2_{\rm CP-odd}\left (
\begin{array}{c} y_2 \\  y_\delta \end{array} \right )\,,
\end{eqnarray}  
where 
\begin{eqnarray}
{\cal M}^2_{\rm CP-odd} = \left ( \begin{array}{cc} 
\frac{\dspl{b_\alpha v_\alpha}}{\dspl{v_u}} &
b_\delta \\[3mm]
b_\gamma  &
(m_{\tilde{\nu}}^2)_{\gamma\delta}  \end{array} \right ) \;.
\label{cpodd}
\end{eqnarray}
For one generation, we obtain two nonzero eigenvalues with the
eigenstates identified as the sneutrino and the CP-odd Higgs,
respectively,
\begin{eqnarray}
m_{\tilde{\nu}_-}^2 &=& \frac{1}{2}\left[ m_{\tilde{\nu}}^2+\frac{2
  \tilde{B}}{\sin2\beta} \right.\nonum \\&& \hspace{-1cm}+
\left.\sqrt{\biggl (m_{\tilde{\nu}}^2-\frac{2
    B}{\sin2\beta}\biggr )^2+4 \widetilde{D}_1^2 \biggl
  (1+\tan^2\beta \biggr )}\;\; \right], \\
M_{A^0}^2 &=& \frac{1}{2}\left[ m_{\tilde{\nu}}^2+\frac{2
  \tilde{B}}{\sin2\beta}\right.\nonum \\
&&\hspace{-1cm}- \left.\sqrt{\biggl (m_{\tilde{\nu}}^2-\frac{2
    B}{\sin2\beta}\biggr )^2+4 \widetilde{D}_1^2 \biggl
  (1+\tan^2\beta \biggr )}\;\; \right ].
\label{cpodd1gen}
\end{eqnarray}
Here $m_{\tilde\nu}$ is the one generation version of Eq.~(\ref{snu}).
Notice the $\tan\beta$ enhancement (reduction) of the sneutrino
(Higgs) mass is due exclusively to an R-parity violating contribution.
For $\widetilde{D}_1\ra 0$ we have $m_{\tilde{\nu}_-}=m_{
\tilde{\nu}}$ and $M_{A^0}^2=\frac{2\tilde{B}}{\sin2\beta}$ as it 
should be.

The generalization of the Higgs mass sum rule $M_{h^0}^2+M_{H^0}^2=
M_{A^0}^2+M_Z^2$ in the RPC case is written here as:
\begin{eqnarray}
\Tr({\cal M}^2_{\rm CP-even})=M_Z^2 + \Tr({\cal M}^2_{\rm CP-odd}) \;.
\label{sr}
\end{eqnarray}
This is easily verified from the matrix forms of ${\cal M}^2_{\rm
CP-even}$ and ${\cal M}^2_{\rm CP-odd}$ given above.  Eq.~(\ref{sr})
leads to the following Higgs mass sum rule in the \rpv-scenario,
\begin{eqnarray}
M_{h^0}^2 \: + \: M_{H^0}^2 \: + \: \sum_{i=1}^3 M_{\tilde{\nu}_+^i}^2 \ =\
\: M_{A^0}^2\: +\: M_Z^2 \: +\: \sum_{i=1}^3 M_{\tilde{\nu}_-^i}^2 \nonumber 
\\  \;.
\label{sr2}
\end{eqnarray} 
This sum rule is valid only at tree level and is altered by radiative
corrections. If the heavy Higgs mass states $A^0$ and $H^0$ are
degenerate and also the sneutrino anti-sneutrino mass difference is small
then the light Higgs boson mass $h^0$ would be very close to the
Z-boson mass.

\subsection{Charged Higgs Bosons-Sleptons}

The charged Higgs bosons mix with the charged sleptons. 
\begin{eqnarray}
{\cal L}=- (h^-_2,\tilde{e}_{L_\gamma},\tilde{e}_{R_k})
 {\cal M}^2_{\rm Charged}\left (
\begin{array}{c} h_2^+ \\  \tilde{e}_{L_\delta}^{*} \\
\tilde{e}_{R_l}^* \end{array} \right )\,.
\end{eqnarray}  
In the basis independent notation, the $8\times 8$ mass matrix is
given by
\begin{widetext}
\begin{eqnarray}
{\cal M}^2_{\rm Charged} = \left ( \begin{array}{ccc} 
(m^2)_{11} + D & b_\delta^*+ D_\delta &
\lam_{\beta\alpha l}\mu_\alpha^* v_\beta \\[3mm]
b_\gamma+D_\gamma^*  & (m^2)_{\delta\gamma}
 +\lam_{\alpha\gamma l}\lam_{\beta\delta l} 
v_\alpha v_\beta + D_{\gamma\delta} & 
h_{\alpha\gamma l} v_\alpha-\lam_{\alpha\gamma l}\mu_\alpha^* v_u
\\[3mm]
\lam^*_{\beta\alpha k} \mu_\alpha v_\beta &
h_{\alpha\delta k}^* v_\alpha -\lam_{\alpha\delta k}\mu_\alpha v_u &
\me_{lk}+\lam_{\alpha\beta k}\lam_{\alpha\gamma l }v_\beta v_\gamma
+D_{lk}
\end{array} \right ) \;,
\label{Charged}
\end{eqnarray}
\end{widetext}
with 
\begin{eqnarray}
(m^2)_{11} &\equiv& m_{H_2}^2+|\mu_\alpha|^2 \;,\quad
 \\
D &\equiv& \frac{1}{4}(g_2^2+g^2)(v_u^2-|v_\alpha|^2)+\frac{1}{2}g_2^2
|v_\alpha|^2 \;,  \\
D_\delta &\equiv& \frac{1}{2}g_2^2 v_u v_\delta \;, \\
(m^2)_{\gamma\delta} &\equiv& (m_{\tilde{\cL}}^2)_{\delta\gamma}+
\mu_\gamma \mu_\delta^* \;, \\
D_{\gamma\delta} &\equiv& \frac{1}{4}(g_2^2-g^2)(v_u^2-v_\alpha^2)
\delta_{\delta\gamma}+\frac{1}{2}g_2^2 v_\gamma v_\delta \;, \\
(D)_{lk} &\equiv& \frac{1}{2} g^2 (v_u^2-v_\alpha^2) \delta_{lk} \;. 
\end{eqnarray}
The remaining parameters are given in Eqs.~(\ref{superpot1}) and
(\ref{softhaber}). Upon diagonalization of the mass matrix
(\ref{Charged}), we obtain the mass eigenstates : $G^\pm,H^\pm,
\tilde{e}_{i=1,\ldots,6}$.  It is not hard to prove that the
determinant of (\ref{Charged}) is zero and the Goldstone boson
corresponds to the eigenvector $(-v_u,v_\alpha,0,0,0) $.

\subsection{Squarks}

\subsubsection{Down Squarks}
The down squark mass eigenstates $\tilde{d}_i, i=1,\ldots,6$ are given
by diagonalizing the following mass matrix 
\begin{eqnarray}
{\cal L}=- (\tilde{d}_{L_i}^*,\tilde{d}_{R_{i+3}}^*) {\cal M}^2_{\rm Down
  } \left ( \begin{array}{c}  \tilde{d}_{L_j} \\
\tilde{d}_{R_{j+3}} \end{array} \right ) \;,
\end{eqnarray}
where in
the $\{\tilde{d}_{L_i},\tilde{d}_{R_{i+3}}\}$ basis we have
\begin{widetext}
\begin{eqnarray}
{\cal M}^2_{\rm Down } = \left (\begin{array}{cc}
\mq_{ij}+\lam^{\prime *}_{\alpha il}\lam^{\prime}_{\gamma jl} 
v_\alpha v_\gamma +
\biggl (\frac{1}{4}g_2^2+\frac{1}{12}g^2 \biggr ) (v_u^2-v_\alpha^2
)\delta_{ij} & 
h_{\alpha ij}^{\prime *} v_\alpha - \lam^{\prime *}_{\alpha
  ij}\mu_\alpha v_u \\[3mm]
* & \md_{ij}+\lam^{\prime *}_{\alpha l j}\lam^{\prime}_{\beta l i}
v_\alpha v_\beta +\frac{1}{6} g^2 (v_u^2 -v_\alpha^2 )\delta_{ij} 
\end{array} \right ) \;. \nonum \\[3mm] \label{downsquarks} 
\end{eqnarray}
\end{widetext}
The $*$ denotes the complex conjugate of the transposed matrix
element, {\it i.e.} in the above case $({\cal M}^2_{\rm Down })
_{ij}^\dagger$.

\subsubsection{Up Squarks}
The up squark mass eigenstates $\tilde{u}_i, \,i=1,\ldots,6$ are
determined by diagonalizing the following mass matrix given in
the $\{\tilde{u}_{L_i},\tilde{u}_{R_{i+3}}\}$ basis
\begin{eqnarray}
{\cal L}=- (\tilde{u}_{L_i}^*,\tilde{u}_{R_{i+3}}^*) {\cal M}^2_{\rm Up
  } \left ( \begin{array}{c}  \tilde{u}_{L_j} \\
\tilde{u}_{R_{j+3}} \end{array} \right ) \;,
\end{eqnarray}
where 
\begin{widetext}
\begin{eqnarray}
{\cal M}^2_{\rm Up } = \left (\begin{array}{cc}
\mq_{ij}+(\yu\yud)_{ji} v_u^2-
\frac{1}{4}\biggl (g_2^2-\frac{1}{3}g^2 \biggr ) (v_u^2-v_\alpha^2
)\delta_{ij} & 
(\hu^*)_{ij} v_u-(\yu^*)_{ij}\mu_\alpha v_1 \\[3mm]
* & \mup_{ij}+(\yud\yu)_{ji} v_u^2
-\frac{1}{3} g^2 (v_u^2 -v_\alpha^2 )\delta_{ij} 
\end{array} \right ) \;. \nonum \\[3mm] \label{upsquarks} 
\end{eqnarray}
\end{widetext}

\subsection{Quarks}

The down quark masses are given by,
\begin{eqnarray}
(\mdee)_{ij}=\lam^\prime_{\alpha i j} v_\alpha \;,
\label{downquarks}
\end{eqnarray}
and the up quark masses are
\begin{eqnarray}
(\myew)_{ij}=(\yu)_{ij} v_u \;,
\label{upquarks}
\end{eqnarray}
and the coupling constants are defined in Eq.~(\ref{superpot1}).

\subsection{Neutrinos-Neutralinos}

The neutrinos mix with the neutralinos resulting in one massive
neutrino at tree level and four massive neutralinos.  The
neutrino-neutralino mass matrix ($7 \times 7$ for three generations of
neutrinos) in the $(-i \bino, -i \wino^{(3)}, \tilde{h_2^0}, \nu_
\alpha)$ basis is given by
\begin{eqnarray}
{\cal L}= -\frac{1}{2} (-i \bino, -i \wino^{(3)}, \tilde{h_2^0},
\nu_\alpha)
{\cal M}_{\rm N} \left ( \begin{array}{c}
-i \bino \\  -i \wino^{(3)} \\ \tilde{h_2^0} \\
\nu_\beta \end{array} \right ) \;, 
\end{eqnarray}
where  \cite{correction}
\begin{widetext}
\begin{eqnarray}
{\cal M}_{\rm N} = \left ( \begin{array}{cccc}
M_1 & 0 & M_Z s_W {\displaystyle\frac{v_u}{\sqrt{v_\gamma^2}}} & -M_Z s_W 
{\displaystyle\frac{v_\beta}{\sqrt{v_\gamma^2}}} \\[7mm]
0 & M_2 &  -M_Z c_W  {\displaystyle\frac{v_u}{\sqrt{v_\gamma^2}}} & M_Z c_W
{\displaystyle\frac{v_\beta}{\sqrt{v_\gamma^2}}} \\[4mm]
M_Z s_W {\displaystyle\frac{v_u}{\sqrt{v_\gamma^2}} }
& -M_Z c_W  {\displaystyle\frac{v_u}{\sqrt{v_\gamma^2}}} & 0 & -\mu_\beta  
\\[7mm]
-M_Z s_W {\displaystyle\frac{v_\alpha}{\sqrt{v_\gamma^2}}} &M_Z c_W
{\displaystyle\frac{v_\alpha}{\sqrt{v_\gamma^2}}}  & -\mu_\alpha & 
0_{\alpha\beta} 
\end{array} \right ) \;,\label{neutralino}
\end{eqnarray}
\end{widetext}
with $M_Z^2$ given in Eq.~(\ref{mz}) and $s_W\equiv\sin\theta_W$, is
the electroweak mixing angle. The matrix (\ref{neutralino}) has five
non-zero eigenvalues, {\it i.e.} four neutralinos and one neutrino.
We denote the mass eigenstates which are obtained upon diagonalization
of the matrix as: $\tilde{\chi}^0_{1,...4},\nu_{i=1,\ldots,3}$, with
the masses $M_{\tilde{\chi}^0_1}<M_{\tilde{\chi}^0_2}<M_{\tilde{\chi
}^0_3} <M_{\tilde{\chi}^0_4}$.

Since $M_1,M_2,M_Z\gg v_i$, the matrix Eq.~(\ref{neutralino}) is
suggestive of the well known sea-saw formula,
\begin{eqnarray}
{\cal M}_{\rm N} =  \left (\begin{array}{cc}
M_{\tilde{\chi}} & m \\[3mm]
m^T  & 0\end{array} \right ) \;,\label{seasaw}
\end{eqnarray}
where $M_{\tilde{\chi}}$ is the $4\times 4$ neutralino mass matrix
with mass eigenvalues typically $M_{{\tilde\chi}_i}\gsim{\cal O}(10\,
{\rm GeV})$ \cite{foot5}. The off-diagonal entry $m$ is a $3\times 4$
matrix with entries of order $g v_i,$ or $\kap_i$. In
Sect.~\ref{results}, we show and below we estimate that $\kap_i\lsim
{\cal O}(1\,{\rm MeV})$ and thus $m\ll M_{\tilde\chi}$. The analogy
with the Majorana see-saw mechanism is then obvious under the
replacements
\begin{eqnarray}
M_{\tilde{\chi}}\equiv M_{\rm SUSY} & \Longleftrightarrow & M_{\rm
Maj}\,,
\nonumber \\
g v_i,\,\kap_i &\Longleftrightarrow & M_{\rm Dirac}\;.
\end{eqnarray}
In addition, the $3\times 3$ zero mass matrix in Eq.~(\ref{seasaw}) can
be filled by finite, loop low energy threshold corrections in the
\rpv-MSSM as opposed to possible Higgs triplet contributions in other
neutrino mass models. Therefore neutrino masses will roughly be given
by
\begin{eqnarray}
m_\nu \sim  \frac{m^2}{M_{\rm SUSY}} \sim \frac{g^2 v_i^2}{M_{\rm
      SUSY}} \lsim 1~{\rm eV}\;.
\end{eqnarray}
For the last inequality, we have imposed the bound from WMAP in
Eq.~(\ref{eq:wmap}). Bearing in mind possible accidental cancellations
(see below) we obtain
\begin{equation}
v_i,\,\kap_i \lsim 1\, {\rm MeV}\quad {\rm for}\quad M_{\rm
SUSY}\lsim 1\, {\rm TeV}.  
\end{equation}
A complete calculation of the one neutrino mass eigenvalue at tree
level reads~\cite{Joshipura,Nowakowski:1995dx}
\begin{eqnarray}
m_\nu = \frac{\mu (M_1 g_2^2 + M_2 g^2) \: \sum_{i=1}^3 \Lam_i^2}{ 
 2 v_u v_d (M_1 g_2^2 + M_2 g^2) - 2 \mu M_1 M_2} \;,
\label{numass}
\end{eqnarray}
with 
\begin{eqnarray}
\Lam_i \equiv v_i - v_d \frac{\kappa_i}{\mu} \label{bigL}\;.
\end{eqnarray}
A redefinition of the phases of the gaugino fields $\widetilde {\cal
  B}$ and $\widetilde{\cal W}$ together with the gaugino universality
assumption $M_1=M_2\equiv M_{1/2}$, can make $M_1$ and $M_2$ real and
positive and so the numerator of Eq.~(\ref{numass}) cannot be fine
tuned to zero (provided $\mu>\mathcal{O}(100\,{\rm GeV})$). According
to the universality assumption, the 1-loop unification gaugino masses
at the electroweak scale are, $M_1=\frac{5}{3}\frac{\alpha_1^2}{\alpha
  _{\rm GUT}^2} M_{1/2}$ and $M_2 =\frac{\alpha_2^2}{\alpha_{\rm
    GUT}^2} M_{1/2}$, where $\alpha_{\rm GUT}=g_{\rm GUT}^2/4\pi\simeq
0.041$ is the grand unified coupling constant. Taking into account
that $v_u v_d \ll \mu M_{1/2}$, which we find in our numerical results
below, we arrive with an excellent approximation at a simple formula
for the tree level neutrino mass:
\begin{eqnarray}
m_\nu = - \frac{16\pi \alpha_{\rm GUT}}{5} \:
\frac{\sum_{i=1}^3 \Lam_i^2}{M_{1/2}} \;.
\label{lownumass}
\end{eqnarray}
This implies $\Lam_i \lsim 1$ MeV for $M_{1/2} \lsim 1$ TeV. One
can obtain a small $\Lam_i$ even with $v_i\sim \kappa_i \sim v$ but
that requires a cancellation of 1 part in $10^5$. So the question
arises how one can naturally obtain a small $\Lam_i$, {\it i.e.} 
$v_i \sim\kappa_i \lsim {\cal O}(1\,{\rm MeV})$? We will come to this
point in Sect.~\ref{results}.

\subsection{Leptons-Charginos}

The charged leptons mix with the charginos. The Lagrangian contains
the ($5 \times 5$ for three generations of leptons) lepton-chargino
mass matrix as \cite{notation}
\begin{eqnarray}
{\cal L}= - (-i \wino^-, 
e_{L_\alpha}^-)
{\cal M}_{\rm C} \left ( \begin{array}{c}
-i \wino^+ \\ \tilde{h}_2^+ \\
e_{R_k}^+ \end{array} \right ) + {\rm h.c.}\,,
\end{eqnarray}
where the mass eigenstates $\tilde{\chi}^\pm_{1,2}, \ell=(e,\mu,\tau)$
are given upon the diagonalization of the matrix
\begin{eqnarray}
{\cal M}_{\rm C} = \left ( \begin{array}{ccc}
M_2 & g_2 v_u & 0_k \\
g_2 v_\alpha & \mu_\alpha & \lam_{\beta\alpha k} v_\beta \end{array}
\right ) \;. \label{chargino}
\end{eqnarray}

\section{Boundary Conditions at \mx}
\label{boundary}

Due to the large number of parameters in the supersymmetry breaking
sector ({\it c.f.} Eq.~(\ref{softhaber})), we shall focus on the case
of minimal supergravity models. These have a much simplified structure
at the high scale, which we assume here to be the unification scale of
the gauge couplings, $M_X= M_{\rm GUT}={\cal O}(10^{16})$.  At this
scale, the soft SUSY breaking scalar masses have a common value,
$M_0$:
\begin{eqnarray}
{\bf m}_{\tilde Q}(M_X) &=& {\bf m}_{\tilde u}(M_X) = {\bf m}_{\tilde
  d}(M_X) = \nonumber \\
{\bf m}_{\tilde L}(M_X) &=& {\bf m}_{\tilde e}(M_X) \equiv M_0
{\bf \hat{1}} \\[2mm]
 m_{H_1}(M_X) &=& m_{H_2}(M_X) \equiv M_0  \;,
\end{eqnarray}
where ${\bf \hat{1}}$ is the $3\times 3$ unit matrix in flavour space.
Motivated by the discussion of Sect.~III, we shall assume that we can
rotate away the $\kappa_i$ terms before supersymmetry breaking and no
$\widetilde{D}_i$ or $\mlh{i}$ terms are generated through
supersymmetry breaking at the unification scale $M_X$,
\begin{eqnarray} 
\kappa_i(M_X) = 0\,, \quad \widetilde{D}_i(M_X)=\mlh{i}(M_X)=0 \;.
\end{eqnarray}
At the scale $M_X$, we shall assume one non-zero \rpv-coupling at a
time, {\it i.e.} one coupling from:
\begin{eqnarray}
\lam_{ijk}(M_X) \ne 0 \;\;,\;\; \lam_{ijk}'(M_X) \ne 0
\;\;,\;\; \lam_{ijk}''(M_X) \ne 0 \;.
\end{eqnarray}
Due to the CKM quark mixing, the $\lam'$ RGEs are coupled. Thus in the
case of a single $\lam'(M_X)\not=0$ we will have more than one $\lam'
(M_Z)\not =0$ at the weak scale. mSUGRA assumptions lead to the same
prefactors, $A_0$ of the supersymmetry breaking trilinear couplings
$h_{ijk} \equiv A_0 Y_{ijk}$:
\begin{eqnarray}
{\bf A_U}(M_X) &=& {\bf A_D}(M_X) = {\bf A_E}(M_X) = \nonumber \\ 
{\bf A_\lam}(M_X) &=& {\bf A_{\lam'}}(M_X) ={\bf
    A_{\lam''}}(M_X)
\equiv A_0 {\bf \hat{1}} \;.
\end{eqnarray}
A common mass, $M_{1/2}$ for the gauginos completes the mSUGRA
boundary conditions at $M_X$,
\begin{eqnarray}   
M_1(M_X)=M_2(M_X)=M_3(M_X) \equiv M_{1/2}\;.
\end{eqnarray}
No assumption for quark or lepton Yukawa unification has been made in
our analysis. We thus have the six parameters :
\begin{equation}
A_0,\; M_0,\; M_{1/2},\;\tan\beta,\; sgn(\mu),\;\{\lam,\lam',
\lam''\}_1.
\label{rpv-parameters2}
\end{equation}  
When determining the mass spectrum, in order to further simplify the
number of input parameters we will restrict ourselves to a particular
supergravity scenario called ``no-scale'' supergravity
\cite{noscale}. This scenario predicts a definite relation
between $A_0$ and $M_0$ namely
\begin{eqnarray}
A_0=M_0=0 ~{\rm GeV}\;.
\end{eqnarray}
The ``no-scale'' scenario, the simplest mSUGRA scenario, is
experimentally excluded in the RPC case, but as we show below allowed
in the \rpv-case. Our results for the bounds on the \rpv-couplings
from neutrino masses should be unaffected by this assumption provided
$(M_0,\,|A_0|)/M_{1/2}<10$. This is because $M_{1/2}$ dominates the
renormalization group behaviour.

In this paper, we only address gravity mediated supersymmetry breaking
and do not consider other scenarios, such as gauge (GMSB)~\cite{GMSB}
or anomaly mediated (AMSB)~\cite{AMSB} supersymmetry breaking. 
Although, the low energy spectrum formul\ae\ we displayed in the
previous section are unchanged, the results for the bounds on the
\rpv-couplings or the LSP content change dramatically from one model
to the other as we will see shortly. We hope that this paper serves as
a basis to study the phenomenology of other SUSY breaking models.

\section{Results}
\label{results} 

In the following numerical analysis, we use a version of {\small \tt
  SOFTSUSY}~\cite{Allanach:2001kg} which has been augmented with
\rpv-couplings.  The beta functions for the \rpv-MSSM couplings and
masses contain the full one-loop \rpv\ and RPC contributions.  The
beta functions for the RPC MSSM couplings and masses also contain the
two-loop pure RPC corrections. As discussed in Sect.V, small neutrino
masses imply that the sneutrino vev's must be small. Although we
derive their values from the minimization of the scalar potential, we
neglect them in the calculation of sparticle masses. This is a good
approximation, valid to $\mathcal{O} (v_i/M_{SUSY}) \ll 1$, when
considering only the spectrum of sparticles and not the small mixing
induced by \rpv-couplings. We have checked that the error induced in
the sparticle masses is much smaller than the current theoretical
uncertainty in the RPC part of the
calculation~\cite{Allanach:2003jw,Azuelos:2002qw,Allanach:2001hm}. The
\rpv\ contribution to the SM Yukawa couplings and fermion masses,
however, is taken into account as described in Sect.~\ref{sec:pspm}.
Radiative electroweak symmetry breaking and the determination of
sneutrino vev's follows the discussion in Sect.~\ref{rad-break}.
{\small \tt SOFTSUSY} adds one-loop RPC threshold corrections to the
sparticle and Higgs masses, and takes one-loop RPC threshold
corrections into account when calculating the Yukawa and gauge
couplings. For further details on the RPC part of the calculation
consult Ref.~\cite{Allanach:2001kg}. Numerical results from the
aumented version of the program {\small \tt SOFTSUSY}, \ie beta
functions, neutrino masses, electroweak breaking, the mass spectrum,
and bounds on the couplings {\it etc} have been carefully checked with
an independent Fortran code~\cite{SUITY}.

We use the input parameters~\cite{pdg} $m_t=174.3$ GeV,
$\alpha_s^{\overline {MS}}(M_Z)=0.1172$ and $m_b(m_b)^{\overline
{MS}}=4.25$ GeV, corresponding to $m_b^{pole}=5.0$ GeV at the 3-loop
level.  Other SM $\overline{MS}$ masses input are: $m_u(2 \gev)=3.0
\times 10^{-3}$ GeV, $m_c(m_c) = 1.2$ GeV, $m_d(2
\gev)=6.75 \times 10^{-3}$, $m_s(2 \gev)= 0.1175$ GeV. The pole lepton
masses are taken as $m_e=5.11 \times 10^{-4}$ GeV, $m_\mu=0.10566$ GeV
and $m_\tau=1.777$ GeV. The Fermi constant $G_F=1.16637 \times
10^{-5}$ GeV$^{-2}$, the fine structure constant
$\alpha(0)^{-1}=137.03599976$ and $M_Z=91.1876$ GeV are used to
determine the electroweak gauge couplings.

\subsection{Bounds on Lepton-number Violating Couplings}
\label{bounds}

\subsubsection{Procedure}
\label{procedure}

We first use the numerical analysis of the RGEs to set bounds upon the
lepton-number violating couplings $(\lam_{ijk},\,\lam'_{ijk})$ from
the cosmological neutrino mass bound and requiring the absence of
negative mass-squared scalars other than the Higgs and sneutrinos.
(This does not refer to the physical mass and thus does not constitute
a tachyon.) Neutrinos contribute to the hot dark matter and as such
can free-stream out of smaller scale fluctuations during matter
domination in the early universe. This changes the shape of the matter
power spectrum and suppresses the amplitude of fluctuations. Combining
the 2dFGRS data \cite{2dFGRS} together with the WMAP measurement
\cite{wmap} one can thus set a bound on the neutrino mass at
95$\%~C.L.$
\begin{equation}
\sum_i m_{\nu_i} < 0.71\,\mbox{eV}.\label{bwmap}
\end{equation}

Scalar mass squared values can be driven negative during the RG
evolution between the GUT- and the weak-scale, as happens to the Higgs
in radiative electroweak symmetry breaking. But if any of the
electrically charged or colour MSSM scalar fields develop negative
mass squared values, QED or QCD would be broken, in conflict with
observation. We therefore reject such values of $\lam,\, \lam'$.

Neutrino mass and charge- and colour-breaking minima bounds depend not
only upon the \rpv-couplings, but also on the RPC SUSY breaking
parameters. For a definite quantitative analysis, we therefore take an
example set of SUSY breaking parameters.  We choose the SPS1a mSUGRA
point~\cite{sps} which has the following parameter values: $M_0$=100
GeV, $M_{1/2}=250$ GeV, and trilinear couplings $A_0=-100$ GeV at
$M_X$. $\tan\beta(M_Z)=10$ and $\mu>0$ are also imposed.

As stated in Sect.~\ref{intro}, a single non-zero \rpv-coupling at
$M_X$ will generate through the coupled RGEs non-zero $\kappa_i(M_Z)$,
$\widetilde{D}_i(M_Z)$ and $\mhl{i}(M_Z)$. This is seen explicitly in
the RGEs in
Eqs.~(\ref{kappa1},\,\ref{LH-mix},\,\ref{smass},\,\ref{mh1lrge}),
where the anomalous dimension $\gamma_{L_i}^{H_1}$ couples $\mu$ and
$\kap_i$ as well as the soft breaking sfermion masses, {\it e.g.} 
${\bf m}_{\tilde{\bf D}}^2$, with $\mhl{i}$. Since the anomalous
dimension
\begin{eqnarray}
\gamma_{L_i}^{H_1}\propto  ({\bf Y}_E\Lam_E+ {\bf Y}_D\Lam_D)\, ,
\label{osama}
\\ [-0.2cm] \nonumber
\end{eqnarray}
$\kappa_i(M_Z),$ $\widetilde{D}_i(M_Z),\,\mhl{i}(M_Z)$ are also
proportional to $({\bf Y}_E\Lam_E+ {\bf Y}_D\Lam_D)$. Through
$\kappa_i,$ $\widetilde{D}_i,\,\mhl{i}\not=0$ at the weak scale, we
obtain non-zero sneutrino vev's, as can be seen from
Eq.~(\ref{minVI}). This in turn gives us a non-zero neutrino mass as
seen in Eq.~(\ref{lownumass}). In order to estimate the resulting
neutrino mass, we na\"{\i}vely integrate the RGEs assuming constant
parameters and insert our result into Eq.~(\ref{lownumass}). We obtain

\begin{widetext}
\begin{eqnarray}
m_\nu \simeq -\frac{16 \pi \alpha_{\rm GUT}}{5 M_{1/2}} \: 
\biggl [\frac{v_d}{16\pi^2}\biggr ]^2 \: 
\biggl [ \ln\frac{M_{\rm GUT}}{M_Z} \biggr ]^2\: 
\Biggl [\sum_{i=1}^3\biggl (
3 \lam'_{ijq} \cdot (Y_D)_{jq} + \lam_{ijq} \cdot (Y_E)_{jq}\biggr )\Biggr ]^2
\: f^2(\frac{\mu^2}{M_0^2} ; \frac{A_0^2}{M_0^2} ;
\frac{\tilde{B}}{M_0^2}
; \tan\beta ) \;,
\label{nuinduced}
\end{eqnarray}
\end{widetext} 
where $f$ is a complicated dimensionless function of the SUSY
parameters with typical values ${\cal O}(10)$. A similar result was
obtained some years ago by Nardi~\cite{Nardi:1996iy}. In
Eq.~(\ref{nuinduced}), we explicitly see the dependence of the induced
neutrino mass on the product of \rpv- and Higgs-Yukawa couplings from
Eq.~(\ref{osama}). Given a neutrino mass bound, {\it e.g.} 
Eq.~(\ref{bwmap}), we can thus derive bounds on the \rpv-couplings. In
the case where the down-like quark or the charged lepton mass matrix
are diagonal, only the \rpv-couplings $\lam'_{ikk}$ or $\lam_
{ikk}$ induce neutrino masses. Thus in the case of the $LL{\bar
E}$-operators, since we do not include lepton mixing, we only obtain
bounds on $\lam_{ikk}$, {\it c.f.} Table~\ref{tab:bounds2}. For the
quarks we include the CKM-mixing and thus obtain bounds on all
$\lam'$, {\it c.f.} Table~\ref{tab:bounds}.

Eq.~(\ref{nuinduced}) works as an order of magnitude estimate. Setting
$\alpha_{GUT}=0.041$, $M_{1/2}=250$ GeV, $\tan\beta=10$, $Y_b=0.18$
and $f=10$ and using the WMAP bound Eq.~(\ref{bwmap}), we obtain
\begin{equation}
\sum_{i=1}^3\biggl (3 \lam'_{ijq} \cdot (Y_D)_{jq} + 
\lam_{ijq} \cdot (Y_E)_{jq}\biggr ) < 2\cdot 10^{-5}\,.
\end{equation}
With $Y_b=0.18$, we thus obtain the single bound $\lam'_{333}<
3\times 10^{-5}$. Full numerical integration shows that $\lam'
_{333}< 6\times10^{-6}$. Note that the only $\tan\beta$ dependence
in Eq.~(\ref{nuinduced}) is in the function $f$.

Another interesting remark arises from Eq.~(\ref{nuinduced}): the
higher the ultraviolet scale is (here denoted as $M_{\rm GUT}$) the
larger the resulting neutrino mass and the stronger the bound on the
$\lam',\lam$. Therefore, for the mSUGRA scenario, $M_{\rm GUT}\simeq
2\cdot10^{16}$ GeV the bounds are stronger than for the GMSB model
where $M_{\rm GUT}$ must be taken at the intermediate energies
$10^{11}$ GeV.

We also have to remark here on another independent source for neutrino
masses in the \rpv-mSUGRA scenario coming from finite threshold
effects involving squark-quark or slepton-lepton loops. The resulting
neutrino masses are given by~\cite{Haber,Davidson:2000uc} 
\begin{eqnarray}
(m_\nu^{\rm loop})_{ij} &=&\frac{1}{32\pi^2} 
\sum_{k,l}\lam_{ikl}\lam_{jlk} m_k^\ell \sin 2\phi_k^\ell
\ln\frac{m_{\tilde{\ell}_{k_1}}^2}{m_{\tilde{\ell}_{k_2}}^2} \nonumber \\
&+& \frac{3}{32\pi^2}
\sum_{k,l}\lam'_{ikl}\lam'_{jlk} m_k^{d} \sin 2\phi_k^{d}
\ln\frac{m_{{{\tilde d}_{l_1}}}^2}{m_{\tilde{d}_{l_2}}^2}
\;,\nonumber \\ \label{loopnu}
\end{eqnarray}
with $m^\ell_k\,(m^d_k)$ the lepton (down-quark) masses, $\phi^\ell(
\phi^d)$ the slepton (squark) mixing angles and $m_{\tilde{l}_i}\,
(m_{\tilde{d}_ {l_i}})$ are the slepton (squark) mass eigenstates
\cite{foot6}. More details are found in Ref.~\cite{Haber,Davidson:2000uc}.  
Since the mixing in the first and second generation is negligible and
also sleptons are almost degenerate the finite neutrino effects of
Eq.~(\ref{loopnu}) are not significant for the heaviest neutrino as
compared to the ones induced from Eq.~(\ref{nuinduced}). For the third
generation we find
\begin{eqnarray}
\frac{m_\nu^{\rm loop}}{m_\nu} =
\frac{\ln\frac{m_{\tilde{b_1}}}{m_{\tilde{b_2}}}}
{ \frac{\alpha_{\rm  GUT}}{M_{1/2}}
\frac{3 m_b}{\pi} \left(\ln\frac{M_{\rm GUT}}{M_Z}\right)^2 
f^2 } \simeq 10^{-2}.
\end{eqnarray}
The above estimate shows that bounds derived from
Eq.~(\ref{nuinduced}) are stronger than those derived from
Eq.~(\ref{loopnu}) \cite{nu-bounds}. Thus the new bounds on the
\rpv-couplings presented in Table II are determined using the
constraint Eq.~(\ref{bwmap}), the full solution to the one-loop RGEs
and an accurate numerical diagonalisation of the neutralino/neutrino
mass matrix.

\subsubsection{Quark Bases}

Before discussing our results, we must insert a discussion on bases.
In our initial parameter set at the GUT scale ({\it c.f.} 
Eq.~(\ref{rpv-parameters2})), the \rpv-couplings are given in the
weak-current eigenstate basis. Similarly the Higgs Yukawa coupling
matrices ${\bf Y}_E,\,{\bf Y}_D,\,{\bf Y}_U,$ and the corresponding
mass matrices are also given in this basis, {\it i.e.} in general they
are not diagonal. The matrices are diagonalized by rotating the left-
and right-handed charged lepton and quark fields from the weak basis
($^{\rm w}$) to the mass basis ($^{\rm m}$)
\begin{eqnarray}
(e_{L,R}^{\rm m})_i &=& ({\bf E}_{\bf L,R})_{ij}\,(e_{L,R}^{\rm w})_j\,,
\\[1.5mm] (u_{L,R}^{\rm m})_i &=& ({\bf U}_{\bf L,R})_{ij}\,(u_{L,R}^{\rm
w})_j \,,\\ [1.5mm](d_{L,R}^{\rm m})_i &=& ({\bf D}_{\bf
L,R})_{ij}\,(d_{L,R}^{\rm w})_j\,.
\end{eqnarray}
In general, the rotation of the left-handed fields ({\it e.g.} ${\bf
U}_{\bf L}$) is different from the right-handed fields (${\bf U}_{\bf
R}$). In the weak basis, due to the non-diagonal elements in ${\bf
Y}_E,\,{\bf Y}_D,\,{\bf Y}_U,$ the RGEs for different \rpv-couplings
are coupled. Thus given one coupling at $M_X$ in the weak basis, we
will in general generate an entire set at $M_Z$ (in the weak basis).
In order to perform this computation, we must know the explicit form
for the Higgs Yukawa matrices. However experimentally, all we know is
the CKM matrix at the weak scale
\begin{equation}
{\bf V}_{\bf CKM}={\bf U}_{\bf L}^\dagger {\bf D}_{\bf L}\,,
\end{equation}
as well as the diagonal matrices in the mass eigenstate basis.
\begin{eqnarray}
[{\mdee}]_{\mbox{diag}}(M_Z)&=&{\rm diag}(
m_d,\, m_s,\,m_b)(M_Z)\,,\\
\left[{\myew}\right]_{\mbox{diag}}(M_Z)&=&{\rm diag}(
m_u,\, m_c,\,m_t)(M_Z)\,.
\end{eqnarray}
For $V_{CKM}$, we use the central values of the
mixing angles in the ``standard'' parameterization detailed in
Ref.~\cite{pdg}
\barr
s_{12} = 0.2195,\quad s_{23} =0.039 ,\quad s_{13} = 0.0031.
\earr
We neglect the CP-violating phase $\delta_{13}=0$.

In order to perform the computation, we shall make the following 
simplifying assumptions.
\begin{enumerate}
\item Due to the uncertainty concerning the neutrino masses and mixings
we shall here assume that ${\bf Y}_E$ is diagonal in the weak current
basis and thus 
\begin{equation}
({\bf E}_{\bf L,R})_{ij}=\delta_{ij}\,.
\end{equation}
We shall return to the discussion of massive neutrinos and their
mixings in our framework in a future publication.
\item We shall assume that ${\bf Y}_{D,U}$ are real and symmetric. Thus
${\bf U}_{\bf L}={\bf U}_{\bf R}$ and ${\bf D}_{\bf L}={\bf D}_{\bf
R}$.
\item When determining bounds below, we consider three extreme cases: 
(a) no-mixing, (b) the mixing is only in the down quark sector, (c)
the mixing is only in the up-quark sector. This corresponds to
\begin{eqnarray}
\begin{array}{clc}
(a)& {\bf D}_{\bf L,R}={\bf 1},&{\bf U}_{\bf L,R}={\bf 1}\,, \\
(b)& {\bf D}_{\bf L,R}={\bf V}_{CKM},\quad & {\bf U}_{\bf L,R}={\bf 1}\,, \\
(c)& {\bf U}_{\bf L,R}={\bf V}_{CKM}, & {\bf D}_{\bf L,R}={\bf 1} \,.
\end{array}
\label{models}
\end{eqnarray}
\end{enumerate}
In these three scenarios, the mass matrices at the weak scale and in
the weak current basis are then given by
\begin{equation}
\begin{array}{cl}
(a) & \mdee(M_Z) = [{\mdee}]_{\mbox{diag}} (M_Z)\,,
 \\ &
\myew(M_Z) = [{\myew}]_{\mbox{diag}} (M_Z) \,,\\[1.7mm]
(b) & \mdee(M_Z) = V_{CKM}^*\cdot [{\mdee}]_{\mbox{diag}} (M_Z)
\cdot V_{CKM}^T\,, \\ & \myew(M_Z) = [{\myew}]_{\mbox{diag}} (M_Z) \,,\\[1.7mm]
(c) & \mdee(M_Z) = [{\mdee}]_{\mbox{diag}} (M_Z)\,, \\
& \myew(M_Z) = V_{CKM}^*\cdot [{\myew}]_{\mbox{diag}} (M_Z)
\cdot V_{CKM}^T \,,
\end{array}
\label{models2}
\end{equation}
Thus in each scenario, the matrices $\mdee(M_Z),\,\myew(M_Z)$ are
determined uniquely in terms of their eigenvalues and the CKM matrix.

The Higgs Yukawa matrices ${\bf Y}_{D,U}$ are proportional to the mass
matrices. Therefore in each scenario of
Eqs.~(\ref{models},\ref{models2}) the RGEs are fully determined. Given
a set of \rpv-couplings at $M_X$ (of which we will here only choose
one to be non-zero), we can then compute the \rpv-couplings (including
$\kap_i$) at the weak scale in the weak current basis. Given the full
set of parameters at $M_Z$ we can diagonalize the neutrino/neutralino
mass matrix in Eq.~(\ref{neutralino}) and compute the neutrino
mass. For a check this neutrino mass should be identical with the one
derived in Eq.~(\ref{numass}).  We can then use the experimental bound
on the neutrino mass, Eq.~(\ref{bwmap}), to determine a bound on the
\rpv-coupling, {\it in the weak current basis}.

For comparison with experiment we must rotate to the quark mass
eigenstate bases in scenarios $(b)$, $(c)$, Eq.~(\ref{models}). To do
this, we follow the procedure of Ref.~\cite{Agashe}. For scenario
$(b)$, with all the mixing in the down quark sector, we obtain the
\rpv-interactions for the superfields in the quark mass eigenbasis
\barr
{\cal W}_{/\!\!\!\!R_p}^{(a)} &\supset& \lam'_{ijk} (V_{CKM}^\dagger)_{mk} 
\biggl [N_i (V_{CKM})_{jl}D_l-E_iU_j \biggr ] \bar{D}_m \nonr \\
&+ &\frac{1}{2}\lam''_{ijk}(V_{CKM}^\dagger)_{mj}
 (V_{CKM}^\dagger)_{nk} \bar{U}_i \bar{D}_m
\bar{D}_n.
\label{downmassbasis}
\earr
Referring to Eq.~(\ref{downmassbasis}), we define the rotation of the
couplings to the quark mass basis (denoted with a tilde)
\barr
\widetilde{\lam}'_{ijk} &=& \lam'_{ijm} (V_{CKM}^*)_{mk}, \label{pdcm}\\
\widetilde{\lam}''_{ijk} &=& \lam''_{imn}(V_{CKM}^*)_{mj}(V_{CKM}^*)_{nk}.
\label{ppdcm}
\earr
For scenario $(c)$, with all mixing in the up-sector, and the
superfields in the quark mass eigenstate basis, the superpotential
terms are
\barr
{\cal W}_{/\!\!\!\!R_p}^{(b)} &\supset& \lam'_{ijk}  \biggl [
N_i D_j-E_iU_l (V_{CKM}^\dagger)_{jl} \biggr ] \bar{D}_k \nonr \\
+& &\frac{1}{2}\lam''_{ijk}(V_{CKM})_{li}  \bar{U}_l \bar{D}_j
\bar{D}_k\,.
\label{upmassbasis}
\earr
This implies the rotation of \rpv-couplings
\barr
\widetilde{\lam}'_{ijk} &=& \lam'_{ilk} (V_{CKM}^*)_{jl}\,, \label{pucm}\\
\widetilde{\lam}''_{ijk} &=& \lam''_{ljk}(V_{CKM})_{il} \,,
\label{ppucm}
\earr
where in the first term we have taken the rotation of the $EUD$ term.

Another set of bounds applied on the \rpv-couplings $\lam'_{ijk}$
arises from the requirement of no sneutrino tachyons, {\it i.e.}  we
require the physical mass $m_{\tilde {\nu}}^2\geq0$. The resulting
bound has been observed first by de Carlos and White~\cite{CW} and can
be estimated as
\begin{eqnarray}
\sum_{jk} \lam_{ijk}^{'2}(M_X) < \frac{m_0^2 +0.5 M_{1/2}^2 +
  \frac{1}{2} M_Z^2 \cos 2\beta}{13 m_0^2 +49 M_{1/2}^2
  -\frac{3}{2}A_0 M_{1/2}-12 A_0^2}\;. \nonumber \\ \label{snbound}
\end{eqnarray}
For the SPS1a benchmark scenario this bound sets {\it all} $\lam'_
{ijk}(M_X)$ to be less than 0.13 in good agreement with the exact
numerical solutions of the RGEs in Table II below.

\subsubsection{Discussion of the Bounds}

\begin{table*}[t]
\begin{center}
\begin{tabular}{|c|c|c|c|c|c|c|}
\hline
 & \multicolumn{2}{c|}{No mixing} &\multicolumn{2}{c|}{Up mixing}
 &\multicolumn{2}{c|}{Down mixing} \\ \hline
 & $M_{GUT}$ &   $M_Z$ & $M_{GUT}$ &   $M_Z$ & $M_{GUT}$ & $M_Z$ \\ \hline
$\;\lam'_{111}\;$ & \ 1.8$\times {10^{-3}}^\nu\;$&\ 6.0$\times {10^{-3}}\;$ 
& \ 1.8$\times
 {10^{-3}}^\nu\;$ & \ 5.9$\times 10^{-3}\;$ & \ 1.0$\times {10^{-3}}^\nu\;$
& \ 3.2$\times 10^{-3}\;$\\
$\lam'_{211}$ & 1.8$\times {10^{-3}}^\nu$&6.0$\times {10^{-3}}$ & 1.8$
\times {10^{-3}}^\nu$ & 5.9$\times 10^{-3}$ & 1.0$\times {10^{-3}}^\nu$
& 3.2$\times 10^{-3}$\\
$\lam'_{311}$ & 1.8$\times {10^{-3}}^\nu$&6.0$\times {10^{-3}}$  & 1.8$
\times {10^{-3}}^\nu$ & 5.9$\times 10^{-3}$ & 1.0$\times {10^{-3}}^\nu$&
 3.2$\times 10^{-3}$\\
$\lam'_{121}$ & 0.13$^t$&0.39 & 0.13$^t$ & 0.38  & 5.0$\times {10^{-4}}^\nu$
& 1.6$\times 10^{-3}$\\
$\lam'_{221}$ & 0.13$^t$&0.39 & 0.13$^t$ & 0.38  & 5.0$\times {10^{-4}}^\nu$
& 1.6$\times 10^{-3}$\\
$\lam'_{321}$ & 0.13$^t$&0.39 & 0.13$^t$ & 0.38  & 5.0$\times {10^{-4}}^\nu$
& 1.6$\times 10^{-3}$\\
$\lam'_{131}$ & 0.15$^t$& 0.40& 0.15$^t$& 0.40 & 9.1$\times {10^{-4}}^\nu$
& 2.6$\times 10^{-3}$\\
$\lam'_{231}$ & 0.15$^t$& 0.40& 0.15$^t$& 0.40 & 9.1$\times {10^{-4}}^\nu$
& 2.6$\times 10^{-3}$\\
$\lam'_{331}$ & 0.15$^t$& 0.40& 0.15$^t$& 0.40 & 9.0$\times {10^{-4}}^\nu$
& 2.6$\times 10^{-3}$\\
$\lam'_{112}$ & 0.13$^t$&0.39 & 0.13$^t$ & 0.38  & 5.0$\times {10^{-4}}^\nu$
& 1.6$\times 10^{-3}$\\
$\lam'_{212}$ & 0.13$^t$&0.39 & 0.13$^t$ & 0.38  & 5.0$\times {10^{-4}}^\nu$&
 1.6$\times 10^{-3}$\\
$\lam'_{312}$ & 0.13$^t$&0.39 & 0.13$^t$ & 0.38  & 5.0$\times {10^{-4}}^\nu$
& 1.6$\times 10^{-3}$\\
$\lam'_{122}$ & 1.0$\times {10^{-4}}^\nu$&3.5$\times 10^{-4}$ & 1.1$\times
 {10^{-4}}^\nu$&  3.4$\times
 10^{-4}$& 1.0$\times {10^{-4}}^\nu$& 3.3$\times 10^{-4}$\\
$\lam'_{222}$ & 1.1$\times {10^{-4}}^\nu$& 3.5$\times 10^{-4}$& 1.1$\times
 {10^{-4}}^\nu$&  3.4$\times 10^{-4}$& 1.0$\times {10^{-4}}^\nu$& 3.3$\times
 10^{-4}$\\
$\lam'_{322}$ & 1.1$\times {10^{-4}}^\nu$& 3.4$\times {10^{-4}}$&
 1.1$\times {10^{-4}}^\nu$&  3.4$\times 10^{-4}$& 1.0$\times {10^{-4}}^\nu$&
 3.3$\times 10^{-4}$\\
$\lam'_{132}$ & 0.15$^t$& 0.40& 2.6$\times{10^{-2}}^\nu$& 7.7$\times 10^{-2}$
 & 7.6$\times {10^{-5}}^\nu$& 2.2$\times 10^{-4}$\\
$\lam'_{232}$ & 0.15$^t$& 0.40& 2.6$\times{10^{-2}}^\nu$& 7.7$\times 10^{-2}$
 &7.6$\times {10^{-5}}^\nu$& 2.2$\times 10^{-4}$\\
$\lam'_{332}$ & 0.15$^t$& 0.40& 2.6$\times{10^{-2}}^\nu$& 7.6$\times 10^{-2}$
 &7.5$\times {10^{-5}}^\nu$& 2.2$\times 10^{-4}$\\ 
$\lam'_{113}$ & 0.13$^t$&0.39 & 5.1$\times{10^{-3}}^\nu$& 1.6$\times 10^{-2}$
 &8.2$\times {10^{-4}}^\nu$& 2.7$\times 10^{-3}$\\ 
$\lam'_{213}$ & 0.13$^t$&0.39 & 5.1$\times{10^{-3}}^\nu$& 1.6$\times 10^{-2}$
 &8.2$\times {10^{-4}}^\nu$& 2.7$\times 10^{-3}$\\ 
$\lam'_{313}$ & 0.13$^t$&0.39 & 5.1$\times{10^{-3}}^\nu$& 1.6$\times 10^{-2}$
 &8.1$\times {10^{-4}}^\nu$& 2.7$\times 10^{-3}$\\ 
$\lam'_{123}$ & 0.13$^t$&0.39 & 7.1$\times{10^{-4}}^\nu$& 2.3$\times 10^{-3}$
 & 6.9$\times {10^{-5}}^\nu$& 2.2$\times 10^{-4}$\\
$\lam'_{223}$ & 0.13$^t$&0.39 & 7.1$\times{10^{-4}}^\nu$& 2.3$\times 10^{-3}$
 &6.9$\times {10^{-5}}^\nu$& 2.2$\times 10^{-4}$\\ 
$\lam'_{323}$ & 0.13$^t$&0.39 & 7.0$\times{10^{-4}}^\nu$& 2.2$\times 10^{-3}$
 &6.8$\times {10^{-5}}^\nu$& 2.2$\times 10^{-4}$\\ 
$\lam'_{133}$ & 3.1$\times {10^{-6}}^\nu$& 8.9$\times {10^{-6}}$& 3.1$
\times {10^{-6}}^\nu$& 8.9$\times 10^{-6}$ &3.1$\times {10^{-6}}^\nu$& 8.9$\times
 10^{-6}$\\
$\lam'_{233}$ & 8.9$\times {10^{-6}}^\nu$& 8.9$\times 10^{-6}$& 3.1$\times
{10^{-6}}^\nu$& 8.9$\times 10^{-6}$ &3.1$\times {10^{-6}}^\nu$& 8.9$
\times 10^{-6}$\\
$\lam'_{333}$ & 3.0$\times {10^{-6}}^\nu$& 8.9$\times 10^{-6}$&
 3.0$\times{10^{-6}}^\nu$& 8.9$\times 10^{-6}$ & 3.0$\times {10^{-6}}^\nu$&
 8.9$\times 10^{-6}$\\ 
\hline
\end{tabular}
\caption{Upper bounds upon trilinear $\lam'$ couplings for
  SPS1a in the quark mass eigenbasis at the weak scale $M_Z$ and in
  the weak eigenbasis at the GUT scale $M_{GUT}$. The quark mixing
  assumption is shown in the first row for each case.  Input
  parameters are given in the text. A superscript of $t,\nu$ denotes
  the fact that the strongest bound comes from the absence of tachyons
  or the neutrino mass constraint respectively.}
\label{tab:bounds}
\end{center}
\end{table*}

Table~\ref{tab:bounds} displays the strongest upper bounds upon
trilinear $\lam'$ couplings coming either from the neutrino mass
constraint or the absence of tachyons at mSUGRA point SPS1a as
described in Sect.~\ref{procedure} above. The different bounds coming
from altering the quark mixing assumption are displayed. In each case,
the upper bound at $M_{GUT}$ is shown in the weak eigenbasis, and the
corresponding bound that is obtained when the couplings and masses of
the MSSM are run down to $M_Z$ and rotated to the quark mass
eigenbasis as in Eqs.~(\ref{pdcm},\ref{ppdcm},\ref{pucm},\ref{ppucm}). 
Neglecting quark mixing we see that some of the bounds come from the
absence of tachyons, and allow large couplings of around 0.4 at
$M_Z$. However, for $\lam'_{ijj}$, the diagonal components of $\yd$
produce a non-zero $\kap$ through the RGEs, which in turn generates a
neutrino mass.  These bounds are much stronger and are of order
$\mathcal{O}(10^{-3} - 10^{-5})$. It should be noted that the neutrino
bounds are sensitive to the down quark mass inputs, because the RGEs
generate $\kap$ proportional to $\yd$. When the CKM mixing is assumed
to be in the up-quark sector, $\lam'_{i23}, \lam'_{i13}$ and
$\lam'_{i32}$ acquire stronger bounds coming from neutrino masses
because the larger up-quark Yukawa couplings in $\yu$ also begin to
mix the $\yd$ through the RGEs. When all down quarks are mixed at
$M_Z$, {\em any} $\lam'_{ijk}$ produces $\kap$ terms and therefore a
non-zero neutrino mass.  In this case, all of the bounds are strong:
$\mathcal{O}(10^{-3} - 10^{-5})$.

\begin{table}[floatfix]
\begin{center}
\begin{tabular}{|c|c|c|}
\hline
 & $M_{GUT}$ & $M_Z$ \\ \hline
$\lam_{121}$ & 0.10$^\nu$&0.15 \\
$\lam_{131}$ & 0.10$^\nu$&0.15 \\
$\lam_{231}$ & 0.55$^t$& 0.61\\
$\lam_{122}$ & 6.3$\times {10^{-4}}^\nu$&9.4$\times 10^{-4}$ \\
$\lam_{132}$ & 0.55$^t$& 0.61\\
$\lam_{232}$ & 6.2$\times {10^{-4}}^\nu$&9.3$\times 10^{-4}$ \\
$\lam_{123}$ & 0.50$^t$&0.58 \\
$\lam_{133}$ & 3.6$\times {10^{-5}}^\nu$&5.4$\times 10^{-5}$ \\
$\lam_{233}$ & 3.6$\times {10^{-5}}^\nu$&5.4$\times 10^{-5}$ \\
\hline
\end{tabular}
\caption{Upper bounds upon trilinear $\lam$ couplings for
  SPS1a at 
  the weak scale $M_Z$ and at the GUT scale $M_{GUT}$. 
  Input parameters are
  given in the text. A superscript of $t,\nu$ denotes the fact that the
  strongest  bound comes from the 
  absence of tachyons or neutrino masses respectively.}
\label{tab:bounds2}
\end{center}
\end{table}

Table~\ref{tab:bounds2} shows the equivalent bounds for the $\lam$
parameters. These bounds are {\em not} sensitive to assumptions about
quark mixing because the RGE generation of $\kap$ proceeds through the
charged lepton Yukawa couplings, which we have assumed to be diagonal
in the weak basis at $M_Z$. Changing this assumption should
drastically change the presented results.  We see that 3 of the 9 $\lam$ 
couplings are not very strongly constrained; they are allowed to be
$\mathcal O(1)$. If the $\ye$ were strongly mixed, this would no
longer be the case and the neutrino mass constraint would provide
stronger constraints, which we expect to be at the level of $\mathcal
{O}(10^{-1})-\mathcal{O}(10^ {-5})$, similar to the 6 couplings that
are constrained by neutrino masses in Table~\ref{tab:bounds2}

We may ask how much the bounds in
Tables~\ref{tab:bounds},\ref{tab:bounds2} depend upon the
supersymmetry breaking parameters. In order to investigate this issue,
we scan over the parameters of the no-scale mSUGRA~\cite{noscale}, a
simple hypersurface of mSUGRA parameter space where $m_0=A_0=0$.
\begin{figure}[h]
\begin{center}
\epsfig{figure=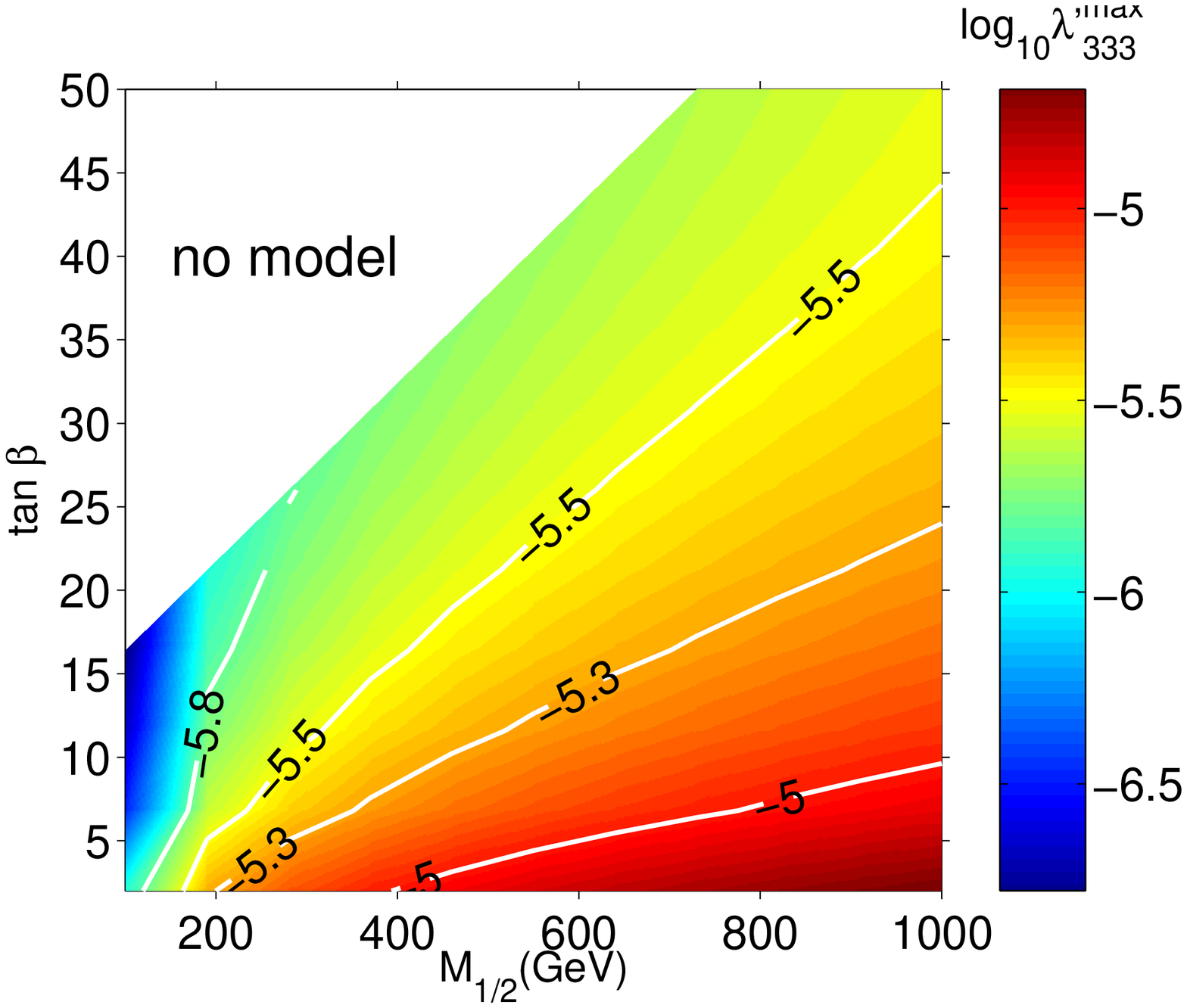,width=8.5cm}
\end{center}
\caption{Upper bound upon $\lam'_{333}(M_{GUT})$ as a function of no-scale
  mSUGRA 
  parameter point, assuming all quark mixing resides in the down sector at the
  weak scale. The background colour displays the bound as measured by the
  bar on the right hand side. Contours of iso-bound are also shown. In the top
  left-hand white region there is no tachyon-free model for any value of the
  coupling.}
\label{fig:lp333}
\end{figure}
The remaining parameters ($\tan \beta$ and $M_{1/2}$) are varied in
Fig.~\ref{fig:lp333} and the maximum possible value of
$\log(\lam'_{333}(M_{GUT}))$ is displayed as the background colour, as
referenced by the bar on the right hand side. The white region marked
``no model'' has tachyons for any value of $\lam'_{333}$ and so is not
valid. White contours of $\lam^\prime_{333}(max)=10^{-5}$, $10^{-5.3}
$, $10^{-5.5}$ and $10^{-5.8}$ are shown from bottom to top
respectively.  The strongest bound comes from the neutrino mass
constraint, and we see a variation of 2 orders of magnitude on the
bound across the parameter space, the strongest bounds coming from the
low $M_{1/2}$ region. The reader should note the $M_{1/2}$ dependence
of the neutrino mass in the simple formula Eq.~(\ref{lownumass}).
This strong variation of the neutrino bound is also apparent for the
case of other $\lam'$ couplings.
\begin{figure}[floatfix]
\begin{center}
\epsfig{figure=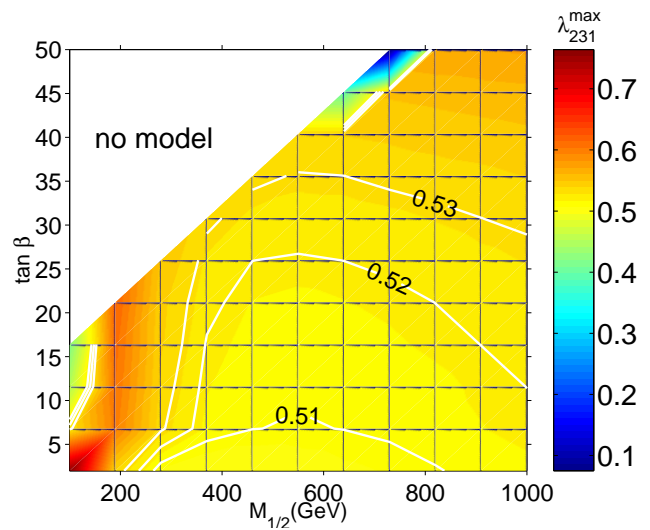,width=8.5cm}
\end{center}
\caption{Upper bound upon $\lam_{231}(M_{GUT})$ as a function of no-scale
  mSUGRA 
  parameter point. The background colour displays the bound as measured by the
  bar on the right hand side. Contours of iso-bound are also shown. In the top
  left-hand white region there is no tachyon-free model for any value of the
  coupling.}
\label{fig:l231}
\end{figure}
Fig.~\ref{fig:l231} shows the variation of the upper bound on
$\lam_{231}(M_{GUT})$ with no-scale mSUGRA parameter point. The
strongest bound comes from the no tachyon constraint, and we see only
a small variation of the bound across the parameter space, the
strongest bounds coming from the high $\tan \beta$ region, at low
$M_{1/2}$. (Recall the $M_{1/2}$ sensitivity in Eq.~(\ref{snbound}).)
The behaviour of small variation in the tachyon bound with
supersymmetry breaking parameters is replicated for other
lepton-number violating couplings. The weak bound of $\approx 0.5$
over much of the parameter space is dependent upon the no-charged
lepton mixing at $M_Z$ assumption.

It is instructive to compare the bounds derived here in a
representative scenario of mSUGRA in Tables~\ref{tab:bounds},
\ref{tab:bounds2} with the 2$\sigma$ bounds at $M_Z$ collected in
Table~1 in Ref.~\cite{Allanach:1999ic} for a rather generic R-parity
violating scenario. For the comparison we choose the no mixing
scenario, \ie case (a) in Eqs.~(\ref{models},\ref{models2}) and squark
and slepton masses of order of 100~GeV in the latter. For the $\lam_
{ijk}' L_i Q_j\bar{D}_k$ couplings, we obtain here one order of
magnitude improvement for $\lam_{211}'$, two orders of magnitude for
$\lam_{311}',\lam_{122}'$, three orders of magnitude for $\lam_{133}
'$, four orders of magnitude for $\lam_{222}',\lam_{322}'$, five and
up to six (!) orders of magnitude for $\lam_{233}',\lam_{333}'$. The
sneutrino tachyon constraint of Eq.~(\ref{snbound}) sets slightly
stronger bounds on the couplings $\lam_{323}',\lam_{223},\lam_{232}',
\lam_{132}', \lam_{331}'$. In the case of the \rpv-couplings $\lam_
{ijk} L_i Lj E_k$ we obtain two order of magnitude stronger bounds
than in Ref.~\cite{Allanach:1999ic} for the couplings: $\lam_{122},
\lam_{322 },\lam_{133},\lam_{233}$. Sneutrino tachyons do not set
better limits in this case. Comparison of the quark mixing cases (b)
or (c) of Eqs.~(\ref{models},\ref{models2}) derived in
Table.~\ref{tab:bounds} with the Table.~IV of
Ref.~\cite{Allanach:1999ic} show similar orders of magnitude, but
stronger bounds for some of the couplings.

\subsection{LSP Content in the No-Scale Model}
As outlined in the introduction, in \rpv-mSUGRA the \rpv-couplings can
affect the weak-scale particle mass spectrum via the RGEs. They can
also affect the interpretation of the resulting spectrum, since with
\rpv\ the LSP is no longer stable, and thus no longer subject to
cosmological constraints on stable relics. In the \rpv-mSUGRA the LSP
need not be electrically and colour neutral.  Before discussing the
\rpv-case we briefly review the RPC case.

\subsubsection{The RPC Case}

\begin{figure}[ht]
\begin{center}
\epsfig{figure=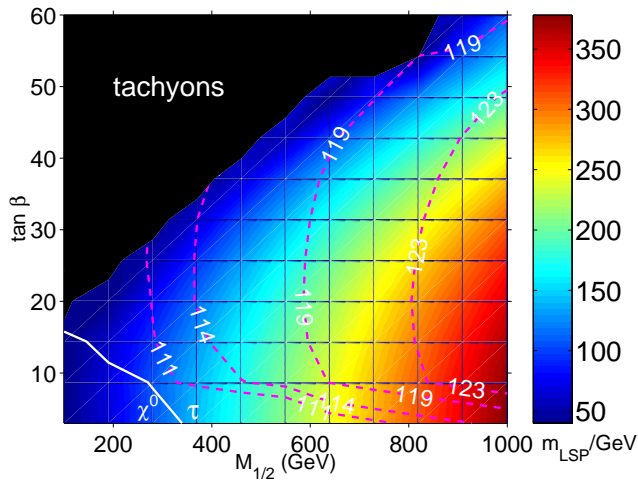,width=8.5cm}
\end{center}
\caption{No-scale supergravity in the R-parity conserved limit. Labeled
  constraints coming from tachyons are shown. The background colour
  displays the LSP mass, which can be read off from the bar on the
  right hand side. Dashed contours are contours of lightest Higgs
  mass. The white line delineates the labeled regions of $\tilde
  \tau$ LSP and $\chi_1^0$ LSP.}
\label{fig:rpc}
\end{figure}
To begin with, we perform the scan in the free parameters $M_{1/2}$
and $\tan \beta$ in R-parity conserved no-scale mSUGRA. The LSP mass
and contours of equal lightest-Higgs mass are displayed in
Fig.~\ref{fig:rpc}.  The background colour displays the LSP mass
according to the scale on the right hand side of the plot. The region
disallowed by tachyons is shown in black.  In the bottom left-hand
side of the plot is a white line which shows the boundary of the LSP
identity. Below the line, the LSP is the lightest neutralino, whereas
above it the LSP is a right-handed stau. A charged LSP is ruled out in
the R-parity conserved scenario from cosmological constraints, and so
the entire region above the white line is ruled out. This bound comes
from limits on abundances of anomalously heavy
isotopes~\cite{Ellis:1983ew}.  LEP2~\cite{higgs} places a lower bound
on the Standard Model Higgs mass of $m_h>114.4$ GeV.  This can also be
applied to the MSSM Higgs when $\sin (\alpha-\beta) \approx 1$, which
is the case in all of our results. The theoretical uncertainty upon
the lightest Higgs mass is estimated to be $\pm 3$
GeV~\cite{Degrassi:2002fi}, so we place a cautious lower bound on
{\small \tt SOFTSUSY}'s prediction of 111 GeV.  Even so, we see from
Fig.~\ref{fig:rpc} that there is no parameter space left with both a
heavy enough Higgs and a neutral LSP. Thus no-scale supergravity is
ruled out for the R-parity conserved MSSM.  However, even a very tiny
\rpv-coupling will make the LSP unstable on cosmological time-scales
and the neutral LSP constraint is then no longer applicable. For small
couplings $< \mathcal O(0.1)$, the spectrum can be approximated by the
R-parity conserved case, and so Fig.~\ref{fig:rpc} can still be
used. We see that the entire region above the Higgs mass contour of
111 GeV would be allowed, for stau LSP masses above 96
GeV~\cite{LEP2bounds}.

\begin{figure}[t!]
\begin{center}
\epsfig{figure=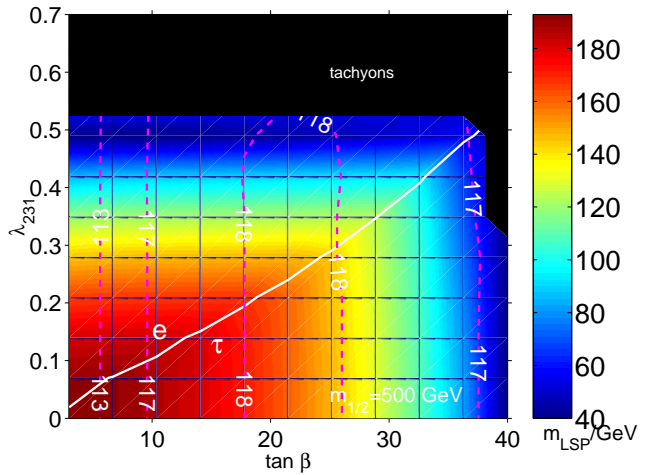,width=8.5cm}
\caption{LSP content of no-scale mSUGRA for $M_{1/2}=500$ GeV, 
  $\lam_{231}$ non-zero at $M_{GUT}$ and weak-scale mixing entirely in
  the down quarks. The mass of the LSP is displayed in the background
  and corresponds to the bar on the right hand side. Regions ruled out
  by the presence of tachyons are in black. The white line delineates
  labeled regions of different LSP content (e for selectrons and
  $\tau$ for staus). The dashed lines display contours of equal
  lightest Higgs mass. \label{fig:lsp231}}
\end{center}
\end{figure}

\subsubsection{The \rpv-Case}

\begin{figure}[t!]
\begin{center}
\epsfig{figure=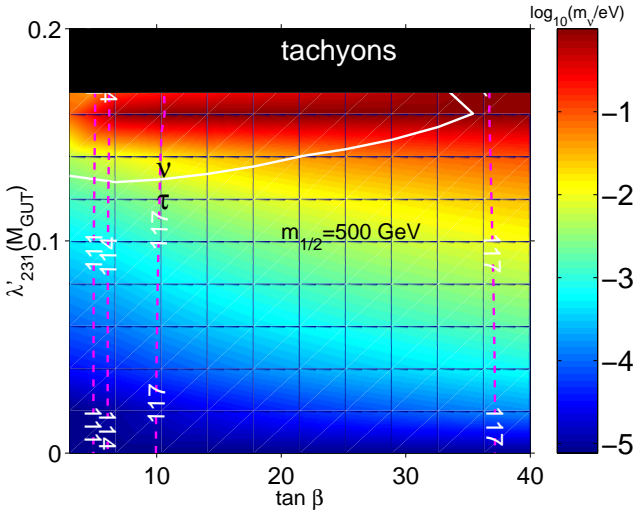,width=8.5cm}
\caption{LSP content of no-scale mSUGRA for $M_{1/2}=500$ GeV, $\lam'_{231}$
non-zero at $M_{GUT}$ and weak-scale mixing entirely in the up quarks. The 
logarithm of the mass of the heaviest neutrino is displayed in the background
and corresponds to the bar on the right-hand side. Regions ruled out by the 
presence of tachyons are in black. The white line delineates labeled regions 
of different LSP content. The dashed lines display contours of equal lightest 
Higgs mass.  \label{fig:lp231}}
\end{center}
\end{figure}

\begin{figure}[ht]
\begin{center}
\epsfig{figure=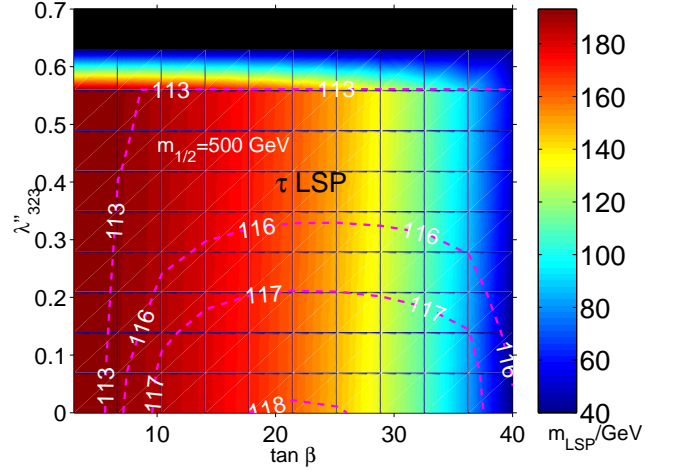,width=8.5cm}
\caption{LSP content of no-scale mSUGRA for $M_{1/2}=500$ GeV,
  $\lam_{323}''$ non-zero at $M_{GUT}$ and weak-scale quark mixing in
  the down sector. The mass of the LSP is displayed in the background
  and corresponds to the bar on the right hand side. Regions ruled out
  by the presence of tachyons are in black. There is a stau LSP
  throughout all of the parameter space.  The dashed lines display
  contours of equal lightest Higgs mass. \label{fig:lpp323}}
\end{center}
\end{figure}

We now map out some parts of no-scale mSUGRA for $M_{1/2}=500$ GeV.
Because we wish to show the effects of R-parity violation on the
spectrum, we pick cases where the upper bound on the \rpv-trilinear
coupling is weak.  This obviously occurs when the tachyon bound is the
stronger of the two we have shown in Tables \ref{tab:bounds} and
\ref{tab:bounds2}. We display one $\lam$-type coupling 
(Fig~\ref{fig:lsp231}), one of type $\lam'$ (Fig~\ref{fig:lp231}) and
one of type $\lam''$ (Fig~\ref{fig:lpp323}).

Fig.~\ref{fig:lsp231} shows the variation of the nature of the LSP
with $\tan \beta$ and $\lam_{231}(M_{GUT})$. The case (b) of
Eqs.~(\ref{models},\ref{models2}) is considered. For $M_{1/2}=500$
GeV, as assumed here, we see from the equal Higgs mass contours, that
the lower bound of 111 GeV on the lightest-Higgs mass does not pose a
very severe constraint for $\tan \beta>3$. The LSP mass varies up to
190 GeV in the plane, but this value is a function of $M_{1/2}$. The
diagonal white line separates regions of selectron LSP (above the
white line) and stau LSP (below the white line). Note that there is an
independent (2$\sigma$) bound for the coupling $\lam_{231}$ from the
known ratios $R_\tau =\Gamma(\tau\to e \nu \bar{\nu})/ \Gamma(\tau\to
\mu \nu \bar{\nu})$ corresponding to~\cite{Allanach:1999ic,Han}: $
\lam_{231}(M_{\rm GUT})< 0.046\times (m_{\tilde{e}_R}/100~{\rm GeV})$.
Comparing this bound with the nature of the LSP in
Fig.~\ref{fig:lsp231} we observe that the scalar tau LSP is favoured
for $\tan\beta \gsim 4$ unless the above laboratory bound is evaded by
taking $M_{1/2}\gg500\,{\rm GeV}$.

In Fig.~\ref{fig:lp231}, we show the variation of the non-zero
neutrino mass in the $\tan \beta-\lam'_{231}(M_{GUT})$ plane. Neutrino
masses provide the upper bound upon $\lam'_{231}$ for mixing in the
up-quark sector [case (c) in Eqs.~(\ref{models},\ref{models2})], as
assumed here.  For larger values of $\lam'_{231}\approx 0.15$,
neutrino masses of $\mathcal O(0.1\,{\rm eV})$ are possible. In this
case, above the white line, the LSP is a tau sneutrino, and below it
the LSP is the stau. The laboratory bound for the coupling $\lam'_{
231}(M_{\rm GUT})$ reads~\cite{Allanach:1999ic,Han} : $\lam_{231}'(
M_{\rm GUT})< 0.057\times (m_{\tilde{b}_L}/100~{\rm GeV})$ and since
we find that for the inputs of Fig.~\ref{fig:lp231} the bottom squark
mass is about 1.2 TeV, the laboratory bound is evaded: the stronger
bound on $\lam'_{231}$comes from the sneutrino tachyon as is shown in
the upper half of Fig.~\ref{fig:lp231}.

Finally, we investigate the case of baryon number violation. The case
(b) of Eqs.~(\ref{models},\ref{models2}) is considered.
Fig.~\ref{fig:lpp323} shows how the no-scale mSUGRA LSP mass varies
with $\tan \beta$ and $\lam''_{323} (M_{GUT})$. There is little
variation with the \rpv-coupling, contrary to the lightest Higgs mass,
which is displayed in the form of contours. The previous bound on
$\lam''_{323}$ (see Table IV of bound~\cite{Allanach:1999ic}) apart
from the theoretical perturbativity bound comes from the leptonic
Z-width ratio and is $\lam''_{323}(M_{\rm GUT})<0.015$ for quark
mixing solely in the down-quark sector, and with a little variation from
$M_{1/2}$. We observe from Fig.~\ref{fig:lpp323} that the stau is
again the LSP.

We have exhibited, in Figs.~\ref{fig:rpc}-\ref{fig:lpp323}, viable
regions of MSSM parameter space where the LSP is the selectron, the
stau or the stau sneutrino.  Different LSP content drastically alters
the collider signatures of the models. The analysis above showed a
preference to the stau being the LSP. We discuss this in some more
detail in Sect.~\ref{non-chi-lsp}, below.

\subsection{Sneutrino-Antisneutrino Mixing with Stau LSP}
\label{Sec:sn-anti-sn-mixing}

Models which violate lepton number by two units ($\Delta L=2$) and
generate neutrino masses, also result in a mass splitting of scalar
neutrinos and anti-neutrinos of the same flavour usually referred in
the literature as sneutrino anti-sneutrino
mixing~\cite{HG1,Hirsch,Chun2}.  If the sneutrino mass difference
$\Delta m_{\tilde{\nu}} = m_{\tilde{\nu}_+}\: -\: m_{\tilde{\nu}_-}$,
is large and the sneutrino branching ratio into a charged lepton is
experimentally significant, then a like sign-dilepton signal in
$e^+\:e^-\to \tilde{\nu}_-\:\tilde{\nu}_+$ with $\tilde{\nu}\to l^-
+X$ could be observed~\cite{HG1}. Like the B-meson mass splitting ,
the observability of the sneutrino mixing effects depend on the ratio
\begin{eqnarray}
x_{\tilde{\nu}} \equiv \frac{\Delta m_{\tilde{\nu}}}{\Gamma_{\tilde{\nu}}}
\;,
\end{eqnarray}
where $\Gamma_{\tilde{\nu}}$ is the total sneutrino decay rate.  As we
have already seen from Figs.~\ref{fig:lsp231}-\ref{fig:lpp323}, in the
no-scale scenario the stau, $\tilde{\tau}$, is the LSP when the
\rpv-couplings are small.  In this (approximately RPC) case the
specific flavour $\ell=(e,\mu)$ sneutrino $\tilde{\nu_\ell}$ decays,
via charginos and neutralinos into $\tilde{\nu_\ell}\:\to\:\ell^-\:
\tilde{\tau }^+\:\nu_\tau$ and $\tilde{\nu_\ell}\:\to\:\nu_\ell\:
\tilde{\tau}^\pm\:\tau^\mp$.  In this case, the probability of tagging
a like-sign dilepton in the process $\tilde{\nu_l}\:\to\: l^-\:
\tilde{\tau}^+\:\nu_\tau$ is $\mathcal {P}(\ell^\pm\ell^\pm)
=\mathcal{P}(\ell^+\ell^+)\:+\:\mathcal{P}(\ell^-\ell^-)$ with~\cite{HG1} 
\begin{eqnarray}
{\cal P}(\ell^\pm\ell^\pm)\ = \ \frac{x_{\tilde{\nu}}^2}{2\:
(1+x_{\tilde{\nu}}^2)} \: \biggl [ {\cal B}(\tilde{\nu_\ell}\:\to\: \ell^-\:
\tilde{\tau}^+\: \nu_\tau) \biggr ]^2  \;.\label{prob}
\end{eqnarray}
We investigate below the magnitude of this probability in the no-scale
model with $M_{1/2}=500$ GeV and $\tan\beta=20$ and with one dominant
\rpv-coupling $\lam'_{122}(M_{\rm GUT})=7.5\times 10^{-5}$.
Furthermore we consider no-quark mixing in determining the relevant
bounds from the neutrino masses. In this model the stau is the LSP. We
first calculate the sneutrino mass squared difference
\begin{eqnarray}
\Delta m_{\tilde{\nu}}=
\frac{\Delta m_{\tilde{\nu}}^2}{2m_{\tilde{\nu}}} 
 = \frac{m_{\tilde{\nu}_+}^2-m_{\tilde{\nu}_-}^2} {2m_{\tilde{\nu}}}
\;,\label{dmsn}
\end{eqnarray} 
where $m_{\tilde{\nu}}$ is the average mass of $m_{\tilde{\nu}_\pm} $.
The sneutrino mass difference has been calculated in Ref.~\cite{HG2}
in a general basis independent manner. With our choice $\lam'_{122}$
we generate at the electroweak scale the non-zero \rpv-parameter set:
${v_1,\kap_1,\widetilde{D_1},\mhl{1}}$. The other \rpv-parameters
remain zero \cite{footx}.  This simplifies our calculation for the
sneutrino mass splitting, since we can use the case of one sneutrino
generation (the other two decouple from the mass matrices
Eqs.~(\ref{cpeven},\ref{cpodd})).  The sneutrino mass splitting
reads~\cite{HG2}:
\begin{eqnarray}
\Delta m_{\tilde{\nu}} = -\frac{2\: \widetilde{B}^2\: M_Z^2 \: 
m_{\tilde{\nu}} \:\sin^2\beta\, \sin^2\delta}
{(M_{H^0}^2-m_{\tilde{\nu}}^2)\:(M_{h^0}^2-m_{\tilde{\nu}}^2)
\:(M_{A^0}^2-m_{\tilde{\nu}}^2)}\: \nonumber \;, \\  \label{split}
\end{eqnarray}
with
\begin{eqnarray}
\cos\delta = \pm \frac{|v_d \widetilde{B} + v_1 
\widetilde{D}_1|}
{(v_d^2+v_1^2)^{1/2}\:(\widetilde{B}^2+\widetilde{D}_1^2)^{1/2}}\;.
\label{cosd}
\end{eqnarray}
Notice that Eq.~(\ref{split}) does not depend on the superpotential
parameters in contrast to the neutrino mass in Eq.~(\ref{numass}). It
is helpful to see the numerical values \cite{footy} for the
parameters at the electroweak scale starting from the no-scale model
defined by: $M_{1/2}=500$ GeV, $\tan\beta=20$ and $\lam'_{122}(M_{\rm
GUT})=7.5 \times 10^{-5}$. We obtain: $\widetilde{B}(M_Z)= 33\,238 ~{
\rm GeV}^2$, $m_{\tilde{\nu}}=357$ GeV, $M_{h^0}=91$ GeV, $M_{H^0}=
816$ GeV, $M_{A^0}=816$ GeV, $v_d(M_Z)=8.7$ GeV, $v_1(M_Z)=-0.0012$
GeV, $\widetilde{D}_1=-0.74~{\rm GeV}^2$, and $\mhl{1}=2.5~{\rm GeV}^2$.
Applying these values to Eqs.~(\ref{split},\ref{cosd}) we obtain
$\sin^2\delta=1.3\times 10^{-8}$ and $\Delta m_{\tilde{\nu}} =2.5$ eV.
The sneutrino mass splitting is of the same order as the neutrino mass
obtained from Eq.~(\ref{numass}), since for $\mu(M_Z)=817$ GeV and
$\kappa_1(M_Z)=3.5\times 10^{-4}$ GeV we have $m_\nu=1.2$ eV
\cite{footz}.

In order to calculate the probability ${\cal P}(\ell^\pm\ell^\pm)$ we
still need the total sneutrino decay rate and the branching ratio
${\cal B}(\tilde{\nu_\ell}\to \ell^-\tilde{\tau}^+\nu_\tau)$.
In the above scenario the right handed selectron of the third
generation (we call it stau here although it is in fact an admixture
of the three charged sleptons with the charged Higgs boson states) is
the LSP with a mass $m_{\tilde{\tau}}=162$ GeV. The rates for the
chargino and neutralino mediated sneutrino decays (which we assume to
be the dominant ones) are~\cite{HG1}:
\begin{eqnarray}
\Gamma (\tilde{\nu}_\ell \to \ell^- \tilde{\tau}^+ \nu_\tau) &=&
\frac{g_2^4 m_{\tilde{\nu}}^3 m_\tau^2 \tan^2\beta f_{\chi^+}
(m_{\tilde{\tau}}^2/m^2_{\tilde{\nu}})} {1536 \pi^3 (M_W^2 \sin 
2\beta-M_2\mu)^2} , \nonumber \\[0.2cm]
\Gamma (\tilde{\nu}_\ell \to \nu_\ell \tilde{\tau}^\pm\tau^\mp) &=&
 \frac{g^4 m_{\tilde{\nu}}^5 f_{\chi^0}(m_{\tilde{\tau}}^2/m^2_{\tilde{\nu}})}
{3072 \pi^3 M_1^4} \;,
\label{rates}
\end{eqnarray}
with 
\begin{eqnarray}
f_{\chi^+}(x)&=&(1-x)\:(1+10 x+x^2)+6 x (1+x) \ln x \nonumber \,,\\
f_{\chi^0}(x)&=&1-8 x+8 x^3-x^4-12 x^2 \ln x \,.
\end{eqnarray} 
In the no-scale model under consideration we obtain: $M_1(M_Z)=206$
GeV, $M_2(M_Z)=411$ GeV with the gauge couplings $g(M_Z)=0.3574$ and
$g_2(M_Z)=0.6525$. Thus from Eq.~(\ref{rates}) we obtain:
$\Gamma(\tilde {\nu}_\ell \to \ell^- \tilde{\tau}^+\nu_\tau)=210$ eV
and $\Gamma (\tilde {\nu}_\ell\to\nu_\ell\tilde{\tau}^\pm\tau^\mp)
=1.2\times 10^5$ eV. So $x_\nu=2\times 10^{-5}$ and
$\mathcal{B}(\tilde{\nu_
\ell}\:\to\: \ell^-\:\tilde{\tau}^+\: \nu_\tau)=1.7\times 10^{-3}$. 
We conclude that in this numerical example the probability for like
sign dileptons, Eq.~(\ref{prob}), is: ${\cal P}(\ell^\pm\ell^\pm)=
6\times 10^{-16}$, far too small to be observable. Of course this
result depends on the parameter space and the probability ${\cal
P}(\ell^\pm\ell^\pm)$ is bigger for smaller values of $M_{1/2}$ and
larger $\tan\beta$ values (see Eq.~(\ref{rates}). However, if we
take into account the current experimental data, $M_{1/2}\gsim 200$,
then ${\cal P}(\ell^\pm\ell^\pm)\lsim 10^{-9}$. We obtain similar
results for the other \rpv-couplings.

The above benchmark computation can be helpful to the reader in order
understand the typical magnitude of the parameters we are dealing with
in this paper.

\section{Stau-LSP Phenomenology}
\label{non-chi-lsp}

As discussed in Sect.~\ref{intro}, in the case of \rpv , the LSP need
not be the lightest neutralino, ${\tilde\chi}^0_1$. In the previous
section we have investigated the nature of the LSP in the mSUGRA
scenario and have found regions in parameter space with different
LSP's.  In Fig.~\ref{fig:lsp231}, we have a selectron or stau LSP, in
Fig.~\ref{fig:lp231} we have found a tau sneutrino or a stau and in
Fig.~\ref{fig:lpp323} we have found a stau LSP. The bounds in
Table~\ref{tab:bounds} imply that if there is any appreciable CKM
mixing in the down-quark sector at the weak scale, $\lam^\prime_{ijk}$
must be very small. We also see some strict bounds upon the
$\lam_{ijk}$ in Table~\ref{tab:bounds2}. If the \rpv-couplings are
very small, the spectrum has negligible perturbation from the R-parity
conserved case, the LSP content of which is displayed in
Fig.~\ref{fig:rpc}. The allowed parameter space with $m_{h^0}>111\,
{\rm GeV}$ in Fig.~\ref{fig:rpc} then leads to a stau LSP. Thus we see
a preference for a stau LSP in many no-scale R-parity violating
scenarios.

In the RPC-MSSM, the collider phenomenology relies crucially on the
${\tilde\chi}^0_1$-LSP, with all produced sparticles decaying in the
detector to ${\tilde\chi}^0_1$ plus other $R_p$-even particles.  This
results in missing transverse energy as a typical signature for all
production processes. In the \rpv-MSSM the RGEs and thus the spectrum,
is altered. This changes the decay chains.  Since typically all decay
chains end in the LSP, the nature of the LSP is essential in
determining the supersymmetric signatures. A detailed investigation is
beyond the scope of this paper. We shall here focus on a
classification of the signatures for the main production processes in
the case of a stau LSP.

\subsubsection{Stau Decays}

The following discussion of the stau-LSP is somewhat analogous to the
discussion in Ref.~\cite{Dreiner:pe} for the ${\tilde\chi}^0_1$-LSP.
In determining the final state signature it is important to know how
the stau-LSP decays. We shall assume that there is a hierarchy among
the \rpv-coupling constants with one dominant coupling, similar to the
SM Yukawa couplings in the mass eigenstate basis. We furthermore
assume the mixing due to $\kappa_ i$ is small as seen in the previous
sections of this paper. Then there are two important distinct cases.
\begin{enumerate}
\item {\it The stau couples to the dominant operator}. The dominant 
operator is in the set $\{L_{e,\mu}L_\tau{\bar E}_{e,\mu,\tau},\,L_e
L_\mu{\bar E}_\tau,\,L_\tau Q_i{\bar D}_j\}$. In this case, the stau 
simply decays via the two-body mode. For the dominant operator 
$L_\tau Q_1{\bar D}_1$ for example we then obtain 
\cite{Dreiner:1999qz}
\begin{equation}
\Gamma({\stau}^-\ra {\bar u}+d)=
\frac{N_c\lam^{\prime2}_{311}M_{\stau}}{16\pi} \,,
\end{equation}
where $N_c=3$ is the number of colours. The complete list of \rpv,
two-body decays is given in Ref.~\cite{Dreiner:1999qz}. For a recent
treatment of two-body stau decays also see \cite{Porod}. For the above
two-body decay mode the decay length is given by
\begin{equation}
c\tau_{\stau} = 
3.3\;10^{-11}{\rm m} \left(\frac{10^{-3}}{\lam^\prime_{311}}\right)^{2} 
\left(\frac{100\,{\rm GeV}}{M_{\stau}}\right)
\end{equation}
which in an experiment must be multiplied by the relevant Lorentz
boost factor $\gamma_L$ of the stau. Only for very small coupling
($\lam'\lsim10^{-7}$) is the decay length relevant.

\item {\it The stau doesn't couple to the dominant \mbox{operator.}} The dominant
operator is in the set $\{L_eL_\mu{\bar E}_{e,\mu},\,L_{e,\mu}Q_i{\bar
D}_j,\,{\bar U}_i {\bar D}_j{\bar D}_k\}$. In this case~the $\stau$
decays via a four-body mode. For the operator $L_\mu Q_1{\bar D}_1$
there are four decay modes via the neutralino
\begin{equation}
{\stau}^-  \ra \tau^- + ({\tilde\chi}^0_1)^*
\ra \tau^-+
\left\{ \begin{array}{l}
\mu^-+ u +{\bar d} \\
\mu^++ {\bar u} +d \\
\nu_\mu + d + {\bar d} \\
{\bar \nu}_\mu + d + {\bar d} 
\end{array}\right.\,,
\end{equation}
and three decay modes via the chargino
\begin{equation}
{\stau}^-  \ra \nu_\tau + ({\tilde\chi}^-_1)^*
\ra \nu_\tau+
\left\{ \begin{array}{l}
\mu^-+ d +{\bar d} \\
\mu^-+ u + {\bar u}  \\
\nu_\mu + d+ {\bar u}  
\end{array}\right.\,.
\end{equation}
As an example we here compute the decay $\stau\ra \tau^-\mu^-u{\bar d}$.
The details of the computation, in particular the four-body phase
space are given in Appendix~\ref{app-stau}. The result is
\begin{widetext}
\begin{equation}
\Gamma({\tilde\tau}^-\ra \tau^- \mu^+{\bar u}d)
=\frac{KN_c\lam^{\prime 2}|a_\tau|^2}
{{2^5\pi^5}M_\chi^2{\tilde m}^4} 
M_{\tilde\tau}^7 (|b_\mu|^2+|b_u|^2+|a_d|^2-b_\mu b_u^\ast+b_\mu a_d^\ast
+b_ua_d^\ast)
\approx
\frac{KN_c\lam^{\prime 2}g^4}{{2^3\pi^5}M_\chi^2{\tilde m}^4} M_{\tilde\tau}^7 
\,,
\label{wid}\end{equation}
\end{widetext}
where $K=1/(720 \times 2^5)=1/23040$. $a_{\tau,d},\,b_{\mu,u}$ are
neutralino coupling constants given in the appendix. $M_\chi$ is the
neutralino mass and ${\tilde m}$ is the universal scalar fermion mass.
We have assumed massless final state particles and neglected the
momenta compared to $M_\chi,\,{\tilde m}$. In the last step we have
set the couplings $a_{\tau,d}=b_{\mu,u}=g$, the weak coupling constant.
If the four-body decay is the dominant decay mode, the decay length
can be estimated as
\begin{widetext}
\begin{equation}
c\tau_{\tilde\tau}= 6.2\,10^{-6}\,{\rm m}\left(\frac{10^{-3}}{\lam'}\right)^2
\left(\frac{M_\chi}{100\gev}\right)^2\left(\frac{{\tilde m}}{100\gev}\right)^4
\left(\frac{100\gev}{M_{\tilde\tau}}\right)^7\,.
\end{equation}
\end{widetext}
 For reasonable supersymmetric masses and couplings
this could lead to detached vertices in the detector. This is a very
promising signature for the stau-LSP.
\end{enumerate}
If the two-body decay is allowed, \ie the relevant coupling is not
suppressed, it usually  dominate over the four-body decay. In order
to estimate the required hierarchy of couplings for the four-body
decay to be relevant we consider the ratio
\begin{equation}
\frac{\Gamma_4({\tilde\tau}^-\ra\tau^-\mu^+{\bar u}d)}
{\Gamma_2({\tilde\tau}\ra{\bar u}d)}= \mathcal{O} \left
(\frac{{\lam'}_{211}^2}{{\lam'}_{3ij}^2} \frac{2Kg^4 
M_{\tilde \tau}^6}{\pi^4 M_\chi^2 {\tilde m}^4}\right)>1\,.
\end{equation}
 Assuming the sparticle masses are roughly equal,
this corresponds to ${\lam'}_{211}/{\lam'}_{3ij} \gsim {\mathcal
O}(10^3)$ for the 4-body decay mode to dominate over the 2-body one. If,
for example, $M_\chi={\tilde m}=2\, M_{\tilde \tau}^2$, we obtain
${\lam'}_{211}/{\lam'}_{3ij} \gsim {\mathcal O}(10^4)$, which is not
an unreasonable hierarchy between generations.

\subsubsection{Collider Signatures}
\begin{figure}[ht]
\begin{center}
\epsfig{figure=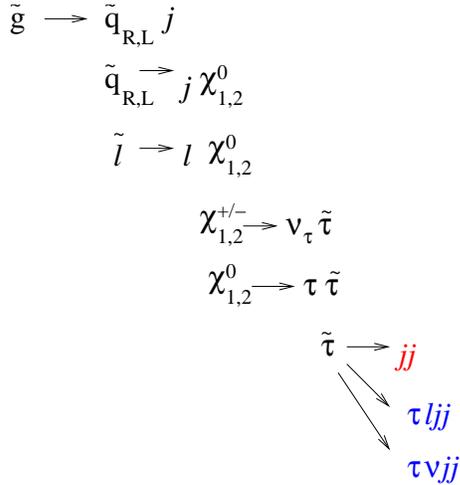,width=6cm}
\caption{Possible dominant links in a sparticle decay chain with a stau 
LSP and R-parity violation. Two-body decay modes of the LSP are shown
in red, and 4-body in blue.\label{fig:decs}}
\end{center}
\end{figure}

At a collider, the main supersymmetric pair production processes are
\begin{equation}
{\widetilde g}{\widetilde g}\,,\quad 
{\widetilde q}{\widetilde q}\,,\quad
{\widetilde\ell}^+{\widetilde\ell}^-,\quad
{\widetilde\chi}^0_i{\widetilde\chi}^0_j\,,\quad 
{\widetilde\chi}^+_i{\widetilde\chi}^-_j,\quad
{\widetilde\chi}^0_i{\widetilde\chi}^\pm_j \,.
\end{equation}
Here we investigate the possible signatures for these processes in the
case of a stau LSP. In order to determine the final state within the
detector, we must know the decay patterns of the particles. This
strongly depends on the supersymmetric spectrum and thus upon which
point in SUSY breaking parameter space is being studied. For this
first study, we shall assume the mass ordering
\begin{equation}
m_{\tilde g}\,>\,
m_{\tilde q} \,>\, m_{\tilde\ell} \,>\, m_{{\tilde\chi}^\pm_1} \,>\, 
m_{{\tilde\chi}^0_1} \,>\,m_{\tilde\tau}\,,
\label{mass-chain}
\end{equation}
which we typically obtain (with or without \rpv) within mSUGRA. If
there are no near-degenerate particles, a produced supersymmetric
particle will dominantly cascade in two-particle decays down the mass
chain (\ref{mass-chain}). We display this decay chain in
Fig.~\ref{fig:decs}. We have added at the end both two- (in red) and
four-particle (in blue) stau decays. Final state quarks are denoted by
``j'' to indicate a jet.
We can use this decay chain to determine a qualitative picture of the
possible final state signatures. Note that due to the strict bounds on
the \rpv-couplings which we have determined we only expect these to be
relevant in the stau-LSP decay. Furthermore, in determining signatures
we shall assume that either the two-body or the four-body stau decay
dominates.  We do not consider the case of comparable partial decay
widths.

At the Tevatron and LHC the largest production cross sections are for
gluinos and squarks. If we consider for example ${\tilde q}_R
{\bar{\tilde q}}_R$ production, then the dominant decay mode for the
squark is
\begin{equation}
{\tilde
q}_R\ra j \chi_1^0 \ra j \tau^\pm {\tilde \tau^\mp}\,,
\end{equation}
and the final state signature will be
\begin{equation}
{\tilde q}_R {\tilde q}_R\ra \left\{ 
\begin{array}{lcl}
6{\rm j}+\tau^+\tau^- &{\rm for} &{\tilde\tau}\ra {\rm j}{\rm j} \\
6{\rm j}+\ell\ell+2(\tau^+\tau^-) & {\rm for}& {\tilde\tau}\ra 
\tau\ell{\rm j}{\rm j} \\
6{\rm j}+\nu\nu+2(\tau^+\tau^-) & {\rm for} & {\tilde\tau}\ra \tau\nu 
{\rm j}{\rm j} \\
6{\rm j}+\nu\ell+2(\tau^+\tau^-) & {\rm for} & 
{\tilde\tau}\ra \tau(\nu,\ell){\rm j}{\rm j} 
\end{array} \right.\,.
\end{equation}
Here, any charge combination for the leptons $\ell$ is allowed, since
they result from the decay of a virtual (Majorana) neutralino ({\it
  c.f.} App. D). This can give us a like-sign di-lepton signature.
Otherwise, we see that we have a large number of jets in the final
state independent of the decay mode of the stau. (This would be
reduced for dominant operators $LL{\bar E}$.)  This makes it more
difficult to observe isolated high $p_T$ charged leptons. We can
obtain missing transverse momentum from the final state neutrinos but
it will be extremely diluted due to the many body-decays. The most
promising signature are like-sign dileptons together with the direct
detection of $\tau$'s \cite{tau-signat}, which is of course difficult.

For ${\widetilde q}_L{\widetilde q}_L$-production, we expect a larger
liklihood for the cascade decay through the heavier neutralinos and
also through the charginos. This can lead to a tri-lepton signature
\cite{trilepton} which can be extended by the {\it additional} $\tau$'s.
This requires a detailed analysis but we expect this to be
more promising than the ${\tilde q}_R{\tilde q}_R$ outlined above.

The gluino decays via the squarks adding an extra jet to the final
state. In this case it might be more promising to consider
non-dominant decay modes, including a possible direct \rpv-decay of
the neutralino. An estimate of the relative rates for a pure wino
neutralino is
\begin{equation}
\frac{\Gamma({\tilde\chi}^0_1\ra\mu+2{\rm j})}
{\Gamma({\tilde\chi}^0_1\ra\tau{\tilde\tau})}\approx 
\frac{3\lam^{\prime 2}}{32\pi^3} 
\left(\frac{M_{{\tilde\chi}^0_1}}{{\tilde m}}\right)^4\lsim\; 3\cdot10^{-7}\,,
\end{equation}
for $\lam'<10^{-2}$, and where we have neglected the stau mass. This
is hopeless, unless the neutralino and the stau are nearly degenerate.

At the Tevatron and LHC the pair production of sleptons is about two
to three orders of magnitude lower than the production of squarks or
gluinos, for equal mass. However, we expect the mass to be lower ({\it
c.f.}  Eq.(\ref{mass-chain})), and also the signal cleaner. At a
future linear collider $e^+e^-$ facility this is typically an ideal
mode for searches or the measurement of \  MSSM parameters. As we can see
from the decay chain in Fig.~\ref{fig:decs}, the slepton dominantly
decays as
\begin{equation}
{\tilde\ell}^+\ra  \chi_1^0 \ell^+ \ra \tau^\pm{\tilde\tau}^\mp\ell^+
\end{equation}
We then obtain the final-state signatures
\begin{equation}
{\tilde \ell}^- {\tilde \ell}^+\ra \left\{ 
\begin{array}{lcl}
4{\rm j}+\ell\ell+
\tau^-\tau^+ &{\rm for} &{\tilde\tau}\ra {\rm j}{\rm j} \\
4{\rm j}+2(\ell\ell)+2(\tau^+\tau^-) & {\rm for}& {\tilde\tau}\ra 
\tau\ell{\rm j}{\rm j} \\
4{\rm j}+\ell\ell+\nu\nu+2(\tau^+\tau^-) & {\rm for} & 
{\tilde\tau}\ra \tau\nu  {\rm j}{\rm j} \\
4{\rm j}+\ell\ell+\nu\ell+2(\tau^+\tau^-) & {\rm for} & 
{\tilde\tau}\ra \tau(\nu,\ell){\rm j}{\rm j} 
\end{array} \right.\,.
\end{equation}
In the second case the sign of the charge of the two leptons from the
stau decays is arbitrary due to the intermediate (Majorana)
neutralino. Thus we can have like-sign tri-leptons, which is a very
promising signature.

Similarly, using the results from Fig.~\ref{fig:decs}, we expect the
as dominant signatures for neutralino pair production
\begin{equation}
{\tilde\chi}^0_1 {\tilde\chi}^0_1\ra \left\{ 
\begin{array}{lcl}
4{\rm j}+
\tau^+\tau^- &{\rm for} &{\tilde\tau}\ra {\rm j}{\rm j} \\
4{\rm j}+\ell\ell+2(\tau^+\tau^-) & {\rm for}& {\tilde\tau}\ra 
\tau\ell{\rm j}{\rm j} \\
4{\rm j}+\nu\nu+2(\tau^+\tau^-) & {\rm for} & 
{\tilde\tau}\ra \tau\nu  {\rm j}{\rm j} \\
4{\rm j}+\nu\ell+2(\tau^+\tau^-) & {\rm for} & 
{\tilde\tau}\ra \tau(\nu,\ell){\rm j}{\rm j} 
\end{array} \right.\,, 
\end{equation}
depending on the decay of the stau-LSP decay which in turn depends on
the dominant \rpv-coupling.  For chargino pair production we have
\begin{equation}
{\tilde\chi}^-_1 {\tilde \chi}^+_1\ra \left\{ 
\begin{array}{lcl}
4{\rm j}+\nu_\tau\nu_\tau
&{\rm for} &{\tilde\tau}\ra {\rm j}{\rm j} \\
4{\rm j}+\ell\ell+ \nu_\tau\nu_\tau +
\tau^+\tau^-& {\rm for}& {\tilde\tau}\ra 
\tau\ell{\rm j}{\rm j} \\
4{\rm j}+\nu\nu+\nu_\tau\nu_\tau +
\tau^+\tau^-& {\rm for} & 
{\tilde\tau}\ra \tau\nu  {\rm j}{\rm j} \\
4{\rm j}+\nu\ell+\nu_\tau\nu_\tau +
\tau^+\tau^-& {\rm for} & 
{\tilde\tau}\ra \tau(\nu,\ell){\rm j}{\rm j} 
\end{array} \right.\,,
\end{equation}
assuming the chargino decays directly to the stau LSP.  If we produce
the heavier electroweak gauginos we can cascade decay through the
lighter gauginos producing more charged leptons. For the neutralino we
have promising multi-lepton sigantures, whereas for the chargino we
expect a significant amount of missing $p_T$.

In summary, as promising signatures in the case of the stau LSP we have
\begin{enumerate}
\item A detached vertex from the long lived stau, particularly in the
case of the four-body stau decay.
\item Multilepton final states.
\item Multi-tau final states, requiring efficient tau tagging.
\end{enumerate}
The four-body decay of the stau results in more final state leptons
than the two-body decay and is thus possibly more promising.

\section{Summary and Conclusions}
\label{summary}
We have investigated for the first time the general \rpv-MSSM in the
context of mSUGRA. We have studied in some detail the origin of
lepton-number violation and have found that with respect to the
dimension-five operators baryon-parity is preferred over R-parity. We
have then shown that in a wide class of models both $\kap_i$ and
${\widetilde D}_i$ are zero after supersymmetry breaking at the
unification scale. We have taken this as our boundary conditions at
$M_X$ in order to investigate the resulting model in considerable
detail. 

In order to embed the model within the unification picture we have
computed the full set of renormalzation group equations in the
appendices. We have used two methods, including a novel method of
Jones {\it et al.}, which is particularly conducive to the numerical
implementation. We then developed an iterative algorithm which solves
the RGEs, minimises the potentail of the five neutral, scalar, CP-even
fields, while implementing weak-scale Yukawa and gauge boundary
conditions. The algorithm is stable and has been checked by an
independent program. This is one of the main technical advances in
this paper. Given the minimum, we determined the complete
supersymmetric spectrum, including also the mass of the heaviest
neutrino.

We have then shown that the \rpv-couplings in this model are severely
constrained by the upper bound on the neutrino masses, as summarized
in Tables \ref{tab:bounds} and \ref{tab:bounds2}. Thus when embedding
the \rpv-MSSM in mSUGRA the neutrino mass bound is the strictest and
most universal, \ie applies to all lepton number violating couplings.
This is one of the main results of this paper.

We have then looked in detail at the nature of the LSP. We have found
solutions with a selectron, tau sneutrino and stau LSP besides the
usual neutralino LSP, with the stau most favoured in the noscale
mSUGRA model. This significantly affects collider phenomenology.  We
present a first discussion of this broad topic in
Sect.~\ref{non-chi-lsp}. We have also studied the phenomenology of
sneutrino-antisneutrino mixing in this model, but do not expect any
significant effect.

We conclude that the \rpv-MSSM is as viable as the RPC-MSSM.  As
we show, it considerably differs both conceptually and
phenomenologically from the RPC. The intimate connection with neutrino
masses is an outstanding feature which we shall discuss in more detail
in a forthcoming publication.

\section*{Acknowledgments}
We would like to thank I Jack, D R T Jones and A F Kord for helping to bring
an error in eq.~(86) to our attention in a previous version of thie paper, 
and for graciously collaborating on 
detailed numerical comparisons~\cite{jj}. 
HKD would like to thank Goran Senjanovic for discussions on GUTs and
R-parity violation and Howie Haber for discussions on the integration
of four-particle phase space.  AD would like to thank M. Drees for
useful discussions on the WMAP neutrino bound. AD and HKD would like
to thank the CERN theory division for hospitality offered while some
of this work was performed.  BCA would like to thank the University of
Bonn for hospitality offered while some of the work contained herein
was carried out.  We also thank S Rimmer for help with a conventions check.
AD acknowledges support in part by the German
Bundesministerium f\"ur Bildung und Forschung under the contract
05HT1WOA3 and the `Deutsche Forschungsgemeinschaft' DFG Project Bu.
706/1-2.

\begin{widetext}
\appendix

\section{Notation and Anomalous Dimensions}
\label{app2}

The chiral superfields of the $R_p$-MSSM and the \rpv-MSSM have the
following $G_{SM}=SU(3)_c\times SU(2)_L\times U(1)_Y$ quantum numbers
\begin{eqnarray}
L:&(1,2,-\half),\quad {\bar E}:&(1,1,1),\qquad\, Q:\,(3,2,\frac{1}{6}),\quad
{\bar U}:\,(3,1,\frac{2}{3}),\nonr\\ {\bar D}:& ~(3,1,-\frac{1}{3}),\quad
H_1:&(1,2,-\half),\quad  H_2:\,(1,2,\half).
\label{fields}
\end{eqnarray}
 The \rpv-MSSM superpotential is then given by
\begin{eqnarray} 
W&=& \eps_{ab} \left[ (\ye)_{ij} L_i^a
H_1^b {\bar E}_j + (\yd)_{ij} Q_i^{ax} H_1^b {\bar D}_{jx} +
(\yu)_{ij} Q_i^{ax} H_2^b {\bar U}_{jx} \right]
- \eps_{ab} \left[\mu H_1^a H_2^b
+\kap^i L_i^a H_2^b \right]
\nonr \\ &&+
\eps_{ab}\left[ \frac{1}{2}(\lame{k})_{ij} L_i^a L_j^b{\bar E}_k +
(\lamd{k})_{ij} L_i^a Q_j^{xb} {\bar D}_{kx} \right]+
\frac{1}{2}\eps_{xyz} (\lamu{i})_{jk} {\bar U}_i^x{\bar
D}_j^y{\bar D}^z_k\,.
\label{superpot} 
\end{eqnarray} 
We denote an $SU(3)$ colour index of the fundamental representation by
$x,y,z=1,2,3$. The $SU(2)_L$ fundamental representation indices are
denoted by $a,b,c=1,2$ and the generation indices by $i,j,k=1,2,3$.
We have introduced the twelve $3\times3$ matrices
\beq 
\ye,\quad \yd,\quad
\yu,\quad \lame{k},\quad \lamd{k},\quad \lamu{i}, 
\label{matrices} 
\eeq
for all the Yukawa couplings. This implies the following conventions
in the Martin and Vaughn~\cite{mv} notation
\begin{eqnarray} 
Y^{L_i^a Q^{bx}_j \bar{D}_{ky}}
&=& Y^{L_i^a \bar{D}_{ky} Q^{bx}_j} = Y^{\bar{D}_{ky} L_i^a Q^{bx}_j}
= Y^{Q^{bx}_j L_i^a \bar{D}_{ky}} \label{a4}\nonum  \\ 
&=& Y^{Q^{bx}_j
\bar{D}_{ky} L_i^a} = Y^{\bar{D}_{ky} Q^{bx}_j L_i^a} =
(\lamd{k})_{ij} \eps_{ab} \delta^{y}_{x}\: \equiv \:\lam_{ijk}'
\eps_{ab}\: \delta^{y}_{x},\\ 
Y^{L_i^a L_j^b\bar{E}_k} &=& Y^{L_i^a \bar{E}_k L_j^b} = 
Y^{\bar{E}_k L_i^a L_j^b} =
(\lame{k})_{ij} \eps_{ab} = -(\lame{k})_{ji} \eps_{ab} \: \equiv\: \lam_{ijk}\:
\eps_{ab},
\\ 
Y^{{\bar U}_{ix}{\bar D}_{jy}{\bar D}_{kz}}&=&Y^{{\bar
D}_{jy}{\bar U}_{ix}{\bar D}_{kz}}= Y^{{\bar D}_{jy}{\bar D}_{kz}{\bar
U}_{ix}}= \eps_{x y z} \left({\bf\Lam}_{U^i}\right)_{jk} =
-\eps_{x y z} \left({\bf\Lam}_{U^i}\right)_{kj}\: \equiv \:\eps_{x y z}\:
\lam_{ijk}^{\prime\prime},
\label{a6} 
\end{eqnarray} 

The soft SUSY breaking Lagrangian is given by,
\begin{eqnarray}
-{\cal L} &=& \mh{1} H_1^\dagger H_1 + \mh{2} H_2^\dagger H_2
+ \widetilde{L}^\dagger \ml \widetilde{L}  
+ \widetilde{L_i}^\dagger  \mlh{i} H_1
+ H_1^\dagger \mhl{i} \widetilde{L}
+\widetilde{Q}^\dagger \mq \widetilde{Q} 
\nonum \\
&+& \widetilde{\bar{E}} \me \widetilde{\bar{E}}^\dagger
+ \widetilde{\bar{D}} \md \widetilde{\bar{D}}^\dagger
+ \widetilde{\bar{U}} \mup \widetilde{\bar{U}}^\dagger 
-\biggl [\widetilde{B} H_1 H_2 + \widetilde{D}_i 
\widetilde{L_i} H_2\; \; + {\rm H.c} \biggr ] 
\nonum \\ &+& 
\biggl [(\he)_{ij} \widetilde{L_i} H_1 \widetilde{\bar{E}_j}
+ (\hd)_{ij} \widetilde{Q_i} H_1 \widetilde{\bar{D}_j}
+ (\hu)_{ij} \widetilde{Q_i} H_2 \widetilde{\bar{U}_j} 
\nonum \\
&+& 
(\lamhe{k})_{ij} \widetilde{L_i} \widetilde{L_j} \widetilde{\bar{E}_k}
+(\lamhd{k})_{ij} \widetilde{L_i} \widetilde{Q_j} \widetilde{\bar{D}_k}
+(\lamhu{i})_{jk} \widetilde{\bar{U}_i} \widetilde{\bar{D}_j} 
\widetilde{\bar{D}_k}
+ \; \; {\rm H.c} \biggr ]
\label{soft}
\end{eqnarray}
where we have introduced the soft SUSY breaking trilinear couplings
\beq 
\he,\quad \hd,\quad
\hu,\quad \lamhe{k},\quad \lamhd{k},\quad \lamhu{i}, 
\label{trilsoft} 
\eeq
defined analogously as the Yukawa couplings in (\ref{a4}-\ref{a6}).

In general the one-loop renormalization group equations for the Yukawa
couplings are given by \cite{mv}
\beq
\frac{d}{dt} Y^{ijk}=Y^{ijp}\left[\frac{1}{16\pi^2}\gamma_p^{k}
\right]+(k\lra i)+(k\lra j)\,,
\label{rgyukawa}
\eeq
and the  anomalous dimensions are
\begin{eqnarray}
\gamma_i^{j}&=& \half Y_{ipq}Y^{jpq}-2\delta_i^j\sum_ag_a^2C_a(i)\,,
 \label{gamone}
\end{eqnarray}
We have denoted by $C_a(f)$ the quadratic Casimir of the
representation $f$ of the gauge group $G_a$.  For details see the
Appendix A of Ref.~~\cite{add}. All equations in this section
are valid in the $\overline{\rm DR}$ renormalisation scheme.

The one-loop anomalous dimensions are given by~\cite{add,typo}:
\begin{eqnarray}
\gamma^{L_i}_{L_j} &=&\left(\ye \ye^\dagg \right)_{ij}
+(\lame{q}\lame{q}^\dagg)_{ij} +3 (\lamd{q}\lamd{q}^\dagg)_{ij}
-\delta^i_j(\frac{3}{10}g_1^2+\frac{3}{2}g_2^2)\,, \label{gamll1}\\
\gamma^{E_i}_{E_j} &=& 2 \left(\ye^\dagg \ye \right)_{ji}
+ \mbox{Tr}(\lame{i}\lame{j}^\dagg) -\delta^i_j(\frac{6}{5}g_1^2)\,,
\label{gamee1}\\
\gamma^{Q_i}_{Q_j} &=& \left(\yd \yd^\dagg \right)_{ij}
+ \left(\yu \yu^\dagg \right)_{ij}
+ (\lamd{q}^\dagg\lamd{q})_{ji}
-\delta^i_j(\frac{1}{30}g_1^2+\frac{3}{2}g_2^2+\frac{8}{3}g_3^2)\,,\\
\gamma^{D_i}_{D_j} &=& 2 \left(\yd^\dagg \yd \right)_{ji}
+2 \mbox{Tr}(\lamd{j}^\dagg\lamd{i})
+2 (\lamu{q}\lamu{q}^\dagg)_{ij}
-\delta^i_j(\frac{2}{15}g_1^2+\frac{8}{3}g_3^2)\,,\\
\gamma^{U_i}_{U_j} &=& 2\left(\yu^\dagg \yu \right)_{ji}
+ \mbox{Tr}(\lamu{i}\lamu{j}^\dagg)
-\delta^i_j(\frac{8}{15}g_1^2+\frac{8}{3}g_3^2)\,,\\
\gamma^{H_1}_{H_1} &=& \mbox{Tr}\left(3\yd\yd^\dagg+\ye\ye^\dagg \right)
-(\frac{3}{10}g_1^2+\frac{3}{2}g_2^2)\,,\\
\gamma^{H_2}_{H_2} &=& 3 \mbox{Tr}\left( \yu\yu^\dagg\right)
-(\frac{3}{10}g_1^2+\frac{3}{2}g_2^2)\,,\\
\gamma^{H_1}_{L_i} &=& {\gamma^{L_i}_{H_1}}^* =-3 (\lamd{q}^*\yd)_{iq}
- (\lame{q}^*\ye)_{iq}\,.
\label{mixedad}
\end{eqnarray}
Note that here, $H_{1,2},L,Q$ represent the fields $H_{1,2}^a$, $L^a$,
$Q^{a}$ where $a$ is the index of the fundamental representation of
$SU(2)$ (\ie no factors of $\eps_{ab}$ are factored). The
$\beta$-functions for the Yukawa couplings~\cite{add} and for the
bilinear superpotential couplings are combinations of the above
anomalous dimensions (\ref{gamll1}-\ref{mixedad}). The two loop
anomalous dimensions in the \rpv -MSSM can be found in \cite{add}. We
present the one-loop beta functions for the superpotential couplings
and masses for completeness.

The RGEs for the Yukawa couplings (including full family dependence)
are  given by
\begin{eqnarray}
16\pi^2 \Dt (\ye)_{ij} &=& (\ye)_{ik}\ggam{E_j}{E_k}
+(\ye)_{ij}\ggam{H_1}{H_1}
-(\lame{j})_{ki}\ggam{H_1}{L_k}
+(\ye)_{kj}\ggam{L_i}{L_k}\,,\label{eq:ye}\\
16\pi^2\Dt (\yd)_{ij} &=& (\yd)_{ik}\ggam{D_j}{D_k}
  +(\yd)_{ij}\ggam{H_1}{H_1}
  -(\lamd{j})_{ki}\ggam{H_1}{L_k}
  +(\yd)_{kj}\ggam{Q_i}{Q_k}\,,\label{eq:yd}\\
16\pi^2\Dt (\yu)_{ij} &=& (\yu)_{ik}\ggam{U_j}{U_k}\label{yu}
+(\yu)_{ij}\ggam{H_2}{H_2}
+(\yu)_{kj}\ggam{Q_i}{Q_k}\,, 
\\
16\pi^2\Dt (\lame{k})_{ij} &=& (\lame{l})_{ij}\ggam{E_k}{E_l}
+(\lame{k})_{il}\ggam{L_j}{L_l}+(\ye)_{ik}\ggam{L_j}{H_1}
-(\lame{k})_{jl}\ggam{L_i}{L_l}-(\ye)_{jk}\ggam{L_i}{H_1}\,, \label{eq:lame}\\
16\pi^2\Dt (\lamd{k})_{ij} &=& (\lamd{l})_{ij}\ggam{D_k}{D_l}\lab{lamd}
+(\lamd{k})_{il}\ggam{Q_j}{Q_l}
+(\lamd{k})_{lj}\ggam{L_i}{L_l}
-(\yd)_{jk}\ggam{L_i}{H_1}\,,\\
16\pi^2\Dt (\lamu{i})_{jk} &=& (\lamu{i})_{jl}\ggam{D_k}{D_l} \lab{lamu}
+(\lamu{i})_{lk}\ggam{D_j}{D_l}
+(\lamu{l})_{jk}\ggam{U_i}{U_l}\,.
\end{eqnarray}
Here $t=\ln(Q)$, and $Q$ is the renormalization scale.  The RGEs for
the bilinear terms are
\begin{eqnarray}
16\pi^2\Dt\mu&=&\mu\left\{ \ggam{H_1}{H_1}+\ggam{H_2}{H_2}\right\}
+\kap^i\ggam{H_1}{L_i},\\[3mm]
16\pi^2\Dt\kap^i&=&\kap^i\ggam{H_2}{H_2}+\kap^p\ggam{L_i}{L_p}
+\mu\ggam{L_i}{H_1}. \lab{kappa}
\end{eqnarray}

\section{A method to derive the soft SUSY breaking RGEs}
\label{app3}

A straightforward way to derive the RGEs for the soft SUSY breaking
couplings and masses is by a direct use of the explicit formul{\ae} at
1-loop given in~\cite{mv}. This is a somewhat tedious job. A very
elegant method which is also very helpful for numerical calculations
is the one described in Ref.~\cite{jones}. All the soft SUSY RGEs can
be derived from the anomalous dimensions (\ref{gamll1}-\ref{mixedad})
by the action of an operator which is given below
\cite{gauginomasses}.  The method works not only at one loop but it
has been proven to all orders in perturbation theory~\cite{jones}.  In
principle one could apply the operators (B1-B5) below to the 2-loop
anomalous dimensions derived in Ref.~\cite{add} and write down the
full two loop coupled RGEs in the most general case. However, here we
restrict ourselves to the one-loop case. In particular the soft
$\beta$-functions for the bilinear ${\bf b}^{ij}$, trilinear ${\bf
h}^{ijk}$ and scalar masses $({\bf m}^2)^i_j$ soft SUSY breaking terms
can be read from
\begin{eqnarray}
16\pi^2 \frac{d {\bf b}^{ij}}{dt} &=& \gamma^i_l {\bf b}^{jl}+\gamma^j_l
{\bf b}^{il}-2 (\gamma_1)^i_l {\bf \mu}^{jl} -2 (\gamma_1)^j_l 
{\bf \mu}^{il} \;, \label{bmass}\\
16\pi^2 \frac{d {\bf h}^{ijk}}{dt} &=& \gamma^i_l {\bf h}^{jkl}+
\gamma^j_l {\bf h}^{ikl} + \gamma^k_l {\bf h}^{jil} 
-2 (\gamma_1)^i_l {\bf Y}^{jkl} -2 (\gamma_1)^j_l {\bf Y}^{ikl}
-2 (\gamma_1)^k_l {\bf Y}^{jil} \;, \label{tril}\\
16\pi^2 \frac{d ({\bf m}^2)^i_j}{dt} &=& \biggl ( 2 {\cal O} {\cal O}^*
+2 M M^* g_a^2 \frac{\partial}{\partial g^2_a} + 
\widetilde{{\bf Y}}_{lmn} \frac{\partial}{\partial {{\bf Y}}_{lmn}}
+
\widetilde{{\bf Y}}^{lmn} \frac{\partial}{\partial {{\bf Y}}^{lmn}}
+ X_a \frac{\partial}{\partial g_a} \biggr )\gamma^i_j \;,\label{smass}
\end{eqnarray}
where 
\begin{eqnarray}
(\gamma_1)_j^i &=& {\cal O} \gamma_j^i\,,\qquad 
{\cal O} = \biggl ( M_a g_a^2 \frac{\partial}{\partial g_a^2} - 
{\bf h}^{lmn} \frac{\partial}{\partial {\bf Y}^{lmn}} \biggr ) \;,
\label{b4}
\\[3mm]
\widetilde{{\bf Y}}^{ijk} &=& {\bf Y}^{ljk} ({\bf m}^2)^i_l +
{\bf Y}^{lik} ({\bf m}^2)^j_l+{\bf Y}^{lji} ({\bf m}^2)^k_l \;, \label{b6}
\end{eqnarray}
and repeated indices are summed over. 
At one loop the last term, $X_a$ ,in Eq.~(\ref{smass}) is not
relevant. Its (scheme dependent form) is given for example in the last
reference of Ref.~\cite{jones} (see their Eq.~(2.11)).  The
RGEs~(\ref{bmass}-\ref{smass}) are valid as long as we do not
eliminate the U(1) Fayet-Iliopoulos (FI) D-term. The RGE running of
the FI-term can then be written independently.  It is known that for
universal boundary conditions this term is not renormalized down to
low energies and we do not discuss its RGE here.  On the other hand if
we eliminate the FI D-term by using its equation of motion then this
renormalization gives rise to additional contributions proportional to
the U(1) gauge coupling (see the ${\cal S}$-term in the RGEs for the
soft SUSY breaking masses in the Appendix~\ref{app4}).
Now from Eq.~(\ref{bmass}) the RGEs for the bilinear soft SUSY
breaking masses in the \rpv-MSSM are
\begin{eqnarray}
16 \pi^2 \frac{d\widetilde{B}}{dt} &=& \widetilde{B}
\biggl [\gamma^{H_1}_{H_1}+
\gamma^{H_2}_{H_2} \biggr ] +\widetilde{D}_i \gamma^{H_1}_{L_i}
-2 \mu \biggl [(\gamma_1)^{H_1}_{H_1}+(\gamma_1)^{H_2}_{H_2}
\biggr ]-2 \kap_i (\gamma_1)^{H_1}_{L_i} \;,\label{b7}\\[3mm]
16 \pi^2 \frac{d\widetilde{D}_i}{dt} &=&
\biggl [\gamma^{L_i}_{L_l}\widetilde{D}^l
+\gamma_{H_2}^{H_2}\widetilde{D}^i
 \biggr]
+\widetilde{B} \gamma_{H_1}^{L_i} 
-2 \biggl [ (\gamma_1)^{L_i}_{L_l} \kap^l +  (\gamma_1)^{H_2}_{H_2}\kap^i
\biggr ] -2 \mu (\gamma_1)_{H_1}^{L_i} \;.
\end{eqnarray}
The RGEs for the trilinear  soft SUSY breaking masses  in the 
\rpv-MSSM can be read from Eq.~(\ref{tril})
\begin{eqnarray}
16 \pi^2 \frac{d (\he)_{ik}}{d t}  &=& 
\gamma_{L_l}^{L_i} (\he)_{lk} +\gamma_{H_1}^{H_1} (\he)_{ik}
+\gamma^{H_1}_{L_l} (\lamhe{k})_{il} + \gamma^{E_k}_{E_l} (\he)_{il}
\nonum \\
&-&
2 (\gamma_1)_{L_l}^{L_i} (\ye)_{lk} -2 (\gamma_1)_{H_1}^{H_1} (\ye)_{ik}
-2 (\gamma_1)^{H_1}_{L_l} (\lame{k})_{il}
-2 (\gamma_1)^{E_k}_{E_l} (\ye)_{il} \;, \\[3mm]
16 \pi^2 \frac{d (\hd)_{ik}}{d t}  &=& 
\gamma_{Q_l}^{Q_i} (\hd)_{lk} +\gamma_{H_1}^{H_1} (\hd)_{ik}
-\gamma^{H_1}_{L_l} (\lamhd{k})_{li} + \gamma^{D_k}_{D_l} (\hd)_{il}
\nonum \\
&-&
2 (\gamma_1)_{Q_l}^{Q_i} (\yd)_{lk} -2 (\gamma_1)_{H_1}^{H_1} (\yd)_{ik}
+2 (\gamma_1)^{H_1}_{L_l} (\lamd{k})_{li}
-2 (\gamma_1)^{D_k}_{D_l} (\yd)_{il} \;, \\[3mm]
16 \pi^2 \frac{d (\hu)_{ik}}{d t}  &=&
\gamma_{Q_l}^{Q_i} (\hu)_{lk} +\gamma_{H_2}^{H_2} (\hu)_{ik}
 + \gamma^{U_k}_{U_l} (\hu)_{il} \nonum \\
&-&
2 (\gamma_1)_{Q_l}^{Q_i} (\yu)_{lk} -2 (\gamma_1)_{H_2}^{H_2} (\yu)_{ik}
-2 (\gamma_1)^{U_k}_{U_l} (\yu)_{il} \;, \\[3mm]
16 \pi^2 \frac{d (\lamhe{k})_{ij}}{d t}  &=&
\gamma^{L_i}_{L_l} (\lamhe{k})_{lj}-\gamma_{H_1}^{L_i}(\he)_{jk}+
\gamma_{L_l}^{L_j}(\lamhe{k})_{il}
+\gamma_{H_1}^{L_j} (\he)_{ik}+\gamma^{E_k}_{E_l}(\lamhe{l})_{ij}
\nonum \\
&-& 
2 (\gamma_1)^{L_i}_{L_l} (\lame{k})_{lj}+2 (\gamma_1)_{H_1}^{L_i}(\ye)_{jk}
-2 (\gamma_1)_{L_l}^{L_j}(\lame{k})_{il} -2 (\gamma_1)_{H_1}^{L_j} (\ye)_{ik}
-2 (\gamma_1)^{E_k}_{E_l}(\lame{l})_{ij} \;, \\[3mm]
16 \pi^2 \frac{d (\lamhd{k})_{ij}}{d t}  &=&
\gamma^{L_i}_{L_l} (\lamhd{k})_{lj}-\gamma_{H_1}^{L_i}(\hd)_{jk}+
\gamma_{Q_l}^{Q_j}(\lamhd{k})_{il}
+\gamma^{D_k}_{D_l}(\lamhd{l})_{ij}
\nonum \\
&-& 
2 (\gamma_1)^{L_i}_{L_l} (\lamd{k})_{lj}+2 (\gamma_1)_{H_1}^{L_i}(\yd)_{jk}
-2 (\gamma_1)_{Q_l}^{Q_j}(\lamd{k})_{il}
-2 (\gamma_1)^{D_k}_{D_l}(\lamd{l})_{ij} \;, \\[3mm]
16 \pi^2 \frac{d (\lamhu{i})_{jk}}{d t}  &=&
\gamma^{U_i}_{U_l} (\lamhu{l})_{jk}
+\gamma_{D_l}^{D_j}(\lamhu{i})_{lk}
+\gamma^{D_k}_{D_l}(\lamhu{i})_{jl}
\nonum \\
&-& 
2 (\gamma_1)^{U_i}_{U_l} (\lamu{l})_{jk}
-2 (\gamma_1)_{D_l}^{D_j}(\lamu{i})_{lk}
-2 (\gamma_1)^{D_k}_{D_l}(\lamu{i})_{jl} \;.
\end{eqnarray}
The RGEs for the soft SUSY breaking masses in the \rpv-MSSM can be
obtained from Eq.~(\ref{smass})
\begin{eqnarray}
16\pi^2 \frac{d \me^{E_i}_{E_j}}{dt}\equiv 16\pi^2 \frac{d
\me_{ji}}{dt} &=& 4 (\hed\he)_{ji} +2 \Tr
(\lamhe{i}\lamhed{j})-\delta_{ij} \biggl (\frac{24}{5} g_1^2 |M_1|^2
\biggr ) \nonum \\ &+& 2 (\yed \widetilde{\ye})_{ji} +\Tr
(\widetilde{\lame{i}}\lamed{j}) +2 (\widetilde{\yed}\ye)_{ji} +\Tr
(\lame{i}\widetilde{\lamed{j}}) \;,\\[3mm]
16\pi^2 \frac{d \ml^{L_i}_{L_j}}{dt}\equiv 
16\pi^2 \frac{d \ml_{ij}}{dt}  &=& 
2 (\he\hed+\lamhe{q}\lamhed{q}+3\lamhd{q}\lamhdd{q})_{ij}
-\delta_{ij}\biggl (\frac{6}{5} g_1^2 |M_1|^2+6 g_2^2 |M_2|^2\biggr )
\nonum \\
&+& (\widetilde{\ye}\yed)_{ij}
+(\widetilde{\lame{q}}\lamed{q})_{ij}
+3 (\widetilde{\lamd{q}}\lamdd{q})_{ij} \nonum \\
&+& (\ye\widetilde{\yed})_{ij}
+ (\lame{q}\widetilde{\lamed{q}})_{ij}
+3 (\lamd{q}\widetilde{\lamdd{q}})_{ij} \;, \\[3mm]
16\pi^2 \frac{d ({\bf m^2})^{H_1}_{L_i}}{dt}\equiv 
16\pi^2 \frac{d \mhl{i}}{dt}  &=&
-6 (\lamhd{q}^*\hd)_{iq}-2(\lamhe{q}^*\he)_{iq} \nonum \\
&-& 3 (\lamds{q}\widetilde{\yd})_{iq}-(\lames{q}\widetilde{\ye})_{iq}
-3 (\widetilde{\lamds{q}}\yd)_{iq} - (\widetilde{\lames{q}}\ye)_{iq}
\,, \label{mh1lrge} \\[3mm]
16\pi^2 \frac{d \mq^{Q_i}_{Q_j}}{dt}\equiv 
16\pi^2 \frac{d \mq_{ij}}{dt}  &=&
2 ( \hd \hdd+\hu\hud)_{ij}+2 (\lamhdd{q}\lamhd{q})_{ji}
-\delta_{ij} \biggl (\frac{2}{15}g_1^2 |M_1|^2+6 g_2^2 |M_2|^2
+\frac{32}{3}g_3^2 |M_3|^2 \biggr ) \nonum \\
&+&(\widetilde{\yd}\ydd)_{ij} + (\widetilde{\yu}\yud)_{ij}
+(\lamdd{q}\widetilde{\lamd{q}})_{ji}  \nonum \\
&+& (\yd\widetilde{\ydd})_{ij} + (\yu\widetilde{\yud})_{ij}
+(\widetilde{\lamdd{q}}\lamd{q})_{ji} \,, \\[3mm]
16\pi^2 \frac{d \md^{D_i}_{D_j}}{dt}\equiv 
16\pi^2 \frac{d \md_{ji}}{dt}  &=&
4(\hdd\hd)_{ji}+4\Tr(\lamhdd{j}\lamhd{i})+4(\lamhu{q}\lamhud{q})_{ij}
\nonum \\
&-& \delta_{ij} \biggl (\frac{8}{15}g_1^2 |M_1|^2+\frac{32}{3}g_3^2
|M_3|^2 \biggr ) \nonum \\
&+& 2 (\ydd\widetilde{\yd})_{ji}+2\Tr(\lamdd{j}\widetilde{\lamd{i}})
+2 (\widetilde{\lamu{q}}\lamud{q})_{ij} \nonum \\
&+& 2(\widetilde{\ydd}\yd)_{ji}+2\Tr(\widetilde{\lamdd{j}}\lamd{i})
+2 (\lamu{q}\widetilde{\lamud{q}})_{ij}\;, \\[3mm]
16\pi^2 \frac{d \mup^{U_i}_{U_j}}{dt}\equiv 
16\pi^2 \frac{d \mup_{ji}}{dt}  &=&
4(\hud\hu)_{ji}+2\Tr(\lamhu{i}\lamhud{j})
-\delta_{ij} \biggl (\frac{32}{15}g_1^2 |M_1|^2+\frac{32}{3}g_3^2
|M_3|^2 \biggr ) \nonum \\
&+& 2 (\yud\widetilde{\yu})_{ji} + \Tr(\widetilde{\lamu{i}}\lamud{j})
+ 2 (\widetilde{\yud}\yu)_{ji} + \Tr(\lamu{i}\widetilde{\lamud{j}})
\;, \\[3mm]
16\pi^2 \frac{d \mh{1}}{dt}  &=&
\Tr (6 \hd\hdd +2\he\hed)-\biggl ( \frac{6}{5} g_1^2 |M_1|^2 +
6 g_2^2 |M_2|^2 \biggr ) \nonum \\
&+& 3 \Tr(\widetilde{\yd}\ydd) +\Tr(\widetilde{\ye}\yed)
+3 \Tr(\yd\widetilde{\ydd}) + \Tr(\ye\widetilde{\yed}) \;,  \\[3mm]
16\pi^2 \frac{d \mh{2}}{dt}  &=&
6 \Tr (\hu\hud)-\biggl ( \frac{6}{5} g_1^2 |M_1|^2 +
6 g_2^2 |M_2|^2 \biggr ) \nonum \\
&+& 3 \Tr(\widetilde{\yu}\yud) + 3 \Tr(\yu\widetilde{\yud}) \;, \label{b22}
\end{eqnarray}
where from Eq.~(\ref{b4}) we have
\begin{eqnarray}
(\gamma_1)^{L_i}_{L_j} &=&-\left(\he \ye^\dagg \right)_{ij}
-(\lamhe{q}\lame{q}^\dagg)_{ij} -3 (\lamhd{q}\lamd{q}^\dagg)_{ij}
-\delta^i_j(\frac{3}{10}M_1g_1^2+\frac{3}{2}M_2g_2^2)\,, \label{g1gamll1}\\
(\gamma_1)^{E_i}_{E_j} &=& -2 \left(\ye^\dagg \he \right)_{ji}
- \mbox{Tr}(\lamhe{i}\lame{j}^\dagg) -\delta^i_j(\frac{6}{5}M_1 g_1^2)\,,
\label{g1gamee1}\\
(\gamma_1)^{Q_i}_{Q_j} &=& -\left(\hd \yd^\dagg \right)_{ij}
- \left(\hu \yu^\dagg \right)_{ij}
- (\lamd{q}^\dagg\lamhd{q})_{ji}
-\delta^i_j(\frac{1}{30}M_1 g_1^2+\frac{3}{2} M_2g_2^2+
\frac{8}{3}M_3 g_3^2)\,,\\
(\gamma_1)^{D_i}_{D_j} &=& -2 \left(\yd^\dagg \hd \right)_{ji}
-2 \mbox{Tr}(\lamd{j}^\dagg\lamhd{i})
-2 (\lamhu{q}\lamu{q}^\dagg)_{ij}
-\delta^i_j(\frac{2}{15}M_1 g_1^2+\frac{8}{3}M_3g_3^2)\,,\\
(\gamma_1)^{U_i}_{U_j} &=& -2\left(\yu^\dagg \hu \right)_{ji}
- \mbox{Tr}(\lamhu{i}\lamu{j}^\dagg)
-\delta^i_j(\frac{8}{15}M_1 g_1^2+\frac{8}{3}M_3g_3^2)\,,\\
(\gamma_1)^{H_1}_{H_1} &=& -\mbox{Tr}\left(3\hd\yd^\dagg+\he\ye^\dagg \right)
-(\frac{3}{10}M_1g_1^2+\frac{3}{2}M_2g_2^2)\,,\\
(\gamma_1)^{H_2}_{H_2} &=& -3 \mbox{Tr}\left( \hu\yu^\dagg\right)
-(\frac{3}{10}M_1g_1^2+\frac{3}{2}M_2g_2^2)\,,\\
(\gamma_1)^{H_1}_{L_i} &=& {(\gamma_1)^{L_i}_{H_1}}^* =3 (\lamd{q}^*\hd)_{iq}
+ (\lame{q}^*\he)_{iq} \;,\label{g1mixedad}
\end{eqnarray}
and from Eq.~(\ref{b6})
\begin{eqnarray}
(\widetilde{\ye})_{ik} &=& (\ye)_{lk}\ml_{il}+(\ye)_{ik} \mh{1}
+(\lame{k})_{il}\mhl{l}+(\ye)_{il}\me_{lk} \;, \label{b31}\\[2mm]
(\widetilde{\yd})_{ik} &=&
(\yd)_{lk}\mq_{il}-(\lamd{k})_{li}\mhl{l}
+(\yd)_{ik}\mh{1}+(\yd)_{il} \md_{lk} \;, \\[2mm]
(\widetilde{\yu})_{ik} &=&
(\yu)_{lk}\mq_{il}+(\yu)_{ik}\mh{2}+(\yu)_{il}
\mup_{lk} \;, \\[2mm]
(\widetilde{\lame{k}})_{ij} &=&
(\lame{k})_{lj}\ml_{il}-(\ye)_{jk}\mlh{i}+(\lame{k})_{il}\ml_{jl}
+(\ye)_{ik}\mlh{j}+(\lame{l})_{ij}\me_{lk} \;,\\[2mm]
(\widetilde{\lamd{k}})_{ij} &=&
(\lamd{k})_{lj}\ml_{il}-(\yd)_{jk}\mlh{i}+(\lamd{k})_{il}\mq_{jl}
+(\lamd{l})_{ij}\md_{lk} \;, \\[2mm]
(\widetilde{\lamu{i}})_{jk} &=&
(\lamu{l})_{jk}\mup_{li}+(\lamu{i})_{lk}\md_{lj}+(\lamu{i})_{jl}\md_{lk} \;.
\label{b36}
\end{eqnarray}
Numerically we follow the following procedure :
\begin{enumerate}
\item[(a)]  Define the anomalous dimensions in Eqs.~(\ref{gamll1}-\ref{mixedad}).
\item[(b)]  Define $(\gamma_1)^i_j$ from Eqs.~(\ref{g1gamll1}-\ref{g1mixedad}).
\item[(c)]  Define Eq.~(\ref{b31}-\ref{b36}).
\item[(d)]  Plug (a,b,c) into Eqs.~(\ref{b7}-\ref{b22}).
\end{enumerate}
This is much simpler than inserting the explicit formul{\ae} of
Appendix~\ref{app4} below.

\section{Explicit RGEs for the soft supersymmetric breaking terms}
\label{app4}

The explicit RGEs for the soft supersymmetric breaking terms have
appeared also before in Refs.\cite{deCarlos:1996du} and
\cite{Besmer:2000rj}. Ref.\cite{Besmer:2000rj} contains the full set (aside
from the aforementioned ${\cal S}$ term), but we
disagree with several terms in the equations for $\mhl{i}$ and $\me_{ij}$.
 Ref.~\cite{deCarlos:1996du} is restricted to
contributions of the third generation quarks and leptons.  We arrange
here the explicit formul{\ae} of the full (not flavour dominance
assumed) RGEs.  As a cross check, we have calculated them by first
using the explicit formulae of Ref.~\cite{mv} and second by using the
method described in Appendix~\ref{app3}.  We found agreement using
both methods. Thus the RGE for the bilinear $\mu$ and $\kap_i$ terms
of the superpotential parameters is given by
\begin{eqnarray}
16 \pi^2 \frac{d \mu}{d t} \ &=& \ \mu \biggl [ 3 \Tr (\yu \yud)+
\Tr (3 \yd \ydd + \ye \yed)-  \frac{3}{5} g_1^2 - 3 g_2^2 
\biggr ] \nonum \\
&-& \kap_p \biggl [ \lames{n} \ye + 3 \lamds{n} \yd \biggr ]_{pn} \;,
 \\[4mm]
16 \pi^2 \frac{d \kap_i}{d t} \ &=& \ \kap_i \biggl [ 3 \Tr ( \yu
\yud) - \frac{3}{5} g_1^2 - 3 g_2^2 \biggr ] \nonum \\  
&+& \kap_p \biggl [\ye \yed + \lame{n} \lamed{n} +3 \lamd{n} \lamdd{n}
\biggr ]_{ip}
- \mu \biggl [ \lame{n} \yes + 3 \lamd{n}{\yds} \biggr ]_{in} \;.
\end{eqnarray}
Similarly, the RGEs for the soft SUSY breaking bilinear terms can
be read from,
\begin{eqnarray}
16 \pi^2 \frac{d \widetilde{B}}{d t} \ &=& \ \widetilde{B}
\biggl [ 3 \Tr (\yud \yu)+3 \Tr (\ydd \yd)+\Tr (\yed \ye) 
-\frac{3}{5} g_1^2 -3 g_2^2 \biggr ] \nonum \\
&+& \mu \biggl [ 6 \Tr(\yud \hu)+6 \Tr(\ydd \hd)+2 \Tr(\yed \he) +
\frac{6}{5} g_1^2 M_1 + 6 g_2^2 M_2 \biggr ] \nonum \\
&-& \widetilde{D}_l \biggl [ \lames{n} \ye + 3 \lamds{n} \yd \biggr ]_{ln}
- \kap_l \biggl [2 \lames{n} \he+6 \lamds{n} \hd \biggr ]_{ln} \; ,
\\[4mm]
16 \pi^2 \frac{d \widetilde{D}_i}{d t} \ &=& \ \widetilde{D}_i \biggl [
3 \Tr(\yu \yud)-\frac{3}{5}g_1^2 -3 g_2^2 \biggr ]
+ 
\kap_i \biggl [ 6 \Tr(\hu \yud)+ \frac{6}{5} g_1^2 M_1 +6 g_2^2 M_2 \biggr ]
\nonum \\
&+& \widetilde{D}_l \biggl [ \ye \yed + \lame{n} \lamed{n} + 3 \lamd{n}
\lamdd{n} \biggr ]_{il} 
+ 2 \kap_l \biggl [ \he \yed + \lamhe{n} \lamed{n} +3 \lamhd{n} \lamdd{n}
\biggr ]_{il} \nonum \\
&-& 2 \mu \biggl [ \lamhe{n} \yes+3 \lamhd{n}\yds \biggr ]_{in}
-\widetilde{B} \biggl [ \lame{n} \yes+3 \lamd{n}\yds \biggr ]_{in} \;.
\label{RG-D}
\end{eqnarray}
The RGEs for the soft SUSY trilinear couplings are given by
\begin{eqnarray}
16 \pi^2 \frac{d (\he)_{ij}}{d t} \ &=& \ (\he)_{il} \biggl [
2 (\yed \ye)_{lj}+ \Tr(\lamed{l}\lame{j}) \biggr ]
\nonum \\ 
&+& (\he)_{lj} \biggl [\ye \yed+\lame{n}\lamed{n}+3 \lamd{n}\lamdd{n}
\biggr ]_{il} \nonum \\
&+& (\he)_{ij} \biggl [ \Tr(\yed \ye)+3 \Tr(\ydd  \yd) -\frac{9}{5}g_1^2
-3 g_2^2 \biggr ]
 \nonum \\ &+& (\lamhe{j})_{il} \biggl [-\lames{n} \ye - 3 \lamds{n} \yd
\biggr ]_{ln} \nonum \\
&+& (\ye)_{il} \biggl [ 4 (\yed \he)_{lj}+2\Tr(\lamed{l}\lamhe{j}) \biggr ]
\nonum \\ 
&+&(\ye)_{lj} \biggl [ 2\he\yed+2\lamhe{n}\lamed{n}+6 \lamhd{n}\lamdd{n}
\biggr ]_{il} \nonum \\
&+& (\ye)_{ij} \biggl [ 2 \Tr(\yed \he)+ 6 \Tr(\ydd \hd)+\frac{18}{5}g_1^2
M_1+6 g_2^2M_2 \biggr ]\nonum \\ &+&
(\lame{j})_{il} \biggl [ -2 (\lames{n}\he)-6(\lamds{n}\hd)\biggr ]_{ln} \;,
\\[4mm]
16 \pi^2 \frac{d (\hd)_{ij}}{d t} \ &=& \ (\hd)_{il} \biggl [
2 (\ydd \yd)_{lj}+2 \Tr(\lamdd{l}\lamd{j})+2(\lamu{n}\lamud{n})_{jl} \biggr
] \nonum \\
&+& (\hd)_{lj} \biggl [ \yd \ydd+\yu\yud \biggr ]_{il}+
(\hd)_{lj} \biggl [\lamdd{n}\lamd{n} \biggr ]_{li}
\nonum \\ 
&+& (\hd)_{ij} \biggl [ \Tr(\yed \ye)+3 \Tr(\ydd\yd) -\frac{7}{15}g_1^2
-3 g_2^2 -\frac{16}{3}g_3^2 \biggr ] \nonum \\
&+& (\lamhd{j})_{li}\biggl [(\lames{n}\ye)+3 (\lamds{n}\yd) \biggr ]_{ln}
\nonum \\ &+& (\yd)_{il} \biggl [ 4 (\ydd\hd)_{lj}+4\Tr(\lamdd{l}\lamhd{j})+
4(\lamhu{n}\lamud{n})_{jl} \biggr ]\nonum \\
&+& (\yd)_{lj} \biggl [ 2 \hd \ydd+2 \hu\yud \biggr ]_{il}
 +(\yd)_{lj} \biggl [ 2 \lamdd{n}\lamhd{n} 
\biggr ]_{li} 
\nonum \\ 
&+& (\yd)_{ij} \biggl [ 2 \Tr(\yed \he)+6 \Tr(\ydd \hd)
+\frac{14}{15}g_1^2 M_1+6 g_2^2M_2+\frac{32}{3}g_3^2M_3  \biggr ]
\nonum \\
&+& (\lamd{j})_{li} \biggl [ 2 (\lames{n}\he)+
6 (\lamds{n}\hd) \biggr ]_{ln} \;,
\\[4mm]
16 \pi^2 \frac{d (\hu)_{ij}}{d t} \ &=& \ 
(\hu)_{il} \biggl [ 2 (\yud \yu)_{lj}+\Tr(\lamud{l}\lamu{j}) \biggr ]
\nonum \\
&+& (\hu)_{lj} \biggl [ (\yu \yud)_{il}+(\yd \ydd)_{il}+(\lamdd{n}
\lamd{n})_{li} \biggr ] \nonum \\
&+& (\hu)_{ij} \biggl [ 3 \Tr(\yud \yu) - \frac{13}{15}g_1^2-3 g_2^2 - 
\frac{16}{3}g_3^2 \biggr ] \nonum \\
&+& (\yu)_{il} \biggl [ 4 (\yud \hu)_{lj}+2 \Tr (\lamud{l}\lamhu{j}) \biggr ]
\nonum \\
&+& (\yu)_{lj} \biggl [ 2 (\hu \yud)_{il} +2 (\hd \ydd)_{il}+
2 (\lamdd{n}\lamhd{n})_{li} 
\biggr  ]
\nonum \\
&+& (\yu)_{ij} \biggl [ 6\Tr(\yud \hu)+\frac{26}{15}g_1^2M_1+6g_2^2M_2
+\frac{32}{3}g_3^2M_3 \biggr ] \;,
\\[4mm]
16 \pi^2 \frac{d (\lamhe{k})_{ij}}{d t} \ &=& \ 
(\lamhe{l})_{ij}\biggl [ 2 (\yed\ye)_{lk}
+ \Tr(\lamed{l}\lame{k})\biggr ] \nonum \\
&+&
(\he)_{jk}\biggl [\lame{n}\yes +3 \lamd{n}\yds \biggl ]_{in}
\nonum \\ &+& (\lamhe{k})_{jl} \biggl [ -\ye\yed -\lame{n}\lamed{n}-
3\lamd{n}\lamdd{n} \biggr ]_{il} \nonum \\
&+& (\lamhe{k})_{il}\biggl [ \ye \yed+\lame{n}\lamed{n}+3\lamd{n}\lamdd{n}
\biggr ]_{jl} \nonum \\ &+& (\he)_{ik} \biggl [-\lame{n}\yes-3\lamd{n}\yds
\biggr ]_{jn} \nonum \\
&+& (\lame{l})_{ij} \biggl [4 (\yed\he)_{lk}+2\Tr(\lamed{l}\lamhe{k}) \biggr ]
\nonum \\
&+&(\ye)_{jk} \biggl [2 \lamhe{n} \yes+6\lamhd{n}\yds \biggr ]_{in}
\nonum \\
&+& (\lame{k})_{jl} \biggl [-2 \he\yed-2\lamhe{n}\lamed{n}-6\lamhd{n}\lamdd{n}
\biggr ]_{il} 
\nonum \\
&+& (\lame{k})_{il} \biggl [2 \he\yed+2\lamhe{n}\lamed{n}+6\lamhd{n}\lamdd{n}
\biggr ]_{jl} 
\nonum \\
&+& (\ye)_{ik} \biggl [-2\lamhe{n}\yes-6\lamhd{n}\yds \biggr ]_{jn}
\nonum \\
&-&(\lamhe{k})_{ij} \biggl [\frac{9}{5}g_1^2+3g_2^2 \biggr] +
(\lame{k})_{ij}\biggl [\frac{18}{5}g_1^2M_1+6g_2^2M_2 \biggr ] \;,
\\[4mm]
16 \pi^2 \frac{d (\lamhd{k})_{ij}}{d t} \ &=& \ 
(\lamhd{l})_{ij}\biggl [2(\ydd\yd)_{lk}+2\Tr(\lamdd{l}\lamd{k})+
2(\lamud{n}\lamu{n})_{lk} \biggr ]
\nonum \\
&+& (\lamhd{k})_{lj} \biggl [\ye\yed+\lame{n}\lamed{n}+3\lamd{n}\lamdd{n}
\biggr ]_{il} 
\nonum \\
&+& (\hd)_{jk} \biggl [ \lame{n}\yes+3 \lamd{n}\yds \biggr]_{in}
\nonum \\
&+& (\lamhd{k})_{il} \biggl [(\yd\ydd)_{jl}+(\yu\yud)_{jl}+
(\lamdd{n}\lamd{n})_{lj} \biggr ]
\nonum \\
&+& (\lamd{l})_{ij} \biggl [4 (\ydd\hd)_{lk}+4\Tr(\lamdd{l}\lamhd{k})+
4(\lamud{n}\lamhu{n})_{lk} \biggr ] 
\nonum \\
&+& (\lamd{k})_{lj} \biggl [2 \he\yed+2\lamhe{n}\lamed{n}+6 \lamhd{n}
\lamdd{n} \biggr ]_{il} 
\nonum \\
&+& (\yd)_{jk} \biggl [2\lamhe{n}\yes+6\lamhd{n}\yds\biggr ]_{in}
\nonum \\
&+& (\lamd{k})_{il} \biggl [2(\hd\ydd)_{jl}+2(\hu\yud)_{jl}+
2(\lamdd{n}\lamhd{n})_{lj} \biggr ]
\nonum \\
&-& (\lamhd{k})_{ij} \biggl [\frac{7}{15}g_1^2+3g_2^2+\frac{16}{3}g_3^2 
\biggr ] 
\nonum \\
&+& (\lamd{k})_{ij} \biggl [
\frac{14}{15}g_1^2M_1+6g_2^2M_2+\frac{32}{3}g_3^2M_3
\biggr ]\;,
\\[4mm]
16 \pi^2 \frac{d (\lamhu{i})_{jk}}{d t} \ &=& \ 
(\lamhu{i})_{jl} \biggl [ 2 (\ydd \yd)_{lk}+2\Tr(\lamdd{l}\lamd{k})
+2(\lamud{m}\lamu{m})_{lk} \biggr]
\nonum \\
&+& (\lamhu{l})_{jk} \biggl [2(\yud\yu)_{li}+\Tr(\lamud{l}\lamu{i}) 
\biggr ]
\nonum \\
&+& (\lamhu{i})_{kl} \biggl [-2(\ydd\yd)_{lj}-2\Tr(\lamdd{l}\lamd{j})
-2(\lamud{n}\lamu{n})_{lj} \biggr ]
\nonum \\
&+&
(\lamu{i})_{jl} \biggl [ 4(\ydd\hd)_{lk}+4\Tr(\lamdd{l}\lamhd{k})+
4(\lamud{m}\lamhu{m})_{lk} \biggr ]
\nonum \\
&+&
(\lamu{l})_{jk} \biggl [4 (\yud \hu)_{li}+2\Tr(\lamud{l}\lamhu{i})\biggr ]
\nonum \\
&+&
(\lamu{i})_{kl} \biggl [ -4(\ydd\hd)_{lj}-4\Tr(\lamdd{l}\lamhd{j})-
4(\lamud{n}\lamhu{n})_{lj}\biggr]
\nonum \\
&-&
(\lamhu{i})_{jk} \biggl [\frac{4}{5}g_1^2+8g_3^2\biggr ]+
(\lamu{i})_{jk} \biggl [\frac{8}{5}g_1^2M_1+16g_3^2M_3 \biggr ].
\end{eqnarray}
The RGEs for the gaugino masses are not affected by the
\rpv-couplings up to 1-loop. The RGEs for the SUSY soft breaking
masses are given by,
\begin{eqnarray}
16\pi^2 \frac{d \me_{ij}}{dt} \ &=& \ 2 (\yed \ye)_{in} \me_{nj} +
\Tr (\lamed{i}\lame{n})\me_{nj} \nonum \\
&+& 2 \me_{in} (\yed \ye)_{nj} + \me_{in}\Tr(\lamed{n}\lame{j}) \nonum \\
&+& 4(\yed\ye)_{ij}\mh{1} + 4(\yed \lame{j})_{ir} \mhl{r} \nonum \\
&+& 4 \Tr[\ml \lamed{i}\lame{j}]+4\mlh{q} (\lamed{i}\ye)_{qj} \nonum \\
&+& 4 [\yed \ml \ye]_{ij}+4(\hed\he)_{ij}+2\Tr(\lamhed{i}\lamhe{j})\nonum \\
&-& \biggl (\frac{24}{5}|M_1|^2 g_1^2-\frac{6}{5}g_1^2 {\cal S} \biggr )
\delta_{ij}\;, \\[4mm]
16\pi^2 \frac{d \ml_{ij}}{dt} \ &=& \
\ml_{in}(\ye\yed)_{nj}-\mlh{i}(\lames{q}\ye)_{jq} \nonum \\
&+& \ml_{in} (\lame{q}\lamed{q})_{nj}+(\ye\yed)_{in}\ml_{nj} \nonum \\
&-& (\lame{q}\yes)_{iq}\mhl{j}+(\lame{q}\lamed{q})_{in}\ml_{nj} \nonum \\
&+& 2(\ye\yed)_{ij}\mh{1}+2(\lame{p})_{ir}(\yed)_{pj}\mhl{r} \nonum \\
&-& 3(\lamd{q}\yds)_{iq}\mhl{j}
+3(\ml\lamd{q}\lamdd{q})_{ij} \nonum \\
&-& 3(\lamds{q}\yd)_{jq}\mlh{i}+3(\lamd{q}\lamdd{q}\ml)_{ij} \nonum \\
&+& 2(\ye)_{ip}(\lamed{p})_{qj}\mlh{q}+
2 (\lame{p})_{ir}\ml_{qr}(\lamed{p})_{qj}
\nonum \\
&+& 2 (\ye)_{ir}\me_{rq}(\yed)_{qj}+2(\lame{r})_{ip}\me_{rq}(\lamed{q})_{pj}
\nonum \\
&+& 6 (\lamd{r})_{ip}\md_{rq} (\lamdd{q})_{pj}+
6(\lamd{k})_{il}\mq_{ml}(\lamdd{k})_{mj} \nonum \\
&+&
2(\he\hed)_{ij}+2(\lamhe{q}\lamhed{q})_{ij}+6(\lamhd{q}\lamhdd{q})_{ij}
 \nonum \\
&-&\biggl (\frac{6}{5}g_1^2 |M_1|^2 + 6 g_2^2|M_2|^2 +
\frac{3}{5}g_1^2 {\cal S} \biggr )\delta_{ij} \;,
\\[4mm]
16\pi^2 \frac{d\mhl{i}}{dt} \ &=& \ 
(\lamed{q}\ye)_{iq}\mh{1}+(\lamed{q}\lame{q})_{in}\mhl{n} \nonum \\
&-& \mhl{n}(\ye\yed)_{ni}-3(\lamds{q}\yd)_{iq}\mh{1} \nonum \\
&+&3(\lamd{q}\lamdd{q})_{ni}\mhl{n}+ [ \Tr(\ye\yed)+3\Tr(\ydd\yd)]\mhl{i}
\nonum \\
&+& \ml_{ni} ( \lamed{q}\ye-3\lamds{q}\yd)_{nq} + 2(\lamed{p})_{iq}
\ml_{qr}(\ye)_{rp} \nonum \\
&+& 2 (\lamed{q}\ye)_{ir}\me_{rq}-6(\lamds{q}\yd)_{ir}\md_{rq}\nonum \\
&-& 6(\lamds{p})_{iq}\mq_{qr}(\yd)_{rp} - 
[ 2\lamhe{q}^*\he+6\lamhd{q}^*\hd ]_{iq} \;,
\\[4mm]
16\pi^2\frac{d \mq_{ij}}{dt} \ &=& \ [\yd\ydd+\yu\yud]_{nj}\mq_{in} 
+(\lamdd{q}\lamd{q})_{jn}\mq_{in}\nonum \\
&+&[\yd\ydd+\yu\yud]_{in}\mq_{nj}+\mq_{nj}(\lamdd{q}\lamd{q})_{ni} \nonum \\
&+&2(\yd)_{ir}\md_{rq}(\ydd)_{qj}+2(\yd\ydd)_{ij}\mh{1} \nonum \\
&+&2(\lamdd{p})_{jq}\ml_{qr}(\lamd{p})_{ri}+2(\yu)_{ir}\mup_{rq}(\yud)_{qj}
\nonum \\
&+&2(\yu\yud)_{ij}\mh{2}+2(\lamdd{q}\lamd{r})_{ji}\md_{rq} \nonum \\
&-& 2 \mhl{l} (\lamd{k})_{li}(\ydd)_{kj}
-2 (\yd)_{iq} (\lamdd{q})_{jk}\mlh{k} \nonum \\
&+&2[\hd\hdd+\hu\hud]_{ij}+2(\lamhdd{q}\lamhd{q})_{ji} \nonum \\
&-&\biggl (\frac{2}{15}g_1^2|M_1|^2+6g_2^2|M_2|^2+\frac{32}{3}g_3^2|M_3|^2-
\frac{1}{5}g_1^2 {\cal S} \biggr )\delta_{ij} \;,
\\[4mm]
16\pi^2\frac{d \md_{ij}}{dt} \ &=& 2(\ydd\yd)_{in}\md_{nj}
+2\Tr(\lamdd{i}\lamd{n})\md_{nj} \nonum \\
&+&2(\lamu{q}\lamud{q})_{ni}\md_{nj}+2(\ydd\yd)_{nj}\md_{in}\nonum \\
&+&2\Tr(\lamd{j}\lamdd{n})\md_{in}+2(\lamu{p}\lamud{p})_{jn}\md_{in}\nonum \\
&+&4(\ydd)_{iq}\mq_{qr}(\yd)_{rj}+4(\ydd)_{ip}(\yd)_{pj}\mh{1}\nonum \\
&+&4(\lamd{j}\lamdd{i})_{rq}\ml_{qr}+4(\lamdd{i}\lamd{j})_{qr}\mq_{qr}\nonum \\
&+&4(\lamud{p})_{qi}(\lamu{p})_{jr}\md_{rq}
+4(\lamu{l}\lamud{q})_{ji}\mup_{lq}\nonum \\
&+&4(\hdd\hd)_{ij}+4\Tr(\lamhdd{i}\lamhd{j})
+4(\lamhu{p}\lamhud{p})_{ji} \nonum \\
&-&4 (\lamds{i}\yd)_{lj}\mhl{l}-4(\lamd{j}\yds)_{li}\mhl{l} \nonum \\
&-&\biggl (\frac{8}{15}g_1^2|M_1|^2+\frac{32}{3}g_3^2|M_3|^2
-\frac{2}{5}g_1^2{\cal S} \biggr )\delta_{ij} \;,
\\[4mm]
16\pi^2\frac{d \mup_{ij}}{dt} \ &=& \
2(\yud\yu)_{in}\mup_{nj}+\Tr(\lamud{i}\lamu{n})\mup_{nj} \nonum \\
&+&2(\yud\yu)_{nj}\mup_{in}+\Tr(\lamu{j}\lamud{n})\mup_{in}\nonum \\
&+&4(\yud)_{iq}\mq_{qr}(\yu)_{rj}+4(\yud\yu)_{ij}\mh{2}\nonum \\
&+&4(\lamud{i}\lamu{j})_{qr}\md_{rq}+4(\hud\hu)_{ij}+2\Tr(\lamhud{i}
\lamhu{j})\nonum \\
&-&\biggl (\frac{32}{15}g_1^2|M_1|^2+\frac{32}{3}g_3^2|M_3|^2
+\frac{4}{5}g_1^2{\cal S} \biggr )\delta_{ij} \;,
\\[4mm]
16\pi^2\frac{d\mh{1}}{dt} \ &=& \
2\Tr(\yed\ye)\mh{1}+6\Tr(\ydd\yd)\mh{1}+(\yed\lame{q})_{qn}\mhl{n} \nonum \\
&-&3(\lamd{k}\yds)_{qk}\mhl{q}-3(\lamds{k}\yd)_{qk}\mlh{q} \nonum \\
&+&(\lamed{q}\ye)_{nq}\mlh{n}+2(\yed\ye)_{qr}\me_{rq} \nonum \\
&+&2(\ye\yed)_{rq}\ml_{qr}+6(\ydd\yd)_{qr}\md_{rq}\nonum \\
&+&6(\yd\ydd)_{rq}\mq_{qr}
+2\Tr(\hed\he)+6\Tr(\hdd\hd) \nonum \\
&-&\biggl (\frac{6}{5}g_1^2|M_1|^2+6g_2^2|M_2|^2+\frac{3}{5}g_1^2{\cal S}
\biggr ) \;,\label{mh1}
\\[4mm]
16\pi^2\frac{d\mh{2}}{dt} \ &=& \
6\Tr(\yud\yu)\mh{2}+6(\yud\yu)_{qr}\mup_{rq} \nonum \\
&+& 6(\yu\yud)_{rq}\mq_{qr}+6\Tr(\hud\hu) \nonum \\
&-&\biggl (\frac{6}{5}g_1^2|M_1|^2+6g_2^2|M_2|^2-\frac{3}{5}g_1^2{\cal S}
\biggr ) \;,\label{mh2}
\end{eqnarray} 
where 
\begin{eqnarray}
\mhl{i}=\mlh{i}^* \;,
\end{eqnarray}
and 
\begin{eqnarray}
{\cal S} = m_{H_2}^2-m_{H_1}^2+\Tr[{\bf m_{\tilde{Q}}}^2-
{\bf m_{\tilde{L}}}^2-2{\bf m_{\tilde{U}}}^2+{\bf m_{\tilde{D}}}^2
+{\bf m_{\tilde{E}}}^2] \;.
\end{eqnarray}




\section{Four-Body $\stau$-Decay}
\label{app-stau}

In this Appendix we compute the four-body decay ${\tilde\tau}^-\ra
\tau^- \mu^+{\bar u}d$ via the \rpv-operator $\lam'L_\mu Q_1{\bar D}
_1$. The relevant Feynman diagrams are given in
Fig.~\ref{intvertices}\,a,b,c. We neglect the contributions from the
heavier neutralinos. Using the notation of Ref.~\cite{Peter} the three
amplitudes corresponding to Fig.~\ref{intvertices} are given by:
\begin{eqnarray}
{\cal M}_a&=& + \frac{2i\lam' b_{\tilde \mu}}
{({\tilde \mu}^2-m^2_{\tilde \mu})(\chi^2-M_\chi^2)}
\left({\bar d}P_L u\right) 
\biggl \{ {\bar\tau} [a_{\tau}P_L+b_{\tau}P_R ]
({\not\!\chi} +M_\chi) P_L \mu \biggr  \} \\
{\cal M}_b&=& -\frac{2i\lam' b_{\tilde u}}{({\tilde u}^2-
m^2_{\tilde\mu})(\chi^2-M_\chi^2)} \left({\bar d}P_L\mu
\right) \biggl \{ {\bar\tau} \{a_{\tau}P_L+b_{\tau}P_R\}
({\not\!\chi} +M_\chi) P_L u\biggr \} \\
{\cal M}_c&=& -\frac{2i\lam' a_{\tilde d}}{({\tilde d}^2-m^2_{\tilde\mu})
(\chi^2-M_\chi^2)} \left({\bar u}P_L\mu\right) 
\biggl \{ {\bar\tau} \{a_{\tau}P_L+b_{\tau}P_R\}
({\not\!\chi} +M_\chi) P_Ld\biggr \} 
\end{eqnarray}
\begin{figure}[t!]
\centerline{\psfig{figure=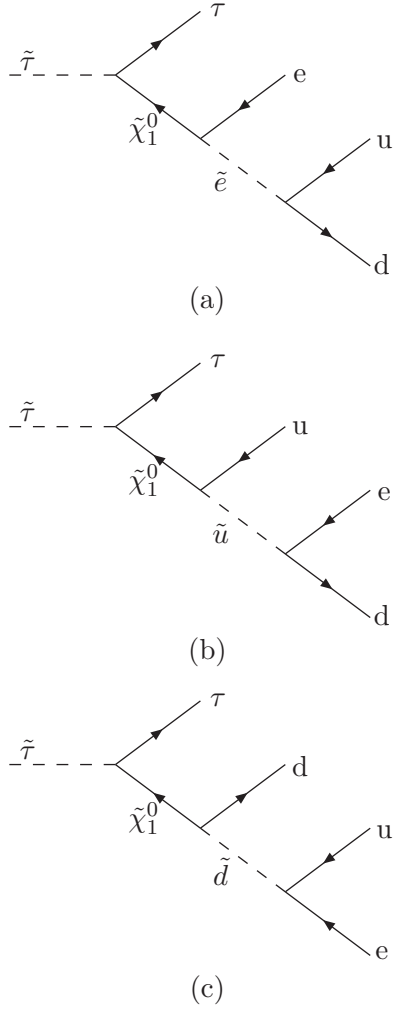}}
\caption{Feynman Diagrams for the decay ${\tilde\tau}\ra 
\tau ({\tilde\chi}^0_1)^*
\ra \tau (\mu u d)$ via  the operator $L_\mu Q_1{\bar D}_1$. }
\label{intvertices}
\end{figure}
Here the four-momenta are denoted by the particle symbol. The momenta
${\tilde \mu},\,{\tilde u},\,{\tilde d},\chi$ flow along the
corresponding propagators from left to right. We use below that
$\chi=\mu+u+d$ as well as the notation $N_p=p^2-m_p^2$ for the
denominators in the propagators.  We have assumed there is no mixing
in the scalar $\mu,\,u,$ and $d$ sectors. However, we allow for mixing
in the stau sector. ${\tilde \mu}_L$, ${\tilde u}_L$, ${\tilde d}_R$
are the only sparticles that couple to the R-parity violating
operators. The coupling constants are given by
\cite{Kane,Haber,Peter}
\begin{eqnarray}
a_\tau&=& L_{21}^\tau \left(eN_{11}^{\prime\ast}-
\frac{g\sin^2\theta_{\rm w}N_{12}^{\prime\ast}}
{\cos\theta_{\rm w}}\right)\,,\qquad 
b_\tau \;=\; -L^\tau_{11}\left(
eN_{11}'+\frac{gN_{12}'(\frac{1}{2}-\sin^2\theta_{\rm w})}
{\cos\theta_{\rm w}}\right)\,,\\
b_\mu&=& -eN_{11}'-\frac{gN_{12}'(\frac{1}{2}-\sin^2\theta_{\rm w})}
{\cos\theta_{\rm w}}\,,\qquad 
b_u\;=\;   ee_uN_{11}'+\frac{gN_{12}'(\frac{1}{2}-e_u\sin^2\theta_{\rm w})}
{\cos\theta_{\rm w}}\,,\\
a_d &=& -ee_dN_{11}^{\prime\ast}+\frac{ge_d\sin^2\theta_
{\rm w}N_{12}^{\prime\ast}} {\cos\theta_{\rm w}}\,.
\end{eqnarray}
The total matrix element squared is given by
\begin{eqnarray}
|{\cal M}|^2&=& N_c \left[
\: |{\cal M}_a|^2+|{\cal M}_a|^2+|{\cal M}_a|^2+ 
2 \: \Re  \left( {\cal M}_a{\cal M}_b^\dagger+{\cal M}_a
{\cal M}_c^\dagger+{\cal M}_b{\cal M}_c^\dagger \right) \right]\,,
\label{amp-sq}
\end{eqnarray}
where $N_c=3$ is the colour factor and 
\begin{eqnarray}
|{\cal M}_a|^2 &=& \frac{16\lam^{\prime 2}|b_\mu|^2}
{N_\chi^2N_{\tilde\mu}^2} \; d\!\cdot \!u 
\left[|a_\tau|^2 M_{\chi}^2  \tau\!\cdot\!\mu
\,+\,|b_\tau|^2 g(\tau,\chi,\mu,\chi)\right]\,,
\\
|{\cal M}_b|^2 &=& \frac{16\lam^{\prime 2}|b_u|^2}
{N_\chi^2N_{\tilde u}^2}\; d\!\cdot\! \mu
\left[|a_\tau|^2 M_{\chi}^2 \tau\!\cdot\! u
\,+\,|b_\tau|^2 g(\tau,\chi,u,\chi)
\right]\,,\\
|{\cal M}_c|^2 &=& \frac{16\lam^{\prime 2}|a_d|^2}
{N_\chi^2N_{\tilde d}^2}\; u\!\cdot\!\mu\left[|a_\tau|^2M_{\chi}^2 
\tau\!\cdot\! d
\,+\,|b_\tau|^2g(\tau,\chi,d,\chi)\right]\,, \\
2 \Re \left( {\cal M}_a{\cal M}_b^\dagger\right) &=& 
-\frac{16\lam^{\prime 2}b_\mu b_u^\ast}{N_\chi^2N_{\tilde\mu}
N_{\tilde u}}\left[ |a_\tau|^2 M_\chi^2 g(\tau,\mu,d,u) 
\,  + \, |b_\tau|^2 f(\tau,\chi,\mu,d,u,\chi) \right]\,,\\
2 \Re \left( {\cal M}_a{\cal M}_c^\dagger\right) &=& 
\frac{16\lam^{\prime 2}b_\mu a_d^\ast}
{N_\chi^2N_{\tilde\mu}N_{\tilde d}}\left[ |a_\tau|^2 M_\chi^2
g(\tau,\mu,u,d) 
\,+ \, |b_\tau|^2 f(\tau,\chi,\mu,u,d,\chi) \right]\,,\\
2 \Re \left( {\cal M}_b{\cal M}_c^\dagger\right) &=& 
\frac{16\lam^{\prime 2}b_ua_d^\ast}
{N_\chi^2N_{\tilde u}N_{\tilde d}}\left[|a_\tau|^2 M_\chi^2
g(\tau,u,\mu,d)  \, + \, |b_\tau|^2 f(\tau,\chi,u,\mu,d,\chi) \right] \,.
\end{eqnarray}
The functions are given by 
\begin{eqnarray}
g(a,b,c,d)&=& a\!\cdot\! b\; c\!\cdot\! d-
a\!\cdot\! c \; b\!\cdot\! d  + a\!\cdot\! d\; b\!\cdot\! c\,,\qquad
f(\tau,\chi,a,b,c,\chi) = -\chi^2 g(\tau,a,b,c)
 \: +\: 2\: \tau\!\cdot\!\chi \;g(\chi,a,b,c)\,.
\end{eqnarray}
The squared amplitude in Eq.(\ref{amp-sq}) can be used in Monte Carlo
simulation programs to generate events with a decaying stau. We are
here interested in an analytic approximation for the total decay
width. To this end, we shall assume $\chi^2\ll M_\chi^2\,$.  This is
equivalent above to setting $b_\tau=0$.  Furthermore we assume that
all scalar propagators are dominated by their mass terms and the
scalar fermion mass is universal: $m_{\tilde \mu}= m_{\tilde u}=m_
{\tilde d}\equiv{\tilde m}$. In this simplified case the amplitude
squared is given by
\begin{eqnarray}
|\mathcal{M}|^2&=&\frac{16\lam^{\prime 2}|a_\tau|^2N_c}{M_\chi^2{\tilde m}^4}
\left[|b_\mu|^2\; d\!\cdot\! u\, \tau\!\cdot\!\mu+ |b_u|^2\; d\!\cdot\!\mu\, 
\tau\!\cdot\!u +\, |a_d|^2\; u\!\cdot\!\mu\, \tau\!\cdot\!d \right. 
\nonumber \\
&&\left. - b_\mu b_u^\ast\; 
g(\tau,\mu,d,u)+\, b_\mu a_d^\ast \;g(\tau,\mu,u,d) 
+b_ua_d^\ast\; g(\tau,u,\mu,d) \right]\,.
\nonumber \\ \label{int}
\end{eqnarray}
The total width is given by \cite{Howie,Singh}
\begin{equation}
\Gamma=\frac{(2\pi)^{-8}}{2M_{\tilde\tau}}\int \prod_{i=1}^4 
\frac{d^3k_i}{2E_i} \; \delta^4({\tilde\tau}-k_1-k_2-k_3-k_4) \;|{\cal M}|^2 \,,
\end{equation}
where $k_1=\tau,\,k_2=\mu,\,k_3=u,\,k_4=d$. After the 
simplification our matrix element squared consists of three kinds of
terms which depend on the final state four-momenta: $(\tau\cdot \mu)
(u\cdot d),\,\;(\tau\cdot u)(\mu\cdot d),$ and $(\tau\cdot d)(\mu\cdot
u)$. As can be seen from the phase space integral, these all contribute
the same, they simply correspond to a relabeling. We thus explicitly
integrate only the first term. Using Eq (4) from Ref.~\cite{Howie} with
$N={\tilde\tau}-k_1-k_2$, we see that 
\begin{eqnarray}
\int\frac{d^3k_u}{2E_u}\frac{d^3k_d}{2E_d}(u\cdot d) \,
\delta^4(N-u-d) = 
\frac{\pi}{4} ({\tilde\tau}-\tau-\mu)^2\,,
\end{eqnarray}
and we thus obtain 
\begin{eqnarray}
{\rm A}_1\equiv \int \prod_{i=1}^4 \frac{d^3k_i}{2E_i}\; \delta^4(N-u-d)
\; (\tau\cdot\mu)\; (u\cdot d)=\frac{\pi}{4}\int\frac{d^3\tau}{2E_\tau}\int
\frac{d^3\mu}{2E_\mu}\; (\tau\cdot \mu)\;({\tilde\tau}-\tau-\mu)^2\,.
\end{eqnarray}
In the rest-frame of the decaying stau with the z-axis in the
direction of the 3-momentum of the $\tau$
\begin{eqnarray}
{\tilde\tau}&=& (M_{\tilde\tau},0,0,0)\,,\qquad 
\tau\;=\; E_\tau(1,0,0,1)\,, \qquad  
\mu\;=\; E_\mu(1,\sin\theta,0,\cos\theta)\,.
\end{eqnarray}
Performing the integrals over $d\Omega_\tau$ and $d\phi_\mu$
\begin{eqnarray}
{\rm A}_1&=& \frac{\pi^3}{2}\int\! dE_\tau \!\int\! dE_\mu \!\int\! d\cos\theta\;
E_\tau^2E_\mu^2(1-\cos\theta) 
\left[M_{\tilde\tau}^2-2M_{\tilde\tau}E_\tau-2M_{\tilde\tau}E_\mu
+2E_\mu E_\tau(1-\cos\theta)\right]\,.
\end{eqnarray}
It is convenient to change to dimensionless variables $E_\mu=
\frac{1}{2} M_{\tilde\tau} z,$ $E_\tau=\frac{1}{2}M_{\tilde\tau}y,$
and $1-\cos \theta=2w\,$~\cite{Howie}. Implementing the integral
boundaries given in Refs.~\cite{Singh,Howie}, this leads to the result
\begin{eqnarray}
{\rm A}_1&=&\frac{\pi^3M_{\tilde\tau}^8}{2^5}
\int_0^1 dz\left[\int_0^{1-z} \!\!\!\! 
dy\int_0^1 dw + \int_{1-z}^1\!\!\!\! 
dy\int_{(y+z-1)/yz}^1 dw\right]
 \left[ z^2y^2w(1-z-y+yzw)\right]
=\frac{\pi^3M_{\tilde\tau}^8}{2^5}\times \frac{1}{720}
\,.
\end{eqnarray}
We thus have for the total width
\begin{eqnarray}
\Gamma({\tilde\tau}^-\ra \tau^- \mu^+{\bar u}d)
=\frac{KN_c\lam^{\prime 2}|a_\tau|^2}
{{2^5\pi^5}M_\chi^2{\tilde m}^4} 
M_{\tilde\tau}^7 \left(|b_\mu|^2+|b_u|^2+|a_d|^2-b_\mu b_u^\ast
+b_\mu a_d^\ast+b_ua_d^\ast\right)\,,
\end{eqnarray}
where $K=1/(720 \times 2^5)=1/23040$.
\end{widetext}


\begin{thebibliography}{99}
\bibitem{foot1}   
  The cases of gauge-mediated and anomaly mediated supersymmetry
  breaking are not discussed here.
\bibitem{foot4}
{From Table~\ref{tab:operators}, it is
obvious that the $B_3=R_3 L_3$ model contains the infamous
non-renormalizable neutrino mass term $g LLH_2H_2/\Lam$ which gives
rise to neutrino masses of order of $\lsim 10^{-3}$ eV if the cut-off
$\Lam$ is bigger than $\sim 10^{16}$ GeV.  In deriving the bounds
on the \rpv couplings $\lam$ and $\lam'$ we shall assume that
neutrino masses arising from this operator are much smaller than 1 eV
. In any case, the addition of the term $LLH_2H_2$ is needed when
trying to make a detailed analysis of the neutrino spectrum.}.
\bibitem{correction} 
  We have corrected the sign error for the $\mu_\alpha$ parameter in
  Eq.~(3.2) of Ref.~\cite{Haber}. We would like to thank H. Haber for
  communication.
\bibitem{foot5}
When the SUSY GUT relation $M_1=\frac{5}{3}\tan^2\theta_WM_2$ is
dropped an arbitrarily light neutralino is still allowed by collider
experiments \cite{light-LSP,supernova}, although constrained by
cosmology \cite{cosmo,supernova}.
\bibitem{foot6}
{These have been computed in the basis where the sneutrino vev's
vanish.  This is sufficient for our order of magnitude estimate.}
\bibitem{footx}
{This is not completely
correct. Even in the case of no mixing one generates, in addition to
$\lam'_{122}(M_Z)=2.5\times 10^{-4}$ (see Table.\ref{tab:bounds}), the
couplings $\lam'_{133}(M_Z)=1.4\times 10^{-7}$, $\lam'_{111}(M_Z)=
1.7\times 10^{-10}$, $\lam_{122}(M_Z)=5.1\times 10^{-9}$, $\lam_{133}
(M_Z)=8.6\times 10^{-8}$. These couplings will induce all the other
vev's, \ie $v_2,v_3$ or soft SUSY breaking masses and couplings.
However, the additional couplings and vev's are so small that they can
safely be neglected.}
\bibitem{footy}
{All the numerical
values given in this paragraph have been evaluated with the code {\tt
SUITY} in \cite{SUITY}. The Higgs masses and the the vacuum
expectation value are given at the tree approximation.}
\bibitem{footz}
{The reader might have noticed that this value for the 
neutrino mass slightly violates the bound of Eq.(\ref{eq:wmap}),
although the input value for the \rpv-coupling is the same at the GUT
scale as in Table~\ref{tab:bounds}. However, here we have adopted a
no-scale scenario, and the bounds obtained in Table~\ref{tab:bounds}
were obtained within the SPS1 point and thus they are on the
conservative side.}
\bibitem{typo}  
  {A typo in $\gamma^{(1)E_j}_{E_i}$ of Ref.~\cite{add} has been
  corrected.}
\bibitem{gauginomasses}  
  For the gaugino masses a knowledge of the $\beta$-function of the
  gauge couplings is also neccecary. Since the \rpv-couplings do not
  affect the gaugino masses at one loop one can use the gaugino RGEs
  directly from Ref.~\cite{mv}.
\end{thebibliography}
\end{document}